# Local Pitch Angles of Galactic Structures in Nearby Galaxies with JWST: II. Dust Lanes and Gas Flows in Bar-Dominated Systems

Stella Faria-Costa[1]⋆, Irapuan Rodrigues[1,2] and Ivânio Puerari[3]
[1] Instituto de Pesquisa e Desenvolvimento, Universidade do Vale do Paraíba, Av. Shishima Hifumi - 2911, 12244-000, São José dos Campos, São Paulo, Brazil
[2] Laboratório Nacional de Astrofísica, Rua Estados Unidos 154, 37504-364, Itajubá, Minas Gerais, Brazil
[3] Instituto Nacional de Astrofísica, Óptica y Electrónica, Calle Luis Enrique Erro 1, 72840, Santa María Tonantzintla, Puebla, México

## ABSTRACT

We present a spatially resolved study of local pitch angles in six barred spiral galaxies observed by PHANGS–JWST, focusing on bar-dominated regions where dust lanes are the main structures. We combine a two-dimensional Fourier transform (2D-FT) with a circular-window correlation method to map pitch angles across multiple wavelengths and spatial scales, extending our previous analysis of spiral arms and interarm substructures in disk-dominated galaxies. The dust lanes show relatively large pitch angles, typically $|\alpha| \simeq 35°–70°$, well above those measured for spiral arms. In several galaxies, the pitch angle changes systematically along the dust lane, with smaller values in the inner regions and larger values farther along the bar, tracing variations in local curvature. The multi-scale analysis shows that larger windows recover the overall geometry of the dust lanes, whereas smaller windows reveal local irregularities and secondary structures. MIRI images generally provide the clearest view of the dust morphology and allow a more precise characterization of local pitch-angle variations, while NIRCam offers a complementary view of the underlying stellar distribution. Unlike spiral-arm pitch angles, which are commonly associated with density-wave patterns, the values measured here characterize the local orientation and curvature of dust lanes associated with bar-driven gas flows. Together with the results of our companion study of disk-dominated galaxies, these findings support a unified, scale-dependent framework in which local pitch angles serve as observational diagnostics of distinct dynamical regimes in nearby galaxies.

## 1 INTRODUCTION

In Faria-Costa et al. (2026), the first of this series, we introduced a methodology to measure local pitch angles in spiral galaxies using high-resolution observations from the *Physics at High Angular resolution in Nearby GalaxieS* survey with the *James Webb Space Telescope* (PHANGS–JWST) (Lee et al. 2023), focusing on SA and SAB systems in which large-scale spiral arms dominate the observed morphology. In that work, we demonstrated that spiral arms and interarm substructures (feathers) exhibit distinct pitch angle distributions, reflecting different physical processes operating across galactic disks. In particular, while the main spiral arms are well described by relatively small pitch angles ($|\alpha| \sim 15°–30°$), the feathers display systematically larger values, often exceeding 50°, highlighting their nature as secondary structures associated with local instabilities and gas dynamics. This paper therefore complements Faria-Costa et al. (2026) by extending the same methodology to a fundamentally different dynamical environment.

However, in strongly barred galaxies (SB), the central regions are not dominated by classical spiral arms, but by bar-driven structures such as dust lanes, nuclear rings, and gas streams. Among these, dust lanes are particularly prominent features that trace shocks and gas inflow along the gravitational potential of the bar (Athanassoula 1984; Knapen et al. 2002; Comerón et al. 2009). In this regime, the physical interpretation of the pitch angle differs from that in spiral arms: instead of tracing quasi-stationary density waves, the measured angles characterize the curvature of dust lanes associated with gas responding to the non-axisymmetric bar potential.

Dust lanes in barred galaxies exhibit distinct morphological classes, commonly described as *straight* and *curved* (Athanassoula 1984). Straight dust lanes are typically aligned with the bar major axis and are often associated with stronger bars, whereas curved dust lanes display a pronounced concavity toward the bar axis and are generally linked to weaker bars or different dynamical configurations (see Figure 1). In some systems, additional filamentary structures or gas streams can intersect the main dust lanes, further increasing the complexity of the observed morphology (e.g. Elmegreen et al. 2009).

These morphological differences are not merely geometric but reflect variations in the underlying gas dynamics and bar properties. As such, a quantitative characterization of dust lane geometry, in terms of local pitch angles, provides a direct way to probe the physical conditions governing gas flows in barred galaxies.

The geometry of dust lanes has been shown to correlate with fundamental bar properties, such as bar strength and pattern speed,

⋆ E-mail: stellafariacosta@gmail.com (SFC)

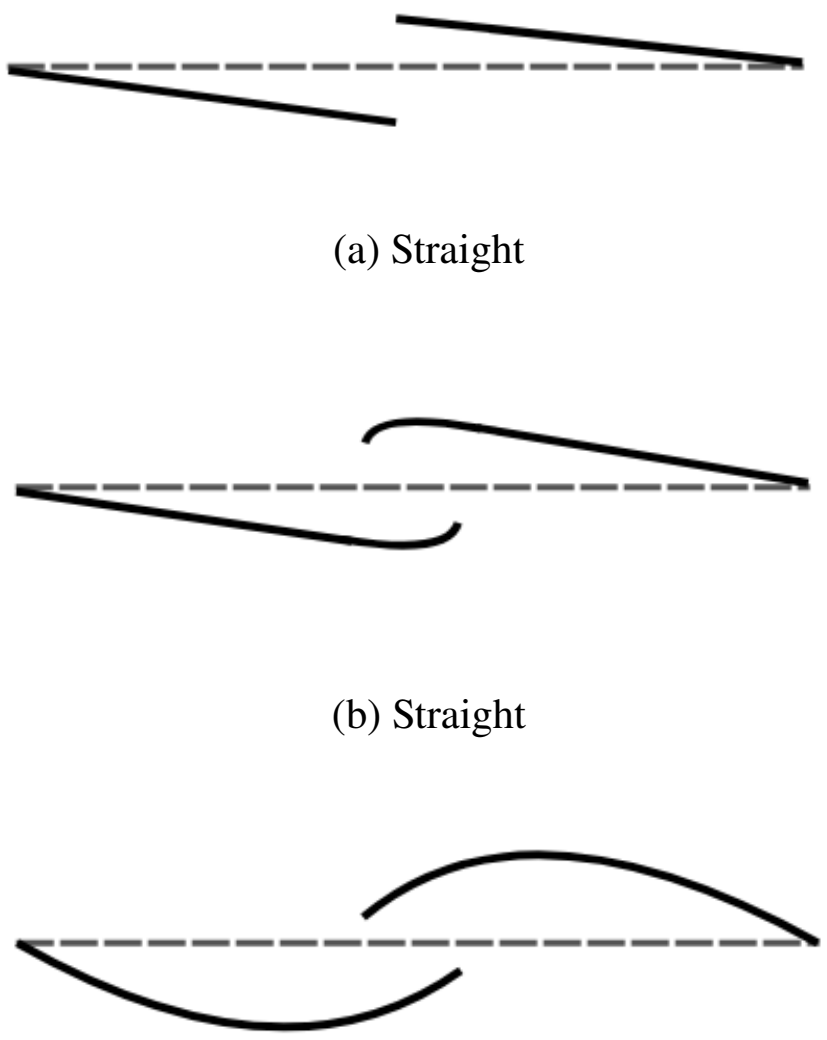

(a) Straight

(b) Straight

(c) Curved

**Figure 1.** Schematic representation of dust lane morphologies in barred galaxies. Straight dust lanes (upper and middle panels) are typically aligned with the bar major axis, while curved dust lanes (lower panel) exhibit concavity toward the bar axis. These morphological classes are associated with different dynamical regimes and bar properties (adapted from Athanassoula (1984)).

with straighter dust lanes typically associated with stronger bars and slower pattern speeds (Athanassoula 1984; Sánchez-Menguiano et al. 2015). Despite their importance as tracers of gas dynamics and angular momentum transport, a systematic, spatially resolved characterization of their curvature in terms of local pitch angles remains relatively unexplored, particularly at the spatial resolution provided by JWST.

The PHANGS–JWST survey provides an unprecedented view of nearby galaxies in the near- and mid-infrared, revealing complex dust and gas structures at sub-kiloparsec scales (Lee et al. 2023; Williams et al. 2024). The combination of NIRCam and MIRI imaging allows us to simultaneously trace the stellar distribution and the dust component, making it possible to investigate bar-driven structures with a level of detail not previously achievable.

In this paper, we extend the methodology developed in Faria-Costa et al. (2026) to a sample of barred spiral galaxies from the PHANGS–JWST survey. Unlike the galaxies analyzed previously, the systems considered here are dominated by their central bar, and the observable structures within the selected radial range correspond primarily to dust lanes rather than to large-scale spiral arms. As a consequence, the pitch angles measured in this work are interpreted as a proxy for the curvature of these dust lanes.

Our main goal is to quantify the local pitch angles of dust lanes in barred galaxies and to investigate how their curvature varies spatially and across different wavelengths. By comparing these results with those obtained for spiral arms and feathers in Faria-Costa et al. (2026), we aim to establish a unified, scale-dependent framework in which the pitch angle serves as a diagnostic of distinct dynamical regimes, ranging from density-wave-driven spiral structure to bar-driven gas flows.

This paper is organized as follows. In Section 2, we briefly summarize the data and preprocessing steps, following the procedure described in detail in Faria-Costa et al. (2026). Section 3 presents a representative full analysis together with the sample-wide results, while the complete diagnostic figure sets for the remaining galaxies are provided in Appendix A. In Section 4, we discuss the implications of our findings in the context of bar dynamics and compare them with the results obtained for spiral structures in Faria-Costa et al. (2026). Finally, Section 5 summarizes our main conclusions.

## 2 DATA PROCESSING AND PITCH ANGLE ANALYSIS

The analysis presented in this work follows the methodology introduced in Faria-Costa et al. (2026), where a detailed description of the data processing pipeline and pitch angle determination is provided. Here, we briefly summarize the main steps, emphasizing the aspects that are specific to barred galaxies.

In contrast to Faria-Costa et al. (2026), where the analysis focused on spiral arms and interarm substructures, the present study targets bar-dominated regions, where the relevant structures correspond primarily to dust lanes. As a consequence, the pitch angles measured here are interpreted as a proxy for the curvature of these features.

### 2.1 Data

As in Paper I, we used the publicly available PHANGS–JWST high-level science products (HLSP) retrieved from the Mikulski Archive for Space Telescopes (MAST). No independent reduction of the raw JWST observations was performed. The original data reduction and processing procedures are described by Williams et al. (2024).

The subsequent geometrical transformations and local pitch-angle measurements were performed using these reduced images, following the methodology described in Faria-Costa et al. (2026).

The galaxy sample analyzed in this work corresponds to the barred (SB) subset of the PHANGS–JWST survey, previously selected in Faria-Costa et al. (2026) from the original sample of 19 nearby spiral galaxies ($d < 20$ Mpc). Six galaxies were excluded due to limited spatial coverage or complex morphology that prevents a reliable application of the adopted methodology. The remaining galaxies were divided into two groups according to their morphological classification and the dominant structures within the observed field: SA/SAB galaxies, analyzed in Faria-Costa et al. (2026), and SB galaxies, which constitute the focus of the present study (see Figure 2, which summarizes the selection process).

In this work, we analyze six barred spiral galaxies (NGC 1300, NGC 1433, NGC 1512, NGC 2835, NGC 3351, and NGC 7496), whose available images predominantly cover the central regions dominated by the bar, with limited or no coverage of the large-scale spiral arms. This characteristic makes them particularly suitable for studying dust lanes and bar-driven structures. The main properties of the galaxies analyzed in this work are summarized in Table 1. The radial limits adopted in the analysis are illustrated in Figure 3, where the values of $R_{\rm min}$ and $R_{\rm max}$ are indicated over the galaxy images. These limits define the bar-dominated region used in the subsequent analysis.

Each galaxy was observed with the JWST using eight filters: four from the Near-Infrared Camera (NIRCam) and four from the Mid-Infrared Instrument (MIRI). The NIRCam $F200W$, $F300M$, and $F360M$ filters primarily trace photospheric stellar emission with relatively low extinction, with additional contributions from nebular emission and hot dust. The $F335M$ band also contains the 3.3 $\mu$m PAH feature. The MIRI filters ($F770W$, $F1000W$, $F1130W$, and $F2100W$) are dominated by PAH emission and/or dust continuum

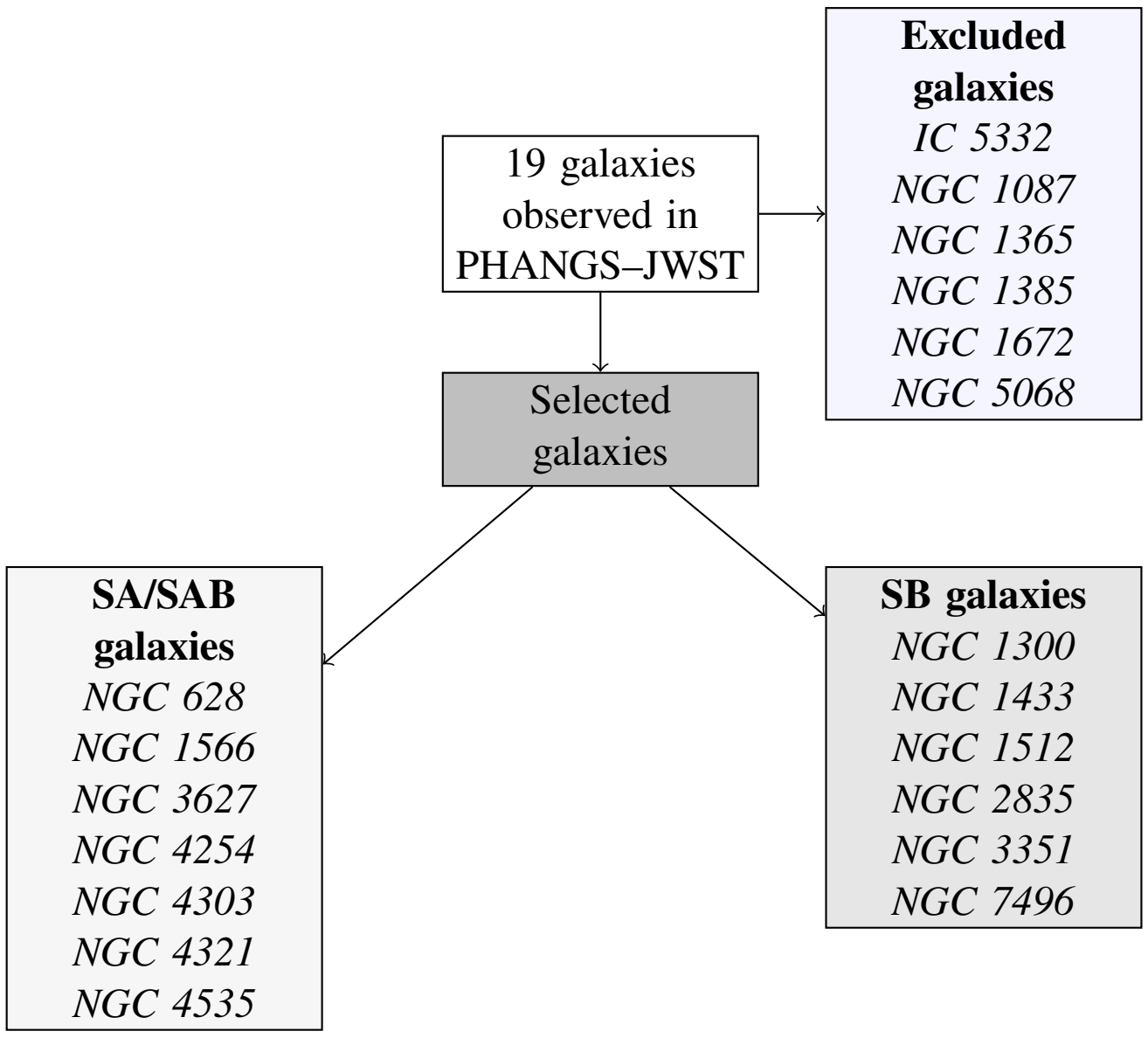

**Figure 2.** Flowchart of the selection and organization of galaxies analyzed from PHANGS–JWST images. Galaxies of types SA/SAB were examined in Part I of this study, while in the present work the galaxies of the SB group are analyzed.

**Table 1.** General properties of the galaxy sample. Columns: (1) Galaxy name; (2) morphological classification; (3) inclination ($i$); (4) position angle (PA); (5) distance ($d$); and (6) physical scale, calculated from the adopted distance.

| Galaxy | Morphology[a] | $i$[b] (deg) | PA[b] (deg) | $d$[c] (Mpc) | Scale (pc/arcsec) |
|---|---|---|---|---|---|
| NGC 1300 | SB(rs)bc | 32 | 278 | 18.99 | 92.07 |
| NGC 1433 | (R')SB(r)ab | 29 | 200 | 18.63 | 90.32 |
| NGC 1512 | SB(r)a | 43 | 262 | 18.83 | 91.29 |
| NGC 2835 | SB(rs)c | 41 | 1 | 12.22 | 59.24 |
| NGC 3351 | SB(r)b | 45 | 193 | 9.96 | 48.29 |
| NGC 7496 | SB(s)b | 36 | 194 | 18.72 | 90.76 |

[a] Morphological classification from the NASA/IPAC Extragalactic Database (NED).
[b] Inclination and position angle adopted from Lang et al. (2020).
[c] Distances adopted from Anand et al. (2021).

and are therefore particularly effective for tracing dusty structures (Lee et al. 2023).

The combination of NIRCam and MIRI data enables a multi-wavelength characterization of the galactic structures. While NIRCam images provide a clearer view of the underlying stellar distribution, MIRI observations reveal the morphology of dust lanes and associated substructures with higher contrast.

The pixel scales are $0.0630''$ pixel$^{-1}$ for NIRCam and $0.1109''$ pixel$^{-1}$ for MIRI, consistent with the values adopted in Faria-Costa et al. (2026).

This dataset provides an ideal basis for investigating the geometry of dust lanes in barred galaxies at high spatial resolution.

## 2.2 Image Preprocessing

The preprocessing steps follow the procedure described in detail in Faria-Costa et al. (2026) and include image deprojection, transformation to the $(\ln(R), \theta)$ coordinate system, and the definition of the radial range used in the analysis. Here, we briefly summarize these steps, emphasizing the aspects specific to barred galaxies.

### 2.2.1 Deprojection

The galaxy images were deprojected to correct for the inclination of the disks with respect to the plane of the sky, using the position angle (PA) and inclination angle ($i$) listed in Table 1. This transformation allows the structures to be analyzed as if the galaxies were observed face-on, minimizing projection effects that could distort their geometry.

### 2.2.2 The $(\ln(R/\mathrm{arcsec}), \theta)$ diagram

To facilitate the analysis of the structures and substructures, the images were transformed into the $(\ln(R/\mathrm{arcsec}), \theta)$ coordinate system. Throughout this work, logarithmic radii are expressed as $\ln(R/\mathrm{arcsec})$, corresponding to the natural logarithm of the numerical value of $R$ when the radius is measured in arcseconds.

In this representation, logarithmic spirals appear as straight lines, which makes the coordinate system particularly convenient for pitch-angle measurements. We stress, however, that the dust lanes analysed here are not assumed to be logarithmic spirals. The transformation is used because it provides the same geometrical framework adopted in Paper I and allows the local orientation of a dust lane to be compared with the family of pitch-angle filters within each circular window. The resulting pitch angle therefore represents a local measure of the orientation and curvature of the dust lane, rather than implying that the entire structure follows a global logarithmic spiral.

### 2.2.3 Radial range selection

For each galaxy, a minimum radius ($R_{\mathrm{min}}$) and a maximum radius ($R_{\mathrm{max}}$) were defined to delimit the region analyzed. In contrast to Faria-Costa et al. (2026), where $R_{\mathrm{min}}$ was chosen to exclude the bulge and isolate the spiral arms, here $R_{\mathrm{min}}$ corresponds to the onset of the bar, while $R_{\mathrm{max}}$ is defined by the extent of the available image coverage, as illustrated in Figure 3.

This radial selection ensures that the analysis focuses on the bar-dominated region, where dust lanes and bar-driven structures are the primary features, avoiding contamination from outer spiral arms or low signal-to-noise regions at the image boundaries.

This preprocessing framework provides a consistent basis for comparing structures across different galaxies and wavelengths, enabling a robust analysis of the local geometry of dust lanes.

## 2.3 Two-dimensional Fourier Transform Method

The two-dimensional Fourier transform (2D-FT) method is applied following the formalism described in detail in Faria-Costa et al. (2026). In brief, the deprojected galaxy images are represented in the $(\ln R, \theta)$ space and decomposed into Fourier modes, allowing the identification of dominant structural components.

This approach follows the classical Fourier decomposition of galactic structures in logarithmic polar coordinates, as originally developed by Considère & Athanassoula (1982) and later applied to spiral galaxies by Puerari & Dottori (1992).

From this decomposition, the Fourier coefficients $A(p, m)$ are computed, where $m$ represents the azimuthal mode and $p$ is related to the pitch angle $\alpha$. The value $p_{\mathrm{max}}$ corresponding to the maximum

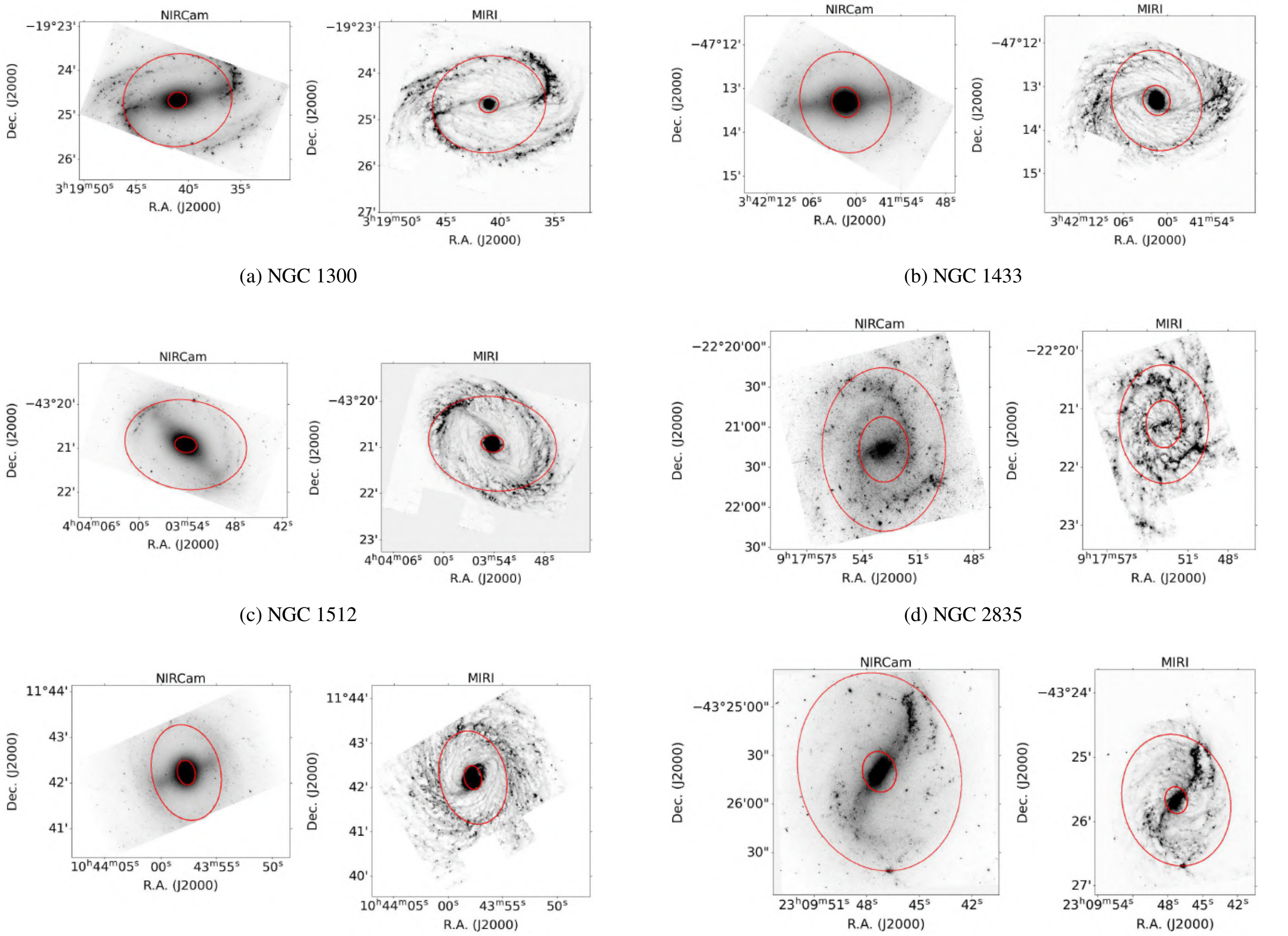

(a) NGC 1300 (b) NGC 1433

(c) NGC 1512 (d) NGC 2835

(e) NGC 3351 (f) NGC 7496

**Figure 3.** Observed $F335M$ (NIRCam) and $F1130W$ (MIRI) images of the selected galaxies. The solid red lines indicate the radial limits $R_{min}$ and $R_{max}$ considered in our analysis.

amplitude for a given mode is used to characterize the dominant structure.

In Faria-Costa et al. (2026), this approach was primarily used to identify spiral arm modes and directly estimate their pitch angles. In the present work, however, the role of the 2D-FT is different. Since the analyzed regions are dominated by the bar rather than by large-scale spiral arms, the Fourier decomposition is used mainly to reconstruct the image and generate masks that separate regions associated with dust lanes from the surrounding structures (see Figure 5).

The reconstructed images obtained from the inverse transform (2D-IFT) allow a systematic and objective definition of the regions used in the subsequent analysis. These masks distinguish between structures associated with the bar (referred to as "arm" regions in the formalism of Faria-Costa et al. (2026), but physically corresponding to dust lanes) and the surrounding interbar regions. Here, we adopt terminology that reflects the physical nature of the structures in barred galaxies.

The application of this formalism to infrared galaxy images has been further developed in works such as Puerari et al. (2014), which demonstrated its effectiveness in isolating dominant structural modes.

This use of the 2D-FT highlights that the same mathematical framework can be applied to different morphological regimes, with its physical interpretation depending on the nature of the structures under analysis.

## 2.4 Circular-Window Correlation Method

The local pitch angle is determined using the circular-window correlation method introduced in Faria-Costa et al. (2026). This technique measures the correlation between the galaxy image and a set of cosine filters with varying pitch angles and window sizes in the $(\ln(R), \theta)$ space. This method builds upon the correlation approach introduced by Puerari et al. (2014), extending it to a spatially resolved, multi-scale analysis of galactic structures.

Each filter is defined within a circular window centered at $(\ln(R_c), \theta_c)$, with a pitch angle $P_w$ and diameter $D_w$. The pitch angle $P_w$ was sampled from $-90°$ to $+90°$ in steps of $1°$, while the window diameter takes the dimensionless values $D_w = 0.25$, $0.50$, and $0.75$ in the $(\ln(R), \theta)$ plane. The three window diameters were adopted to probe different structural scales: the smaller window is more sensitive to localized variations and substructures, the larger window emphasizes coherent extended structures, and the intermediate value $D_w = 0.50$ provides sensitivity to both regimes. Examples of the cosine filters used in this analysis are shown in Figure 6, and their relative sizes with respect to the structures being measured are illustrated in Figure 7.

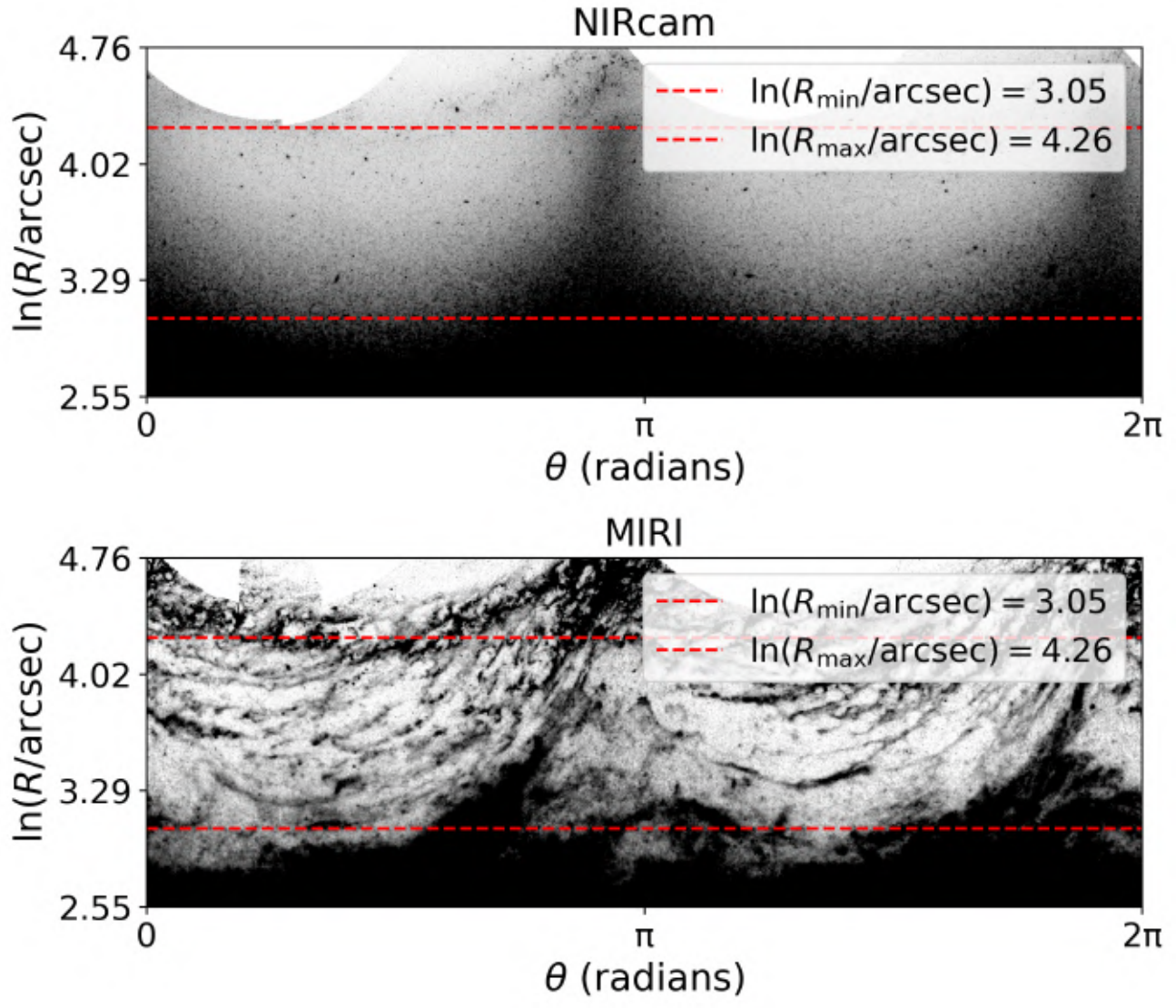

**Figure 4.** Diagrams $\ln(R/\text{arcsec})$ vs. $\theta$ for the $F335M$ (NIRCam) and $F1130W$ (MIRI) images of NGC 1433. The dashed red lines indicate the values of $\ln(R_{min}/\text{arcsec})$ and $\ln(R_{max}/\text{arcsec})$ adopted in the analysis.

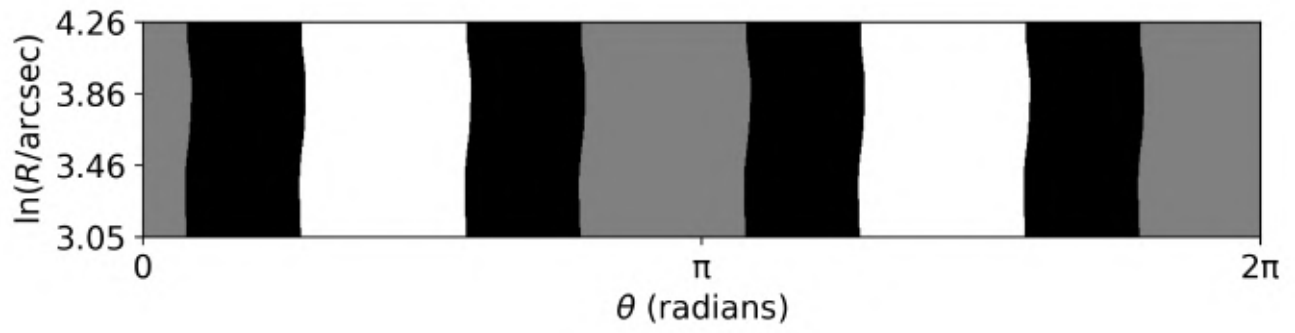

**Figure 5.** Mask generated from the 2D-FT and 2D-IFT for the definition of the bar and interbar regions of the galaxy NGC 1433. The gray area corresponds to the bar region, the white area represents the interbar region, and the black areas indicate the separations between these regions.

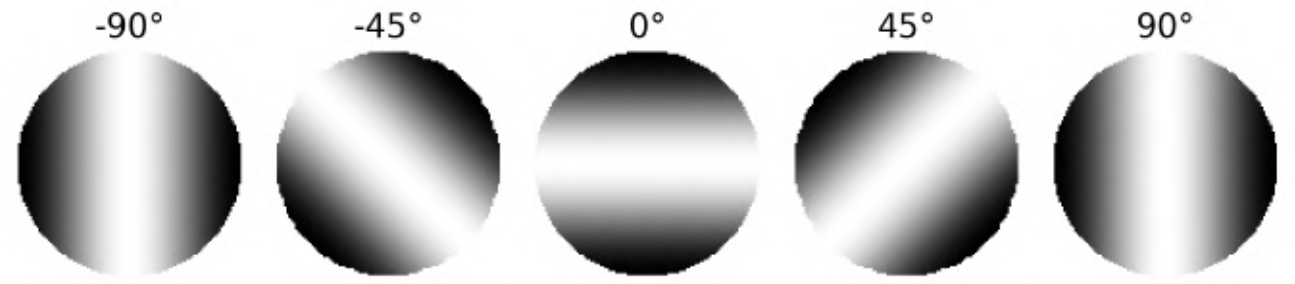

**Figure 6.** Cosine filters on circular windows with $P_w$ = -90°, -45°, 0°, 45° and 90°. Filters like these were used, with $P_w$ varying from -90° to 90° to calculate the correlation between the filter and the galaxy image at a position $(\ln(R), \theta)$.

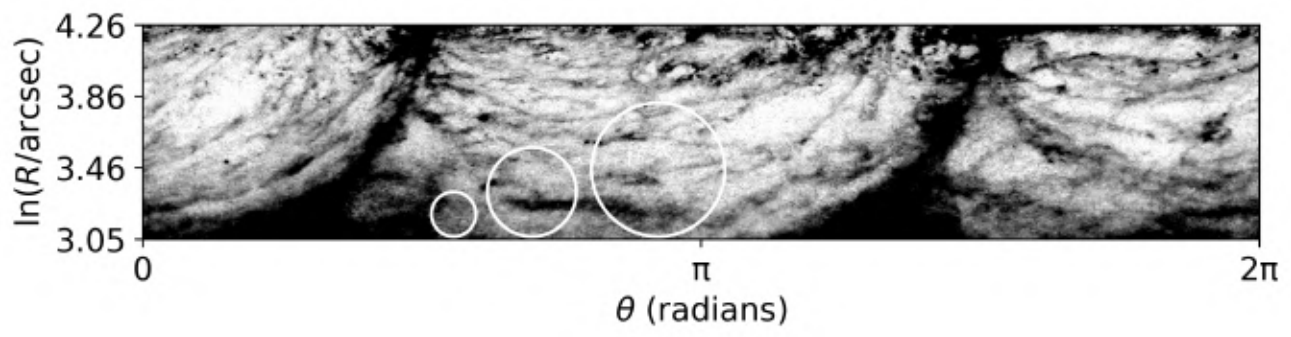

**Figure 7.** Image of NGC 1433 in the $F1130W$ filter, presented in the $\ln(R/\text{arcsec})$ vs. $\theta$ diagram, with three circles representing the different dimensionless window diameters used in the analysis: $D_w$ = 0.25, 0.5, and 0.75, shown from left to right.

At each position in the $(\ln(R), \theta)$ plane, the correlation between the image $I(\ln(R), \theta)$ and the filter $F(\ln(R_c), \theta_c, P_w, D_w)$ is computed as

$$C(\ln(R_c), \theta_c, P_w, D_w) = \sum_{i=1}^{N} F(\ln(R_c), \theta_c, P_w, D_w)\, I(\ln(R), \theta), \quad (1)$$

where the maximum correlation identifies the pitch angle that best matches the local structure. The pitch angle associated with this maximum is then assigned to that position, providing a spatially resolved map of local pitch angles.

The representative pitch-angle components quoted for each galaxy were identified by examining the summed-correlation spectra together with the spatial distribution of the corresponding local pitch angles. Candidate components were checked for recurrence across the filters and were retained when the associated pitch-angle values formed a spatially coherent pattern along the morphological structure of interest. When several nearby peaks were present, the corresponding pitch-angle intervals were inspected in the spatial maps to determine whether they traced distinct structures or local variations along the same dust lane. In the latter case, a representative value or range was adopted. No parametric profile fitting or formal statistical significance criterion was applied to the summed-correlation peaks; the final association between a spectral component and a morphological feature was based on visual inspection of the spectra together with the spatial maps.

In Faria-Costa et al. (2026), this method was used to characterize the geometry of spiral arms and interarm substructures such as feathers, where the measured pitch angles are directly related to logarithmic spiral patterns. In the present work, however, the interpretation is different. Since the analyzed regions are dominated by the bar, the structures traced by the correlation correspond primarily to dust lanes and gas flows rather than to classical spiral arms.

In this context, the pitch angle associated with the maximum correlation should be interpreted as a measure of the local orientation and curvature of the dust lanes. The scale dependence described above therefore allows the extended dust-lane geometry to be distinguished from localized variations and secondary structures within the bar-dominated region.

This scale-dependent approach allows us to characterize both the global shape and the local curvature of the dust lanes, providing a quantitative description of bar-driven structures. More generally, it illustrates how the same correlation framework can be applied to different dynamical regimes, with the physical meaning of the pitch angle depending on the nature of the structures being analyzed.

This interpretation links the measured pitch angles to the underlying gas dynamics, making them a proxy for the response of the interstellar medium to the bar potential.

## 3 RESULTS

The results for the barred-galaxy sample are presented in terms of the local pitch-angle distributions obtained with the circular-window correlation method. For each galaxy, we analysed eight JWST images (four NIRCam and four MIRI filters) and all three window diameters, $D_w$ = 0.25, 0.50, and 0.75. Following the procedure described in Section 2.4, the representative values discussed below correspond to spectral components whose spatial distributions are coherently associated with the dust lanes or other structures identified in the images.

The radial limits adopted for each galaxy are defined in the $(\ln(R/\mathrm{arcsec}), \theta)$ plane. The representative diagram for NGC 1433 is shown in Figure 4, while the corresponding diagrams for the remaining galaxies are collected in Appendix A. In the barred systems, $R_{\mathrm{min}}$ marks the onset of the bar-dominated region and $R_{\mathrm{max}}$ is set by the useful spatial coverage of the images. The bar (dust-lane) and interbar regions were defined using masks derived from the 2D-FT/2D-IFT reconstruction described in Section 2.

To avoid presenting the same sequence of diagnostic plots for every galaxy, we show the complete analysis for NGC 1433 as a representative example in the main text. The equivalent correlation spectra and local pitch-angle maps for NGC 1300, NGC 1512, NGC 2835, NGC 3351, and NGC 7496 are presented in Appendix A.

### 3.1 Representative example: NGC 1433

NGC 1433 illustrates the sequence of diagnostics used throughout the barred sample. Figure 8 shows the summed correlation spectra for the three window diameters, while Figures 9 and 10 show the corresponding spatial distributions of the locally selected pitch angles in the $F335M$ and $F1130W$ images. Candidate components in the spectra were cross-checked against these maps, and values near 43° and 58° were retained because their spatial distributions trace coherent portions of the dust lane. These values describe a relatively modest change in local orientation along the structure compared with several other galaxies in the sample (Figure 11).

### 3.2 Global behaviour of pitch angles

Across the sample, the dust lanes generally occupy a pitch-angle range distinct from that of the large-scale spiral arms analysed in Paper I. Representative values for the dust lanes are typically $|\alpha| \sim 35°$–$70°$, although the local angle can vary substantially within an individual galaxy. The quantitative values adopted for each system are summarized in Table 2.

A recurring behaviour is a change in local orientation along the dust lane. In several systems, smaller absolute pitch angles occur closer to the central region, whereas larger values characterize straighter segments farther along the bar. NGC 1300, for example, shows representative values of approximately 35° and 70° for different portions of its dust lanes, while NGC 3351 shows values around −46° and −62°. The complete diagnostic plots for these galaxies are shown in Appendix A.

NGC 2835 is the principal exception to this behaviour. Within the analysed region, spiral-like structures are present and the measured values (∼ 15° and ∼ 27°) are closer to the regime found for spiral arms in Paper I than to the large angles typical of the dust lanes in the other barred systems.

### 3.3 Scale-dependent structure

The three values of $D_w$ reveal structure on different spatial scales. The largest window, $D_w = 0.75$, produces smoother pitch-angle distributions and emphasizes the overall geometry of extended features, whereas $D_w = 0.25$ retains more localized variations and secondary structures. The intermediate window, $D_w = 0.50$, provides a transition between these regimes and often preserves both the overall geometry and part of the local structure. This behaviour can be seen

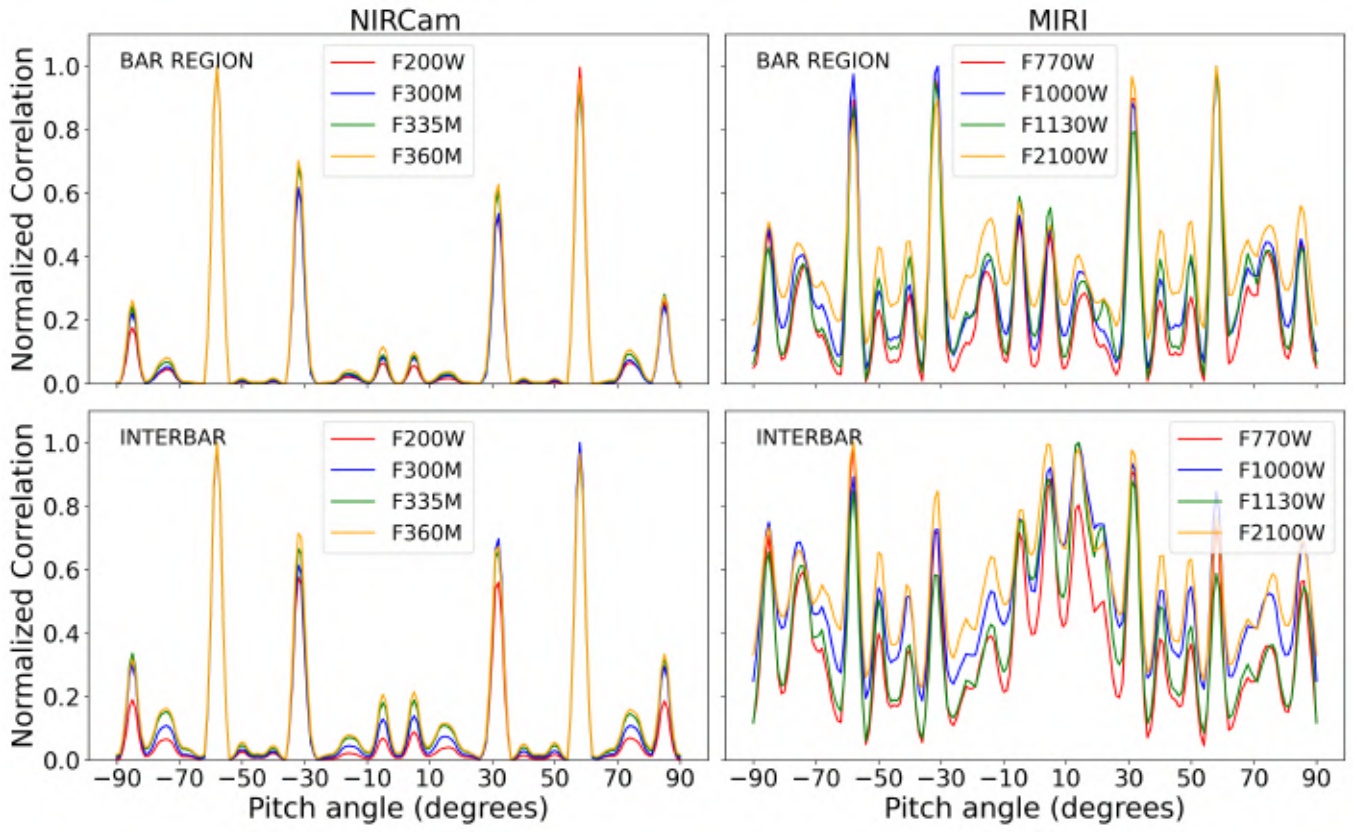

(a) Sum of the correlations calculated with $D_w = 0.25$ for the bar and interbar regions of NGC 1433.

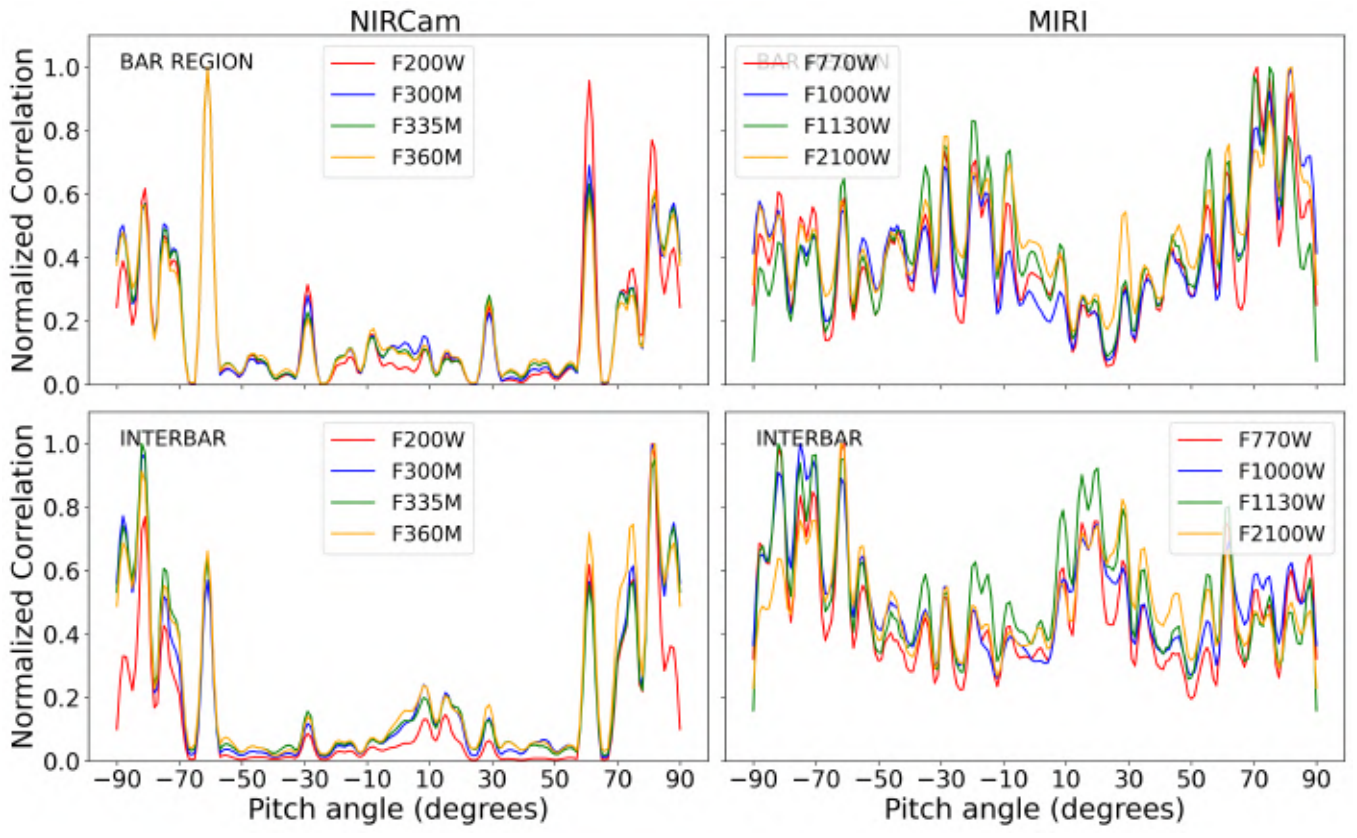

(b) Sum of the correlations calculated with $D_w = 0.5$ for the bar and interbar regions of NGC 1433.

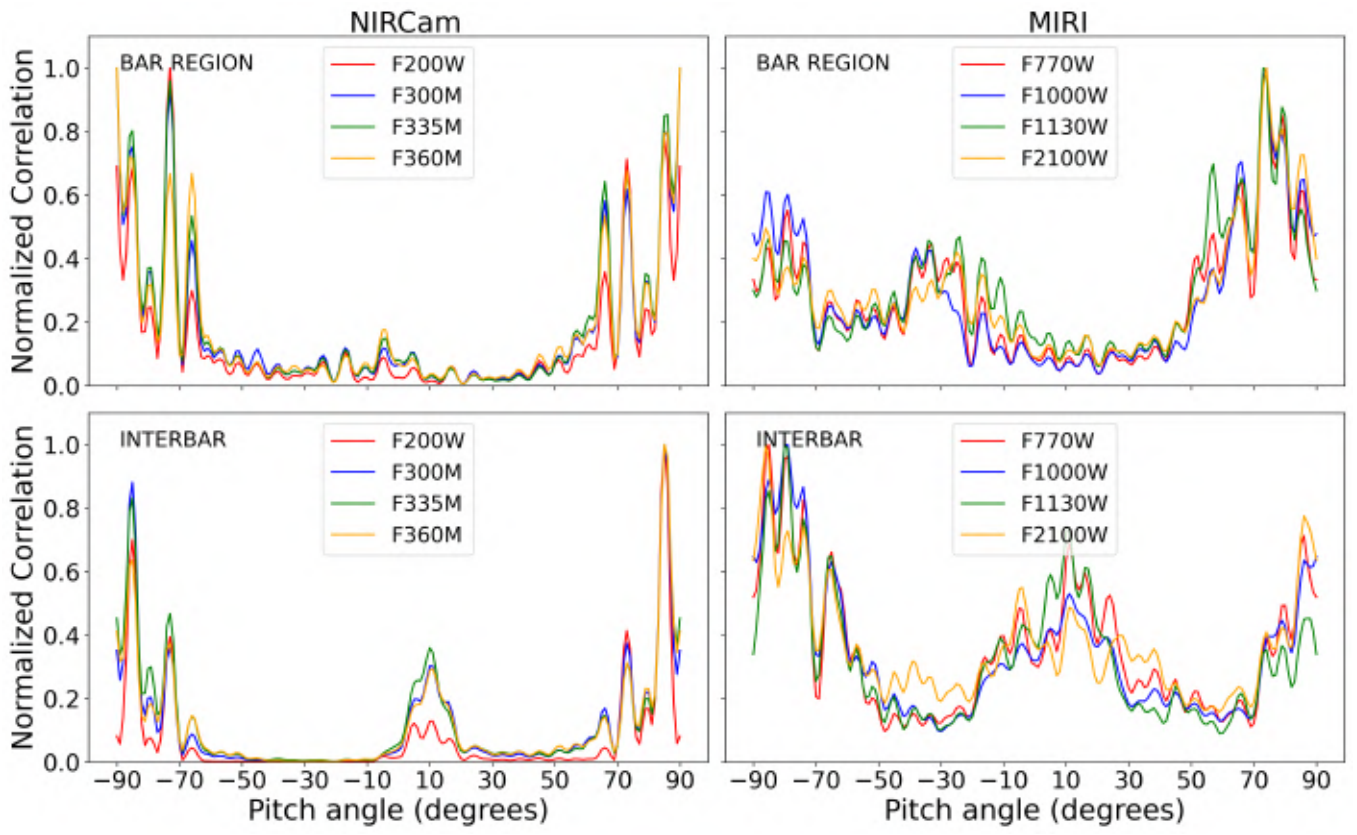

(c) Sum of the correlations calculated with $D_w = 0.75$ for the bar and interbar regions of NGC 1433.

**Figure 8.** Results for the sum of the correlations in the analyzed regions of NGC 1433. On the left, plots from NIRCam images with filters $F200W$ (red), $F300M$ (blue), $F335M$ (green), and $F360M$ (yellow), where the upper panel corresponds to the bar region and the lower panel to the interbar region. On the right, plots from MIRI images with filters $F770W$ (red), $F1000W$ (blue), $F1130W$ (green), and $F2100W$ (yellow), following the same division between bar (upper) and interbar (lower) regions.

**Table 2.** Representative pitch angles ($\alpha$) measured for the galaxies in the barred sample. The $\ln(R_{\rm min}/{\rm arcsec})$ and $\ln(R_{\rm max}/{\rm arcsec})$ values indicate the radial limits adopted in the analysis. The values summarize pitch-angle components identified from the summed-correlation spectra and confirmed by their spatial association with the corresponding structures in the local pitch-angle maps. For the bar-dominated systems, these angles describe the local orientation and curvature of the dust lanes; variations within a galaxy reflect changes in their geometry along the bar. NGC 2835 is listed separately in terms of its spiral-like structures.

| Galaxy | $\ln(R_{\rm min}/{\rm arcsec})$ | $\ln(R_{\rm max}/{\rm arcsec})$ | Dust lanes (°) | Main arms (°) | Notes |
|---|---|---|---|---|---|
| NGC 1300 | 2.55 | 4.31 | ∼ 35, ∼ 70 | – | Straight dust lanes; smaller angles near the bulge and larger ones toward the bar ends. |
| NGC 1433 | 3.05 | 4.26 | ∼ 43, ∼ 58 | – | Curved dust lane; representative values of ∼ 43° and ∼ 58° trace a modest variation in local orientation along the structure. |
| NGC 1512 | 2.7 | 4.43 | ∼ 45, ∼ 60 | – | Straight dust lane; ∼ 60° describes the main segment, while ∼ 45° traces an additional local component. |
| NGC 2835 | 3.2 | 4.12 | – | ∼ 15, ∼ 27 | The pitch angle of the main arms varies radially, with larger angles at inner radii and smaller angles at outer radii. |
| NGC 3351 | 2.8 | 4.16 | ∼ −46, ∼ −62 | – | Curved dust lanes; variation from −46° (inner) to −62° (outer) along the structure. |
| NGC 7496 | 2.55 | 4.13 | ∼ 48, ∼ 72 | – | Dust lane starts straight (∼ 72°) and becomes more curved (∼ 48°) at the onset of the spiral arms. |

for NGC 1433 in Figures 9 and 10, with equivalent plots for the rest of the sample in Appendix A.

The effect is particularly relevant for systems with more complex morphology. In NGC 1512, multiple pitch-angle components become more apparent at smaller window sizes, while NGC 7496 shows a broader distribution associated with several structures contributing within the bar-dominated region. These cases illustrate why a multi-scale analysis is useful for separating the overall dust-lane geometry from localized irregularities.

### 3.4 Comparison between NIRCam and MIRI

NIRCam and MIRI provide complementary views of the structures under analysis. The NIRCam images give a smoother representation of the underlying stellar distribution, whereas the MIRI images trace the dust morphology with higher contrast and reveal finer spatial variations, particularly at smaller window sizes.

For the representative case NGC 1433, this comparison can be made directly between Figures 9 and 10. The same comparison for the remaining galaxies is provided in Appendix A. Across the sample, the broad pitch-angle behaviour is generally consistent between the two instruments, while the relative prominence of individual components and the visibility of smaller-scale features vary with wavelength. The MIRI maps usually provide the clearest morphological association with the dusty structures.

### 3.5 Individual galaxies and notable features

The main quantitative results are summarized in Table 2; here we highlight the principal galaxy-to-galaxy differences. NGC 1300 shows a clear change from a more curved inner portion ($\alpha \sim 35°$) to a straighter segment along the bar ($\alpha \sim 70°$). NGC 1512 contains multiple components, with representative values near 45° and 60°, and its smaller-scale structure is especially apparent in the MIRI data. NGC 3351 shows a moderate variation along its curved dust lanes, from approximately −46° to −62°. NGC 7496 has a comparatively complex morphology, with values around 48° and 72° associated with different portions of the bar/spiral transition. The corresponding diagnostic figures are collected in Appendix A.

As noted above, NGC 2835 differs from the rest of the barred sample because spiral-like structures dominate the analysed region; its smaller pitch angles therefore provide an internal example of the transition between the bar-dominated and spiral-dominated morphological regimes considered across Papers I and II.

### 3.6 Summary of results

Overall, the barred sample shows that local pitch angles provide a scale-dependent description of dust-lane geometry. The dust lanes generally display larger and more spatially variable angles than the main spiral arms analysed in Paper I, while the use of multiple window sizes separates the overall geometry from localized variations. The NIRCam and MIRI measurements are complementary, with MIRI providing the clearest view of the dust morphology and smaller-scale structure.

## 4 DISCUSSION

### 4.1 Pitch angle as a diagnostic of bar-driven gas dynamics

The results presented in this work show that dust lanes in barred galaxies are characterized by relatively large and spatially variable pitch angles, typically in the range $|\alpha| \sim 35°$ to $70°$. These values are generally larger than those measured for spiral arms in Faria-Costa et al. (2026), indicating that the structures analysed here occupy a different morphological and dynamical context.

These values are summarized in Table 2, which compiles the dominant pitch angles identified for each galaxy. The table highlights both the systematically large pitch angles associated with dust lanes and the variations of $\alpha$ within individual systems, reflecting changes in curvature along the bar.

In barred galaxies, dust lanes are understood as tracers of shocks and gas inflow along the bar potential (Athanassoula 1984; Knapen et al. 2002; Comerón et al. 2009). In this context, the measured pitch angles do not correspond to logarithmic spiral patterns, but instead reflect the local orientation and curvature of gas streams responding to the non-axisymmetric gravitational field of the bar.

The observed variation of pitch angle along the dust lanes, with smaller values in the inner regions and larger values along the bar, is consistent with a gradual transition from strongly curved gas flows near the central regions to straighter trajectories aligned with the bar. This behaviour provides a quantitative morphological description of the dust-lane geometry associated with bar-driven gas flows.

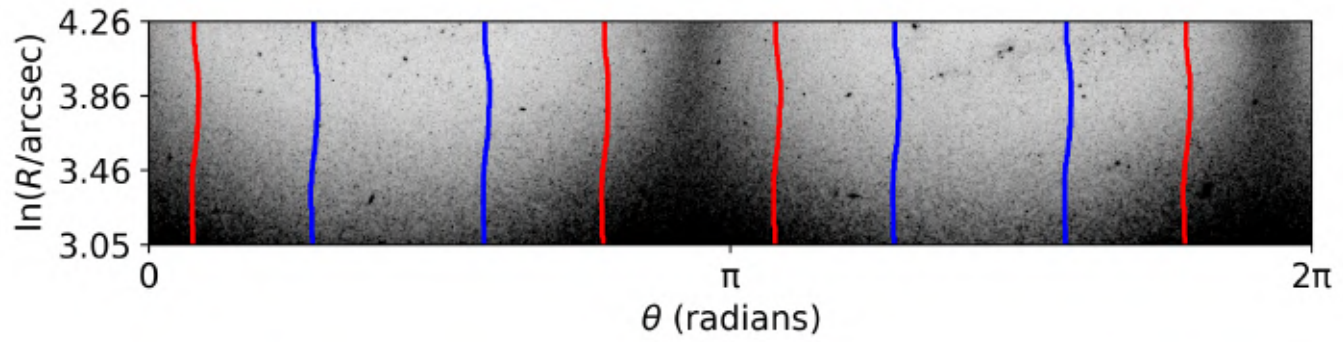

(a) Regions delimited in the $F335M$ image of the galaxy NGC 1433 from the 2D-FT/2D-IFT mask. The bar regions are located between the red lines, and the interbar regions are between the blue lines.

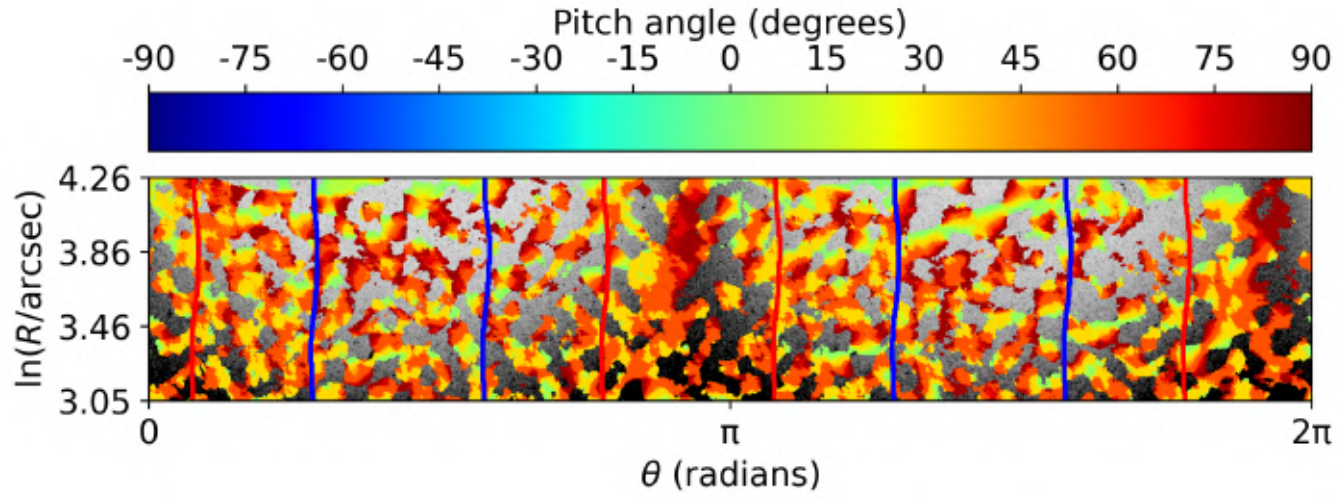

(b) Distribution of pitch angles obtained with the $D_w = 0.25$ filter.

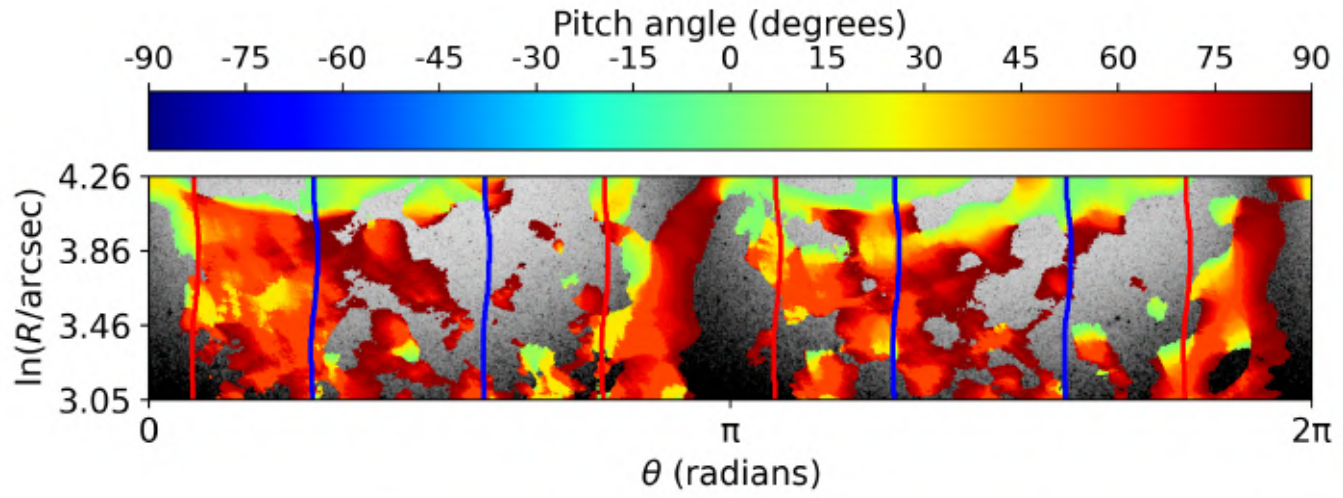

(c) Distribution of pitch angles obtained with the $D_w = 0.5$ filter.

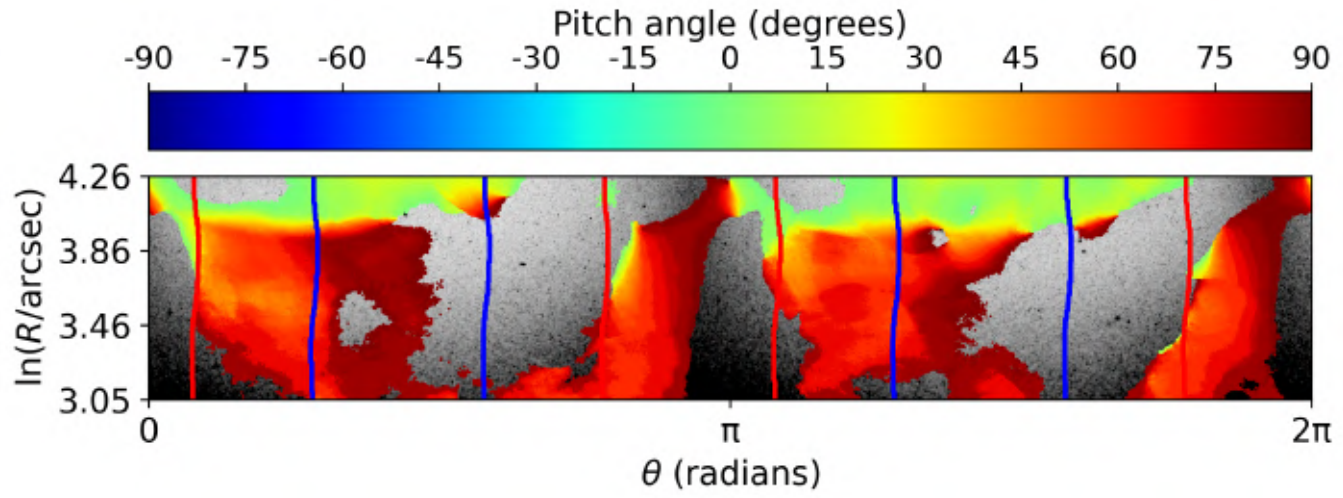

(d) Distribution of pitch angles obtained with the $D_w = 0.75$ filter.

**Figure 9.** Distribution of pitch angles obtained from the highest correlations at each position $(\ln(R/\mathrm{arcsec}), \theta)$ in the $F335M$ image of NGC 1433. The colorbar represents pitch angle values, ranging from $-90°$ (dark blue) to $90°$ (dark red). The colors of the points indicate the pitch angle associated with the maximum correlation, based on the idea that the strongest correlations occur when the structure has the same pitch angle as the filter at the corresponding position.

## 4.2 Comparison with spiral arms and interarm structures

In Faria-Costa et al. (2026), we showed that spiral arms are characterized by relatively small pitch angles ($|\alpha| \sim 15°$–$30°$), consistent with their interpretation as large-scale density-wave structures. In contrast, interarm substructures such as feathers exhibit substantially larger pitch angles, often exceeding $50°$, consistent with their association with local instabilities and shearing processes.

The results obtained here for dust lanes overlap partially with the pitch angle range of feathers, but arise from a fundamentally different physical mechanism. While feathers are secondary structures emerging from spiral arms, dust lanes are primary features associated with bar-driven gas dynamics.

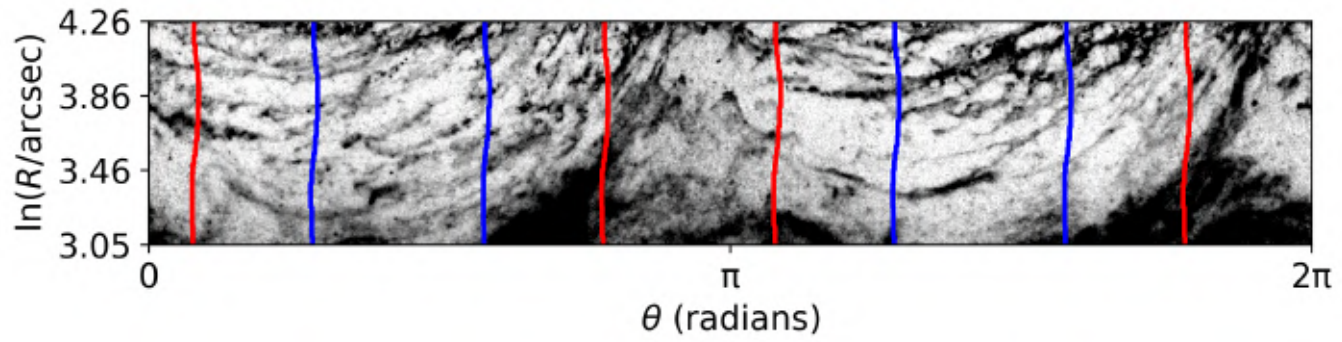

(a) Regions delimited in the $F1130W$ image of the galaxy NGC 1433 from the 2D-FT/2D-IFT mask. The bar regions are located between the red lines, and the interbar regions are between the blue lines.

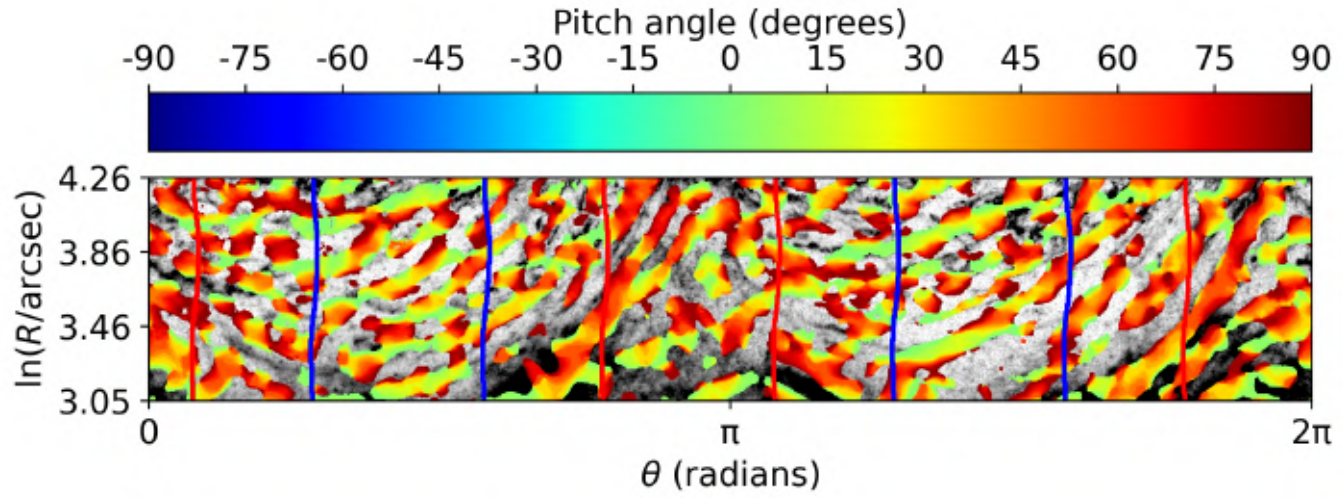

(b) Distribution of pitch angles obtained with the $D_w = 0.25$ filter.

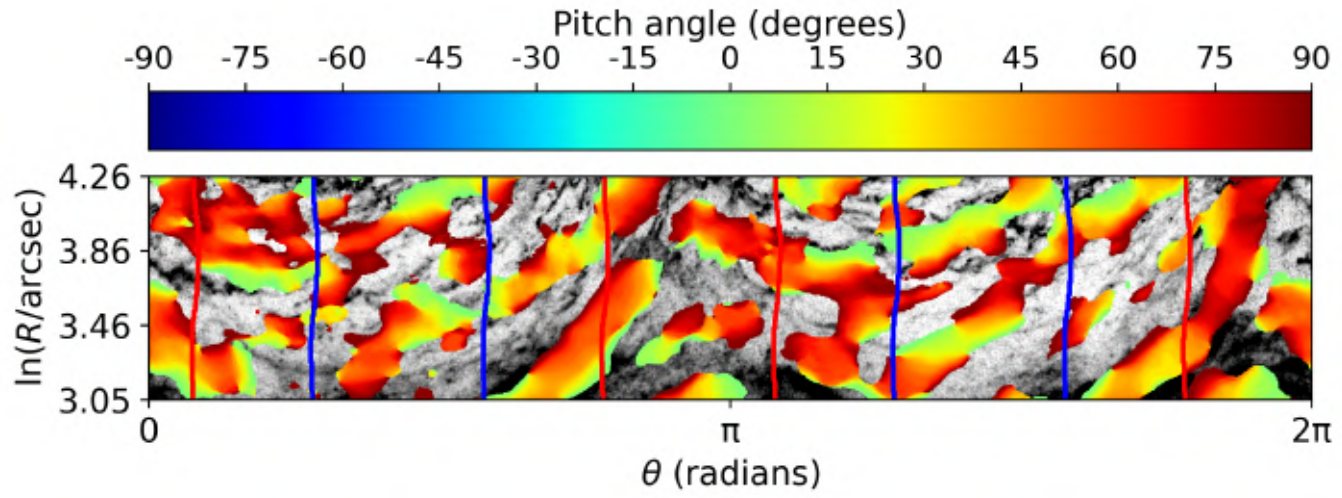

(c) Distribution of pitch angles obtained with the $D_w = 0.5$ filter.

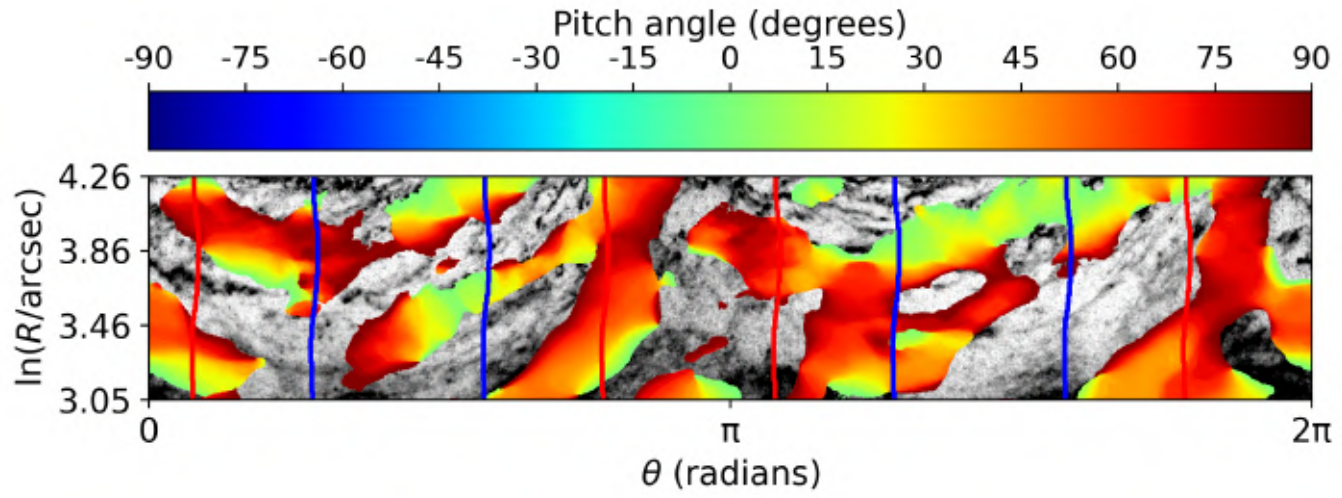

(d) Distribution of pitch angles obtained with the $D_w = 0.75$ filter.

**Figure 10.** Distribution of pitch angles obtained from the highest correlations at each position $(\ln(R/\mathrm{arcsec}), \theta)$ in the $F1130W$ image of NGC 1433. The colorbar represents pitch angle values, ranging from $-90°$ (dark blue) to $90°$ (dark red). The colors of the points indicate the pitch angle associated with the maximum correlation, based on the idea that the strongest correlations occur when the structure has the same pitch angle as the filter at the corresponding position.

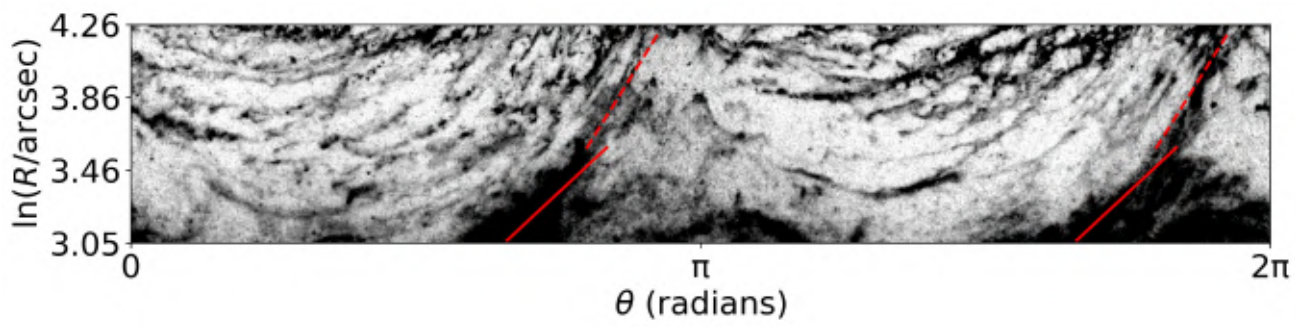

**Figure 11.** Lines overlaid on the $F1130W$ image of NGC 1433, indicating the curvature of the dust lanes. The solid red line corresponds to $43°$, and the dashed red line to $58°$.

This comparison highlights that similar pitch angle values can correspond to different physical processes, depending on the morphological context. Therefore, the interpretation of pitch angle measurements requires careful consideration of the nature of the structures being analyzed.

### 4.3 A unified, scale-dependent framework

Taken together, the results from Faria-Costa et al. (2026) and the present work suggest that the pitch angle is not a universal descriptor of spiral structure, but rather a scale-dependent and structure-dependent quantity.

At large scales in spiral-dominated systems, the pitch angle reflects the geometry of density waves and the global organization of the galactic disk. At smaller scales, it captures the orientation of substructures such as feathers, which are associated with local instabilities. In bar-dominated systems, as shown here, the local pitch angle characterizes the curvature of dust lanes associated with bar-driven gas flows, thereby providing a morphological proxy for the response of the interstellar medium to the bar potential.

The results summarized in Table 2, when considered in conjunction with those obtained in Faria-Costa et al. (2026), clearly illustrate the systematic shift in pitch angle values between spiral arm-dominated and bar-dominated regimes. This comparison reinforces the interpretation of pitch angle as a diagnostic that depends on both spatial scale and physical context.

This unified view indicates that the same mathematical framework, based on local pitch angle measurements, can be applied across different galactic environments, providing a consistent way to probe distinct dynamical regimes.

### 4.4 Implications for the use of pitch angle

The results of this study have important implications for the interpretation of pitch angle measurements in galaxies. Traditionally, the pitch angle has been treated as a global parameter characterizing spiral structure. However, our analysis shows that, when measured locally, it provides a richer and more nuanced description of galactic morphology and dynamics.

In particular, local pitch angle measurements can be used to: (i) distinguish between different structural components within galaxies, (ii) identify transitions between dynamical regimes, and (iii) trace the response of the interstellar medium to gravitational perturbations such as bars.

This perspective opens new possibilities for comparing observations with theoretical models and numerical simulations, where the detailed geometry of gas flows and spiral structures can be directly related to the underlying physical processes.

The most important result emerging from Papers I and II is that local pitch angles are not universal descriptors of galactic morphology, but quantities whose physical meaning depends on the dynamical regime under consideration.

## 5 CONCLUSIONS

In this work, we analyzed the local pitch angles of structures in a sample of barred spiral galaxies observed with the PHANGS–JWST survey, focusing on the bar-dominated regions where dust lanes are the primary features. Using a combination of two-dimensional Fourier transform (2D-FT) and circular-window correlation methods, we derived spatially resolved measurements of the pitch angle across multiple scales.

We find that dust lanes in barred galaxies are characterized by relatively large pitch angles, typically in the range $|\alpha| \sim 35°$ to $70°$, generally exceeding the values measured for spiral arms in Faria-Costa et al. (2026). In many cases, the pitch angle varies systematically along the dust lanes, with smaller values in the inner regions and larger values along the bar, reflecting changes in curvature.

The comparison between NIRCam and MIRI data shows that MIRI observations provide a clearer and more detailed view of the dust structures, allowing a more precise characterization of local variations in pitch angle. The use of multiple spatial scales further reveals the complexity of the dust lane morphology, distinguishing between global geometry and local substructures.

In contrast to spiral arms, where the pitch angle is associated with logarithmic spiral patterns, the values measured in this work describe the local orientation and curvature of dust lanes associated with bar-driven gas flows. This demonstrates that the physical interpretation of the pitch angle depends on the nature of the structures being analyzed.

Taken together with the results from Faria-Costa et al. (2026), our findings support a unified, scale-dependent framework in which the pitch angle serves as a diagnostic of distinct dynamical regimes in galaxies. While small pitch angles characterize large-scale spiral arms, larger values are associated with interarm substructures and bar-driven dust lanes, reflecting different physical processes.

This approach highlights the potential of local pitch angle measurements as a tool for probing galactic dynamics, providing a direct link between observed morphology and the underlying physical mechanisms governing gas flows and structure formation in galaxies. The methodology presented in this series opens the possibility of using local pitch-angle distributions as a quantitative observational diagnostic of galactic dynamics across multiple spatial scales.

## ACKNOWLEDGEMENTS

Stella Faria Costa acknowledges the support received from Coordenação de Aperfeiçoamento de Pessoal de Nível Superior - Brasil (CAPES) under process numbers 88887.823252/2023-00 and 88881.935622/2024-01; Irapuan Rodrigues acknowledges support from Conselho Nacional de Desenvolvimento Científico e Tecnológico (CNPq/Brazil, Grant No. 314267/2021-2). We thank the anonymous referee for the constructive comments and suggestions, which were particularly helpful in restructuring the manuscript and improving the presentation of the results. The JWST imaging data analysed in this work are publicly available as part of the PHANGS–JWST high-level science products hosted by the Mikulski Archive for Space Telescopes. The data can be accessed at `https://doi.org/10.17909/ew88-jt15`. The derived products generated in this study are available from the corresponding author upon reasonable request.

## DATA AVAILABILITY

The images used in this study were obtained from the Mikulski Archive for Space Telescopes (MAST), available at https://mast.stsci.edu/portal/Mashup/Clients/Mast/Portal.html.

## APPENDIX A: DIAGNOSTIC FIGURES FOR THE INDIVIDUAL GALAXIES

This appendix contains the complete correlation spectra and local pitch-angle maps for the galaxies whose diagnostic figure sets are not shown in the main text. The panels follow the same format as the representative NGC 1433 figures in Section 3.1. For each galaxy, we show the summed-correlation spectra for the three window diameters, the spatial pitch-angle distributions in $F335M$ and $F1130W$, and the adopted representative orientations overlaid on the $F1130W$ image.

### A1 NGC 1300

NGC 1300 shows a clear variation in the local orientation of its dust lanes. The representative components near $\alpha \sim 35°$ and $\alpha \sim 70°$ correspond to different portions of the structure, from the more curved inner region to the straighter segment along the bar. The spectra and spatial pitch-angle maps below illustrate how these components are recovered across the three window scales.

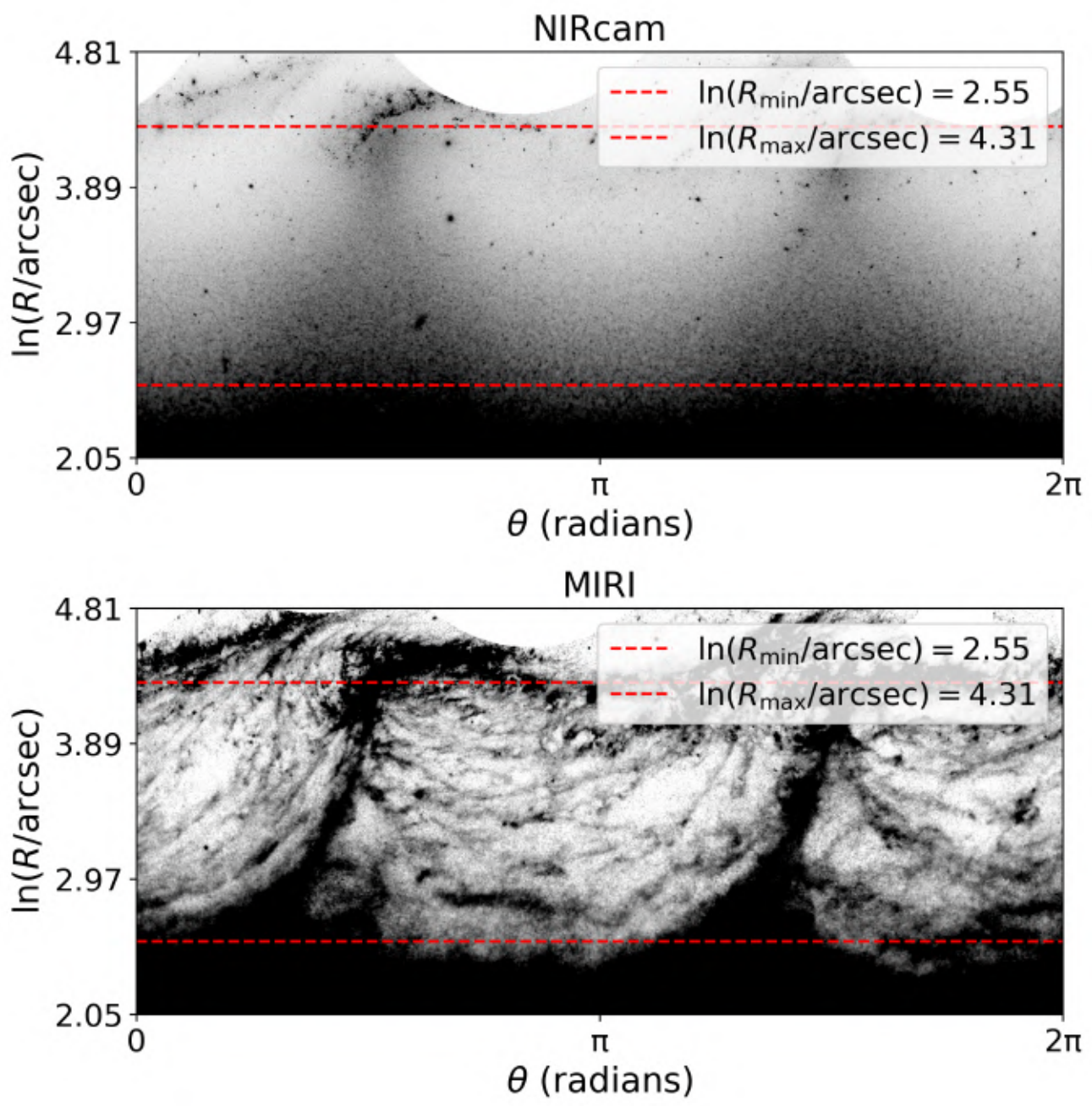

**Figure A1.** Diagrams $\ln(R/\text{arcsec})$ vs. $\theta$ for the $F335M$ (NIRCam) and $F1130W$ (MIRI) images of NGC 1300. The dashed red lines indicate the values of $\ln(R_{min}/\text{arcsec})$ and $\ln(R_{max}/\text{arcsec})$ adopted in the analysis.

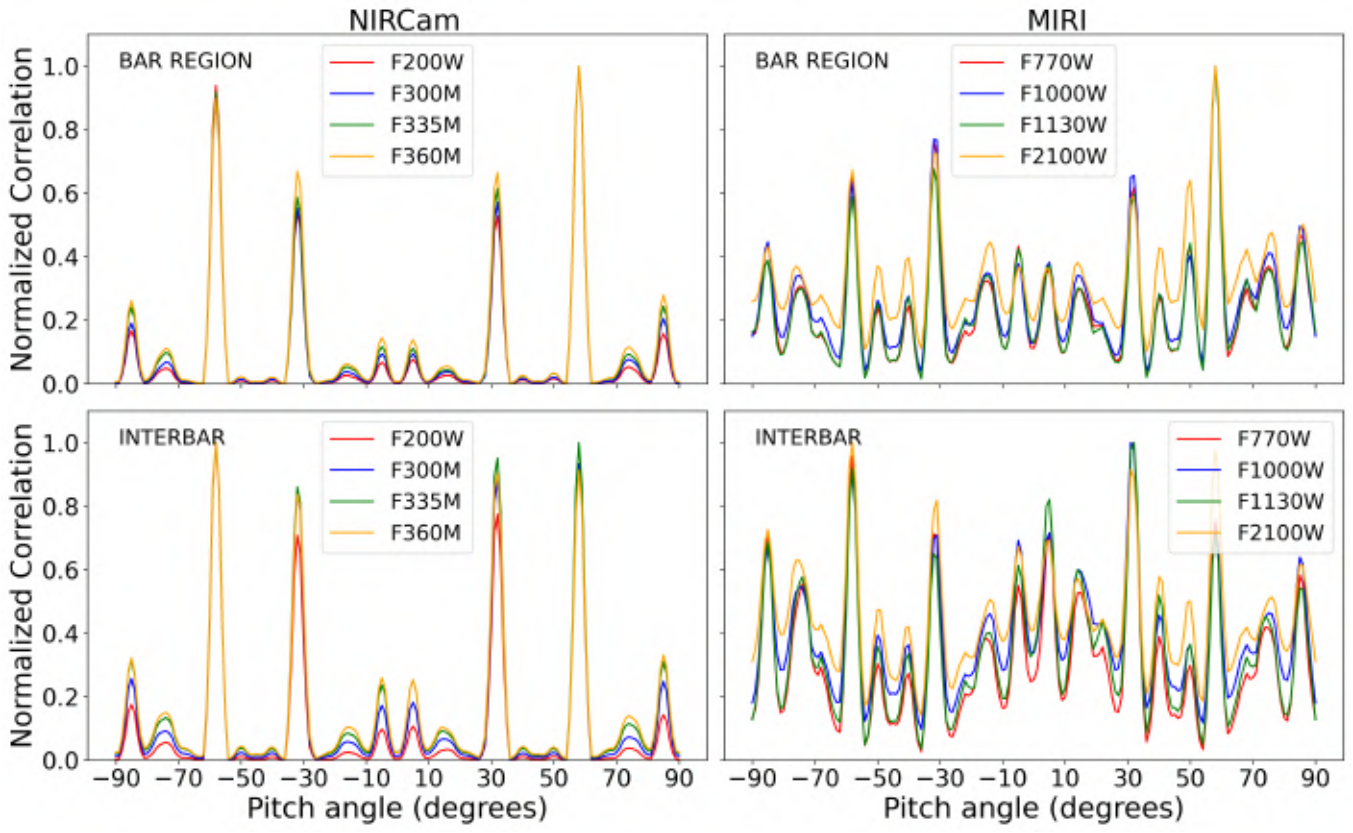

(a) Sum of the correlations calculated with $D_w$ = 0.25 for the bar and interbar regions of NGC 1300.

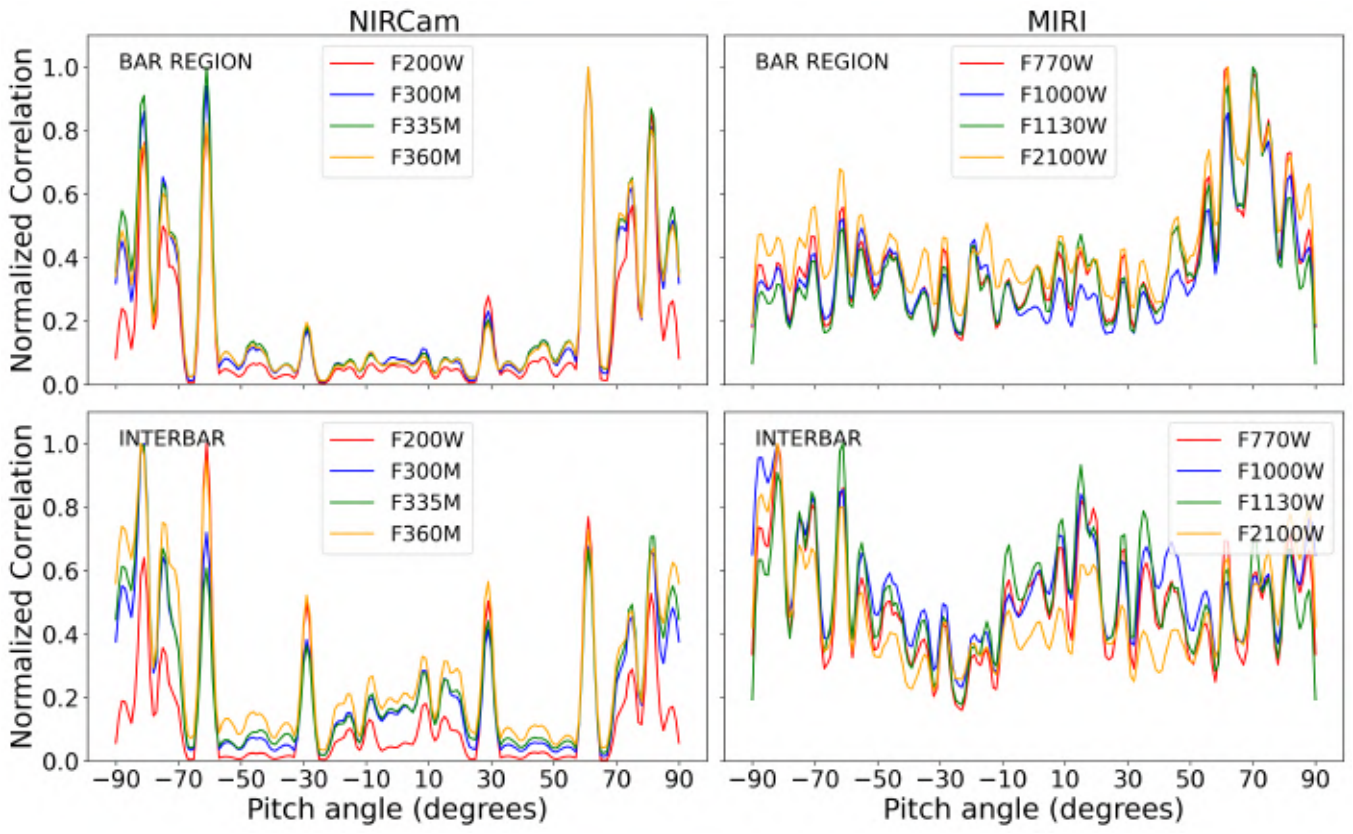

(b) Sum of the correlations calculated with $D_w$ = 0.5 for the bar and interbar regions of NGC 1300.

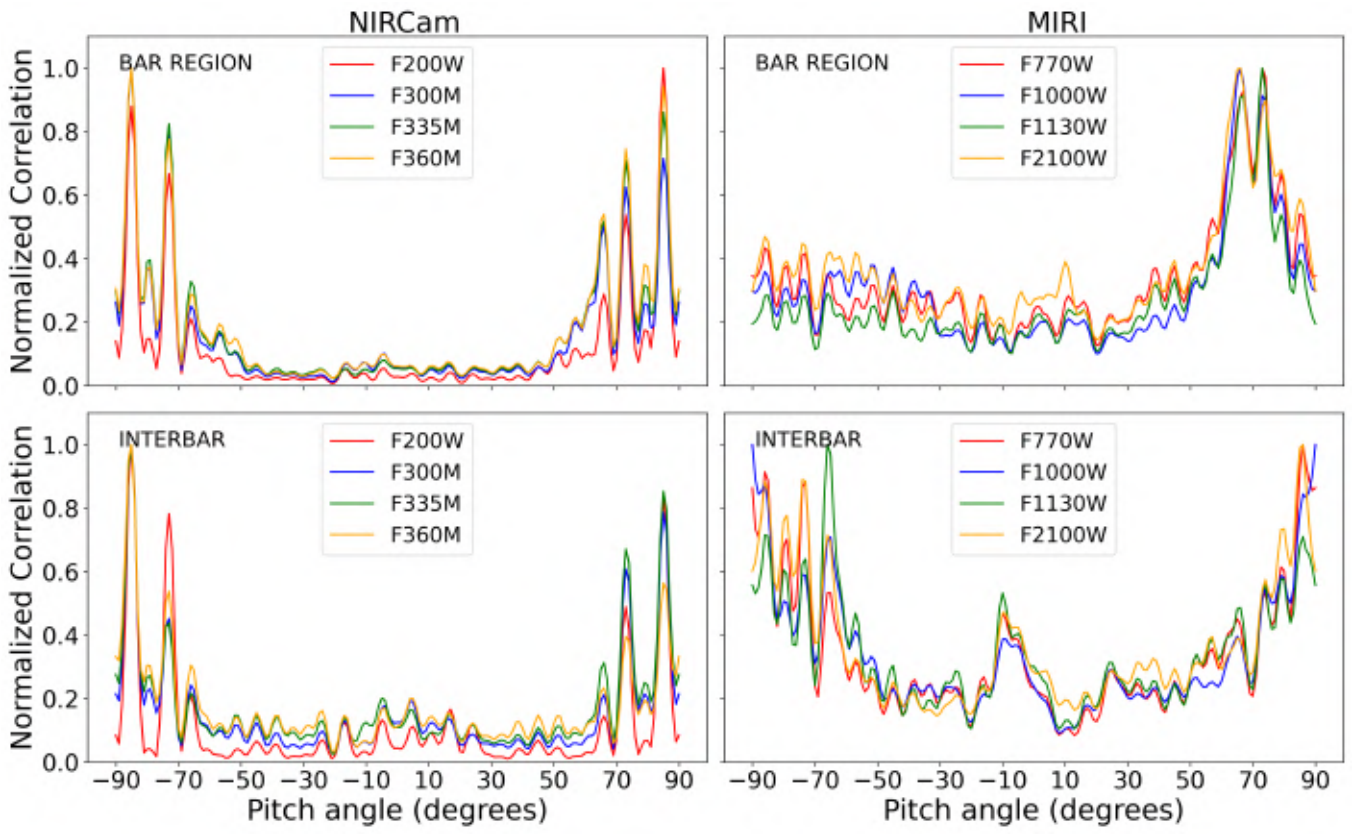

(c) Sum of the correlations calculated with $D_w$ = 0.75 for the bar and interbar regions of NGC 1300.

**Figure A2.** Results for the sum of the correlations in the analyzed regions of NGC 1300. On the left, plots from NIRCam images with filters $F200W$ (red), $F300M$ (blue), $F335M$ (green), and $F360M$ (yellow), where the upper panel corresponds to the bar region and the lower panel to the interbar region. On the right, plots from MIRI images with filters $F770W$ (red), $F1000W$ (blue), $F1130W$ (green), and $F2100W$ (yellow), following the same division between bar (upper) and interbar (lower) regions.

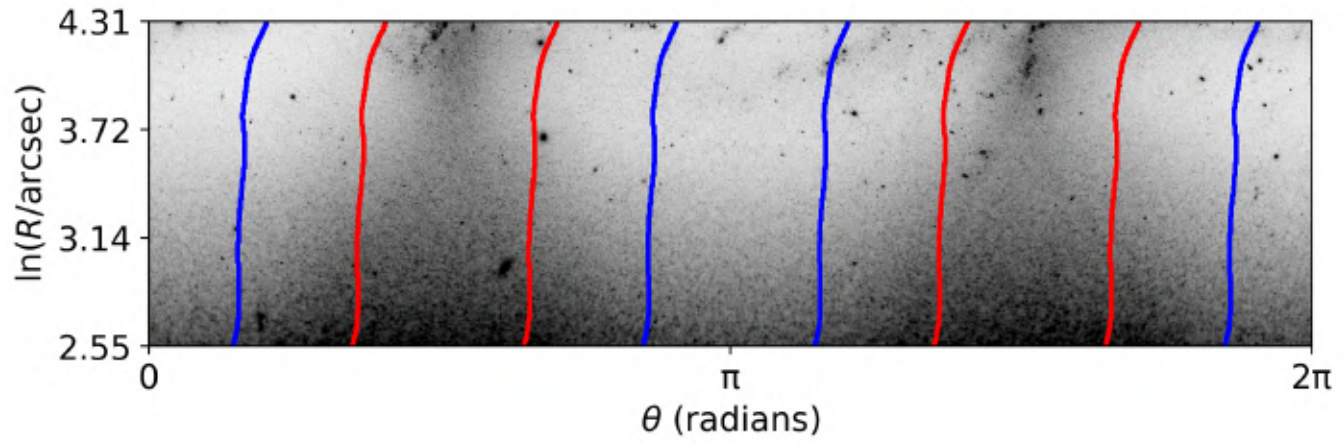

(a) Regions delimited in the $F335M$ image of the galaxy NGC 1300 from the 2D-FT/2D-IFT mask. The bar regions are located between the red lines, and the interbar regions are between the blue lines.

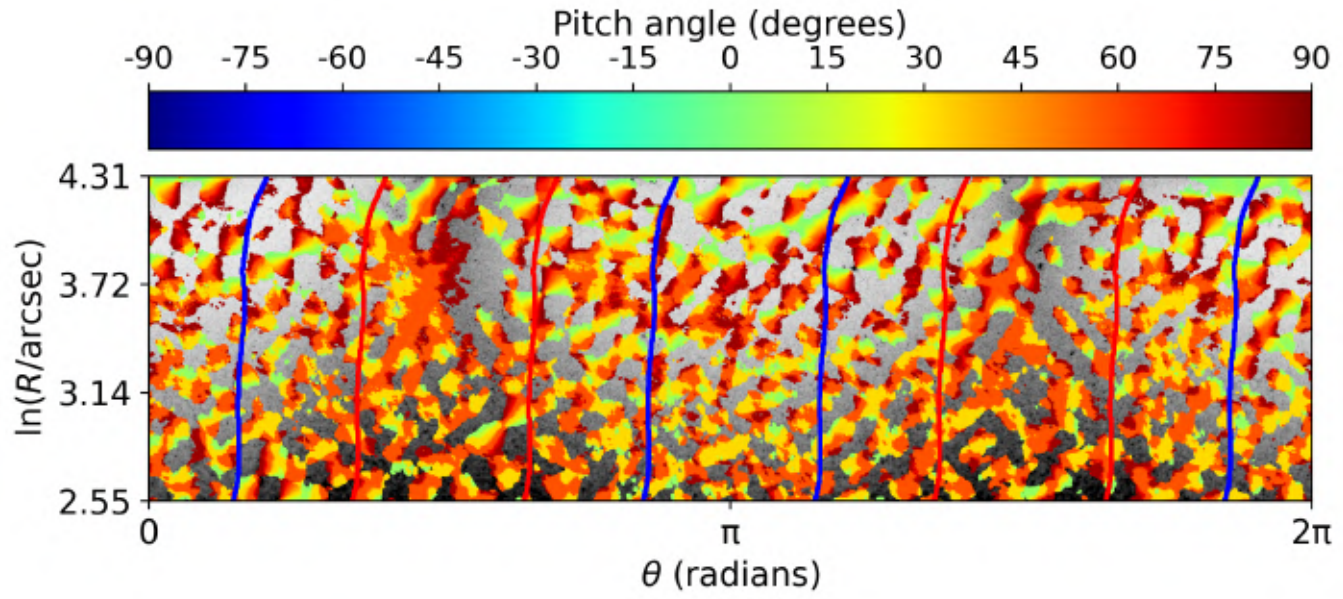

(b) Distribution of pitch angles obtained with the $D_w = 0.25$ filter.

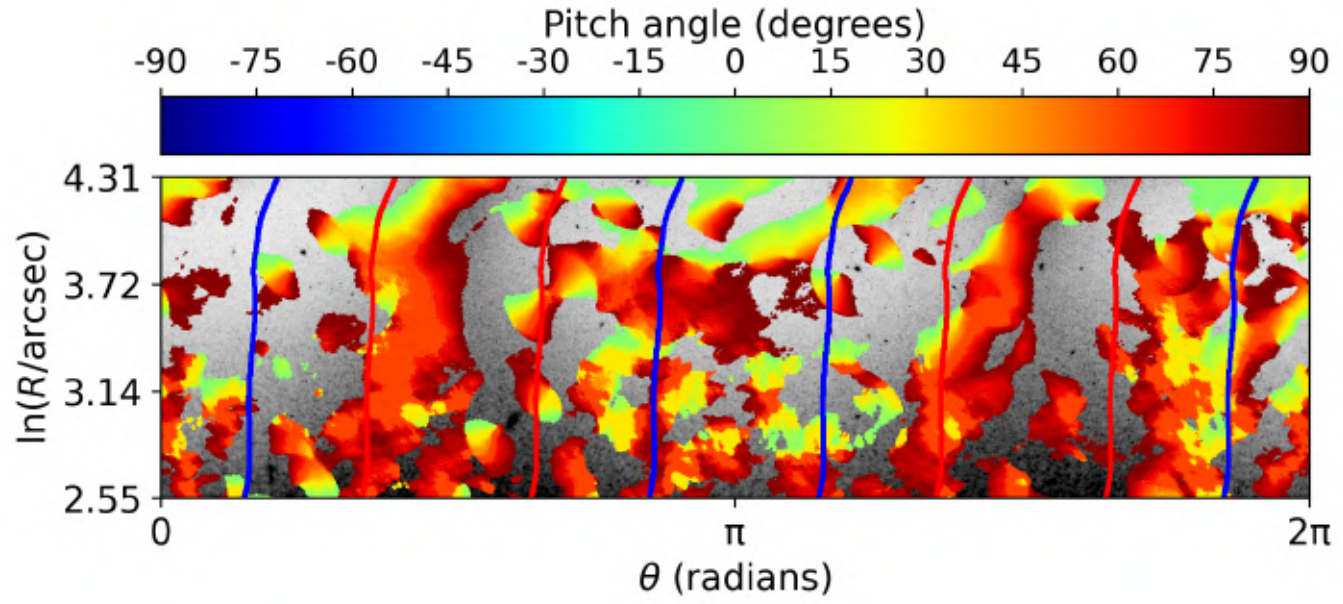

(c) Distribution of pitch angles obtained with the $D_w = 0.5$ filter.

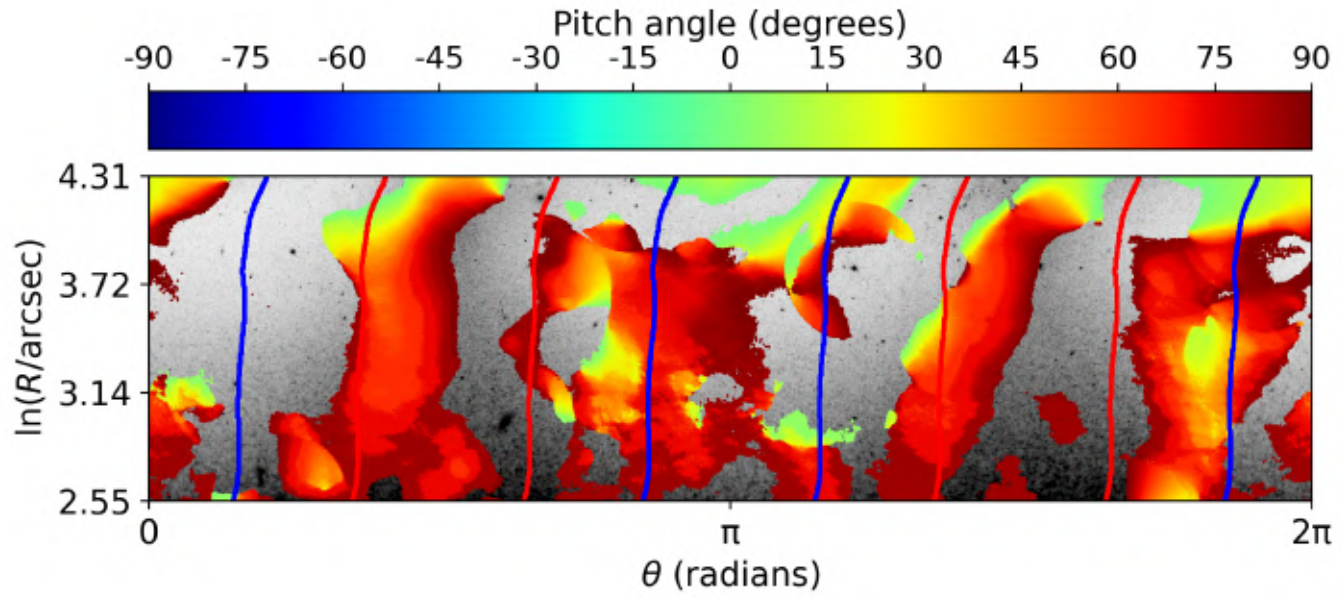

(d) Distribution of pitch angles obtained with the $D_w = 0.75$ filter.

**Figure A3.** Distribution of pitch angles obtained from the highest correlations at each position $(\ln(R/\mathrm{arcsec}), \theta)$ in the $F335M$ image of NGC 1300. The colorbar represents pitch angle values, ranging from −90° (dark blue) to 90° (dark red). The colors of the points indicate the pitch angle associated with the maximum correlation, based on the idea that the strongest correlations occur when the structure has the same pitch angle as the filter at the corresponding position.

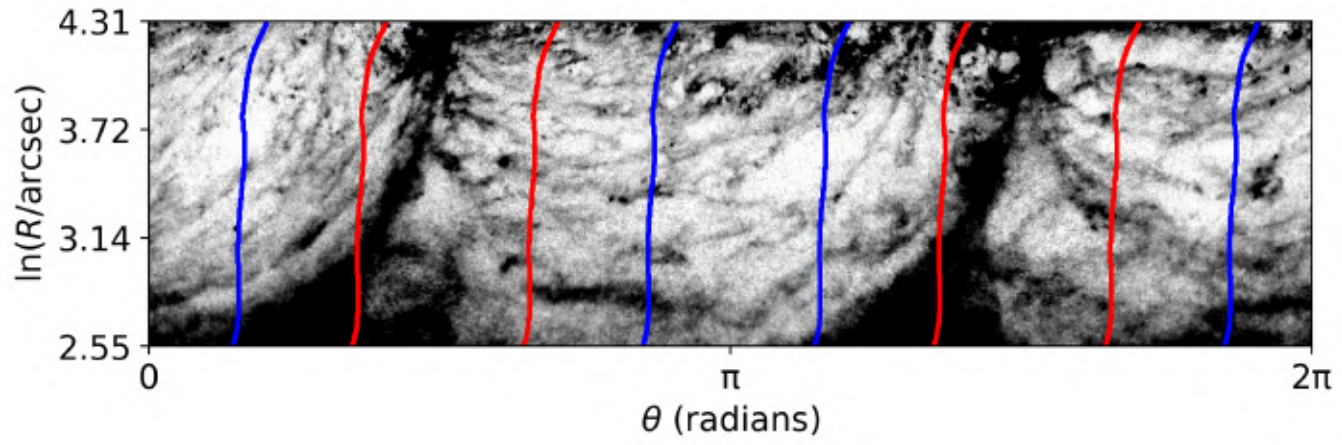

(a) Regions delimited in the $F1130W$ image of the galaxy NGC 1300 from the 2D-FT/2D-IFT mask. The bar regions are located between the red lines, and the interbar regions are between the blue lines.

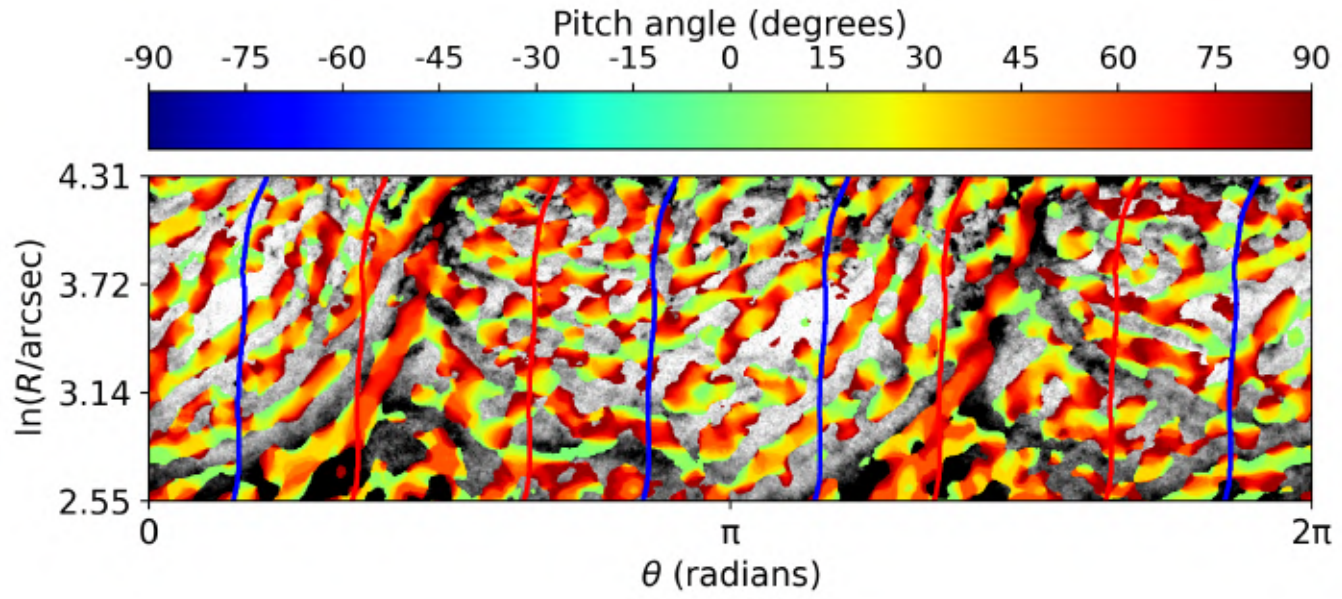

(b) Distribution of pitch angles obtained with the $D_w = 0.25$ filter.

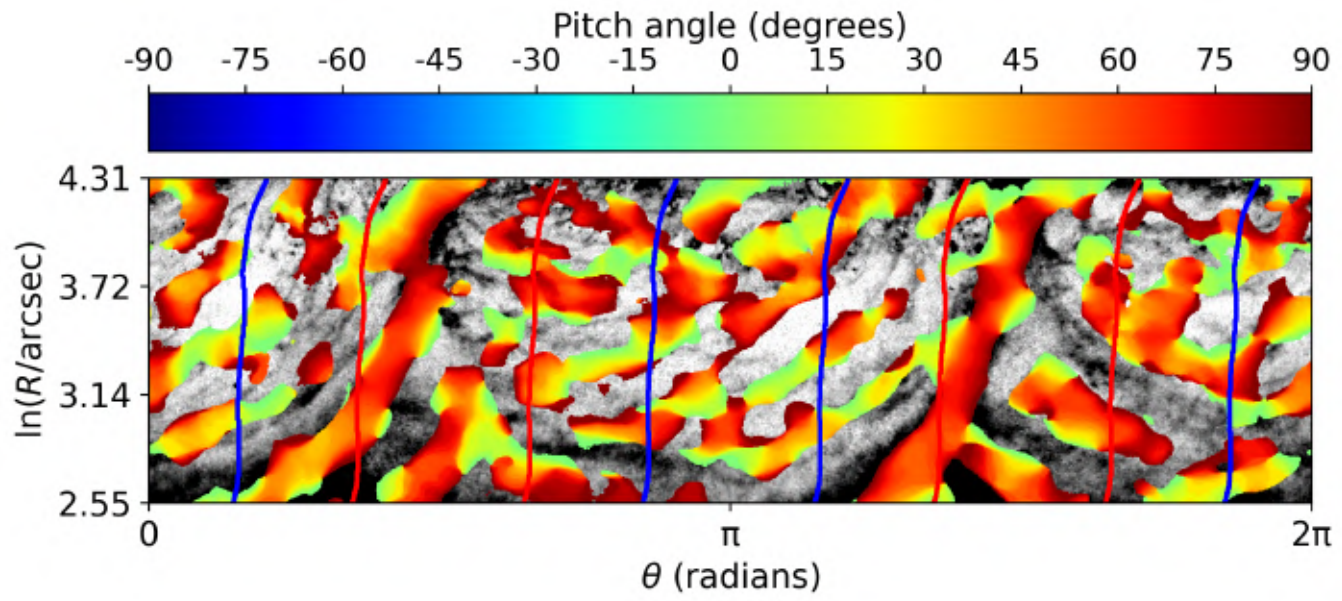

(c) Distribution of pitch angles obtained with the $D_w = 0.5$ filter.

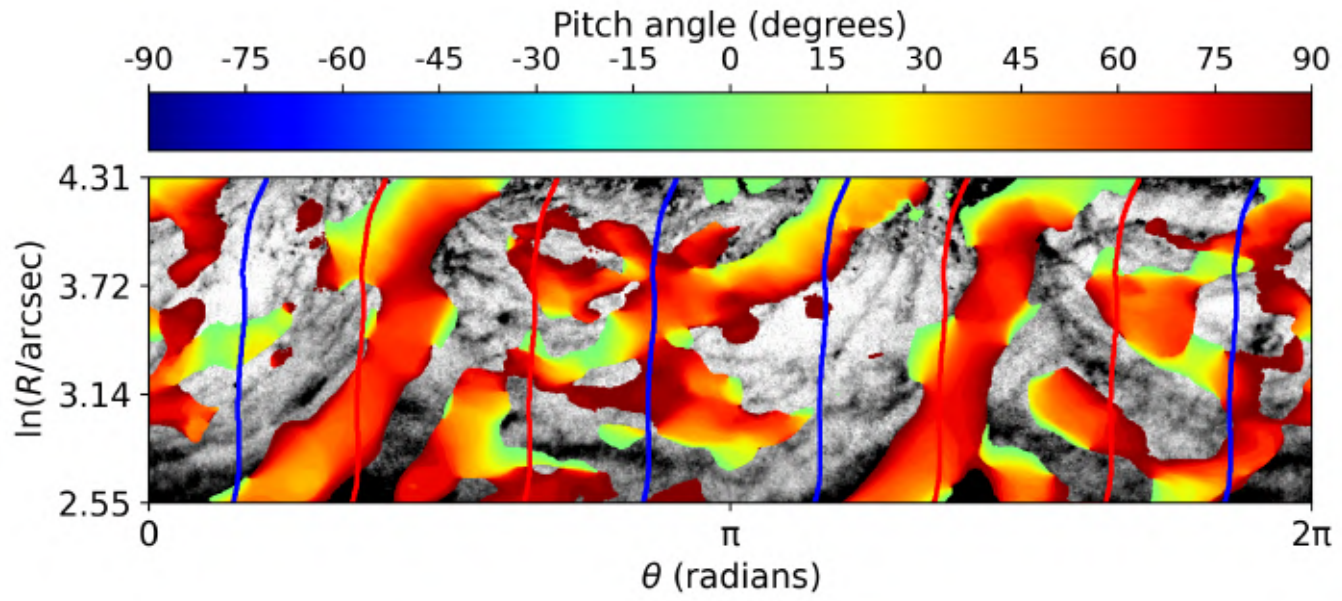

(d) Distribution of pitch angles obtained with the $D_w = 0.75$ filter.

**Figure A4.** Distribution of pitch angles obtained from the highest correlations at each position $(\ln(R/\mathrm{arcsec}), \theta)$ in the $F1130W$ image of NGC 1300. The colorbar represents pitch angle values, ranging from −90° (dark blue) to 90° (dark red). The colors of the points indicate the pitch angle associated with the maximum correlation, based on the idea that the strongest correlations occur when the structure has the same pitch angle as the filter at the corresponding position.

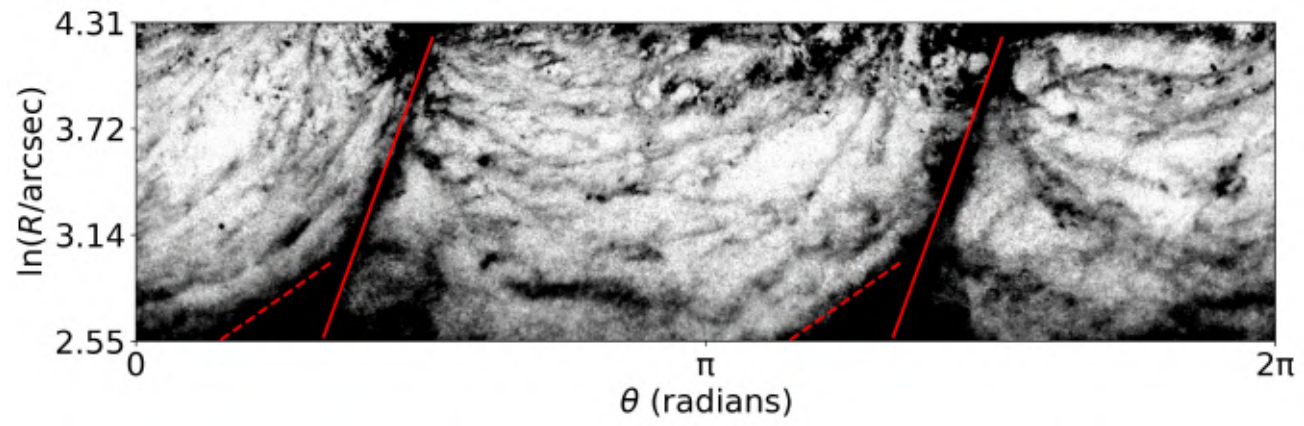

**Figure A5.** Lines overlaid on the $F1130W$ image of NGC 1300, indicating the curvature of the dust lanes. The solid red line corresponds to 70°, and the dashed red line to 35°.

## A2 NGC 1512

NGC 1512 presents multiple pitch-angle components, with representative values near $\alpha \sim 45^\circ$ and $\alpha \sim 60^\circ$. The value near $60^\circ$ describes the main segment of the straight dust lane, whereas the $\sim 45^\circ$ component traces an additional local feature. The smaller window sizes reveal additional local structure, particularly in the MIRI maps, while the larger windows emphasize the overall geometry of the dust lane.

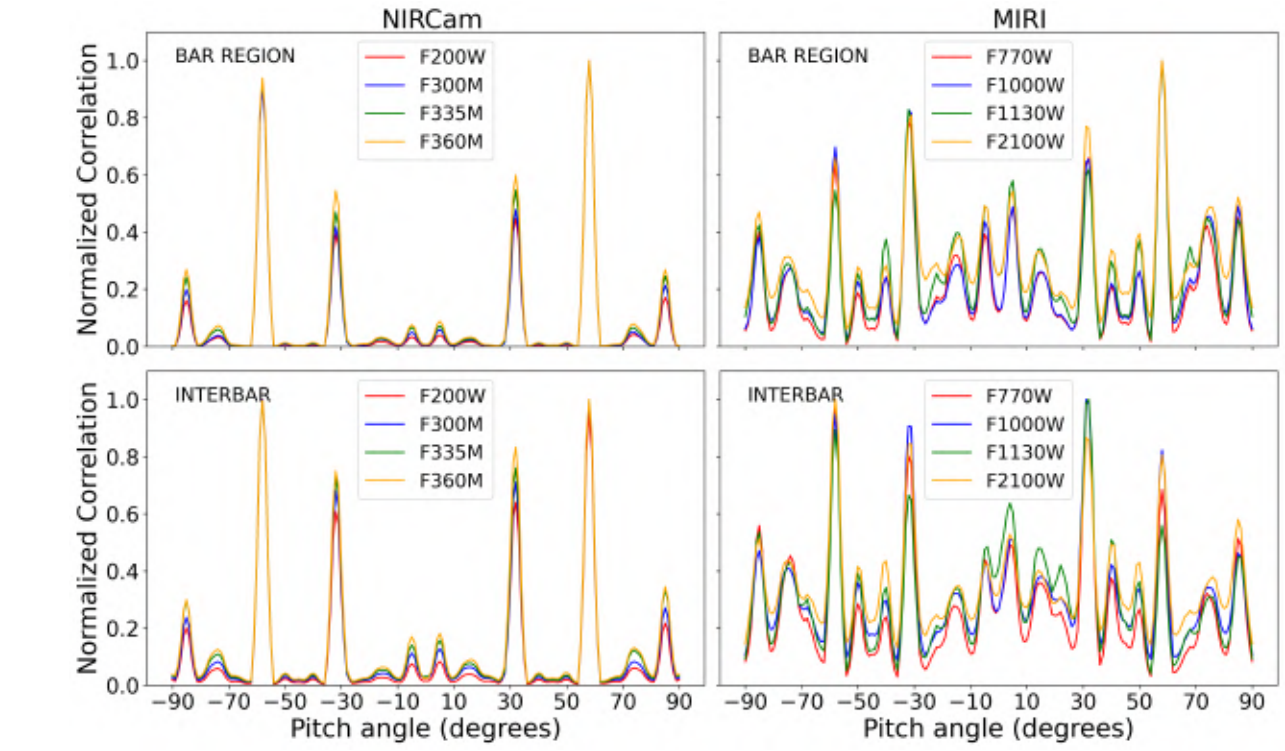

(a) Sum of the correlations calculated with $D_w$ = 0.25 for the bar and interbar regions of NGC 1512.

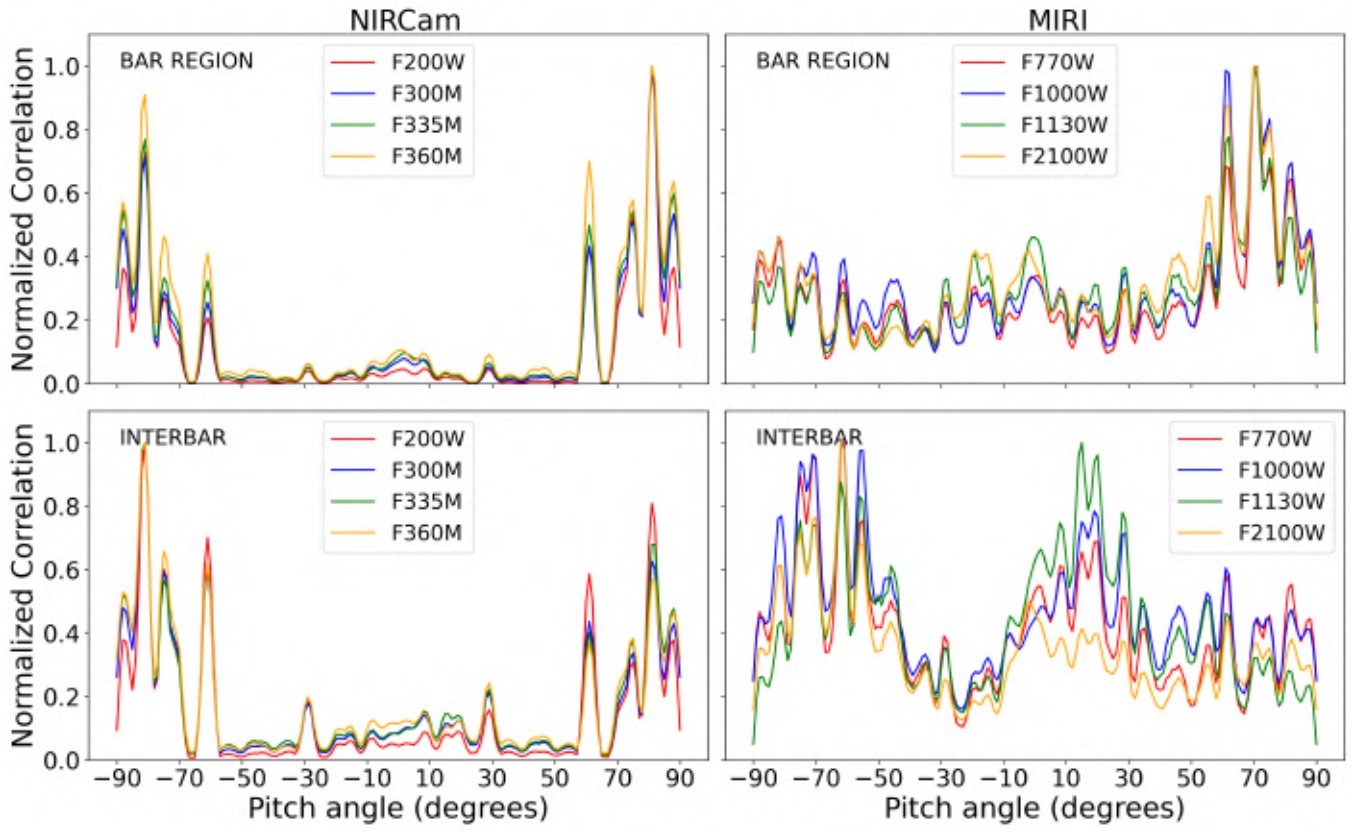

(b) Sum of the correlations calculated with $D_w$ = 0.5 for the bar and interbar regions of NGC 1512.

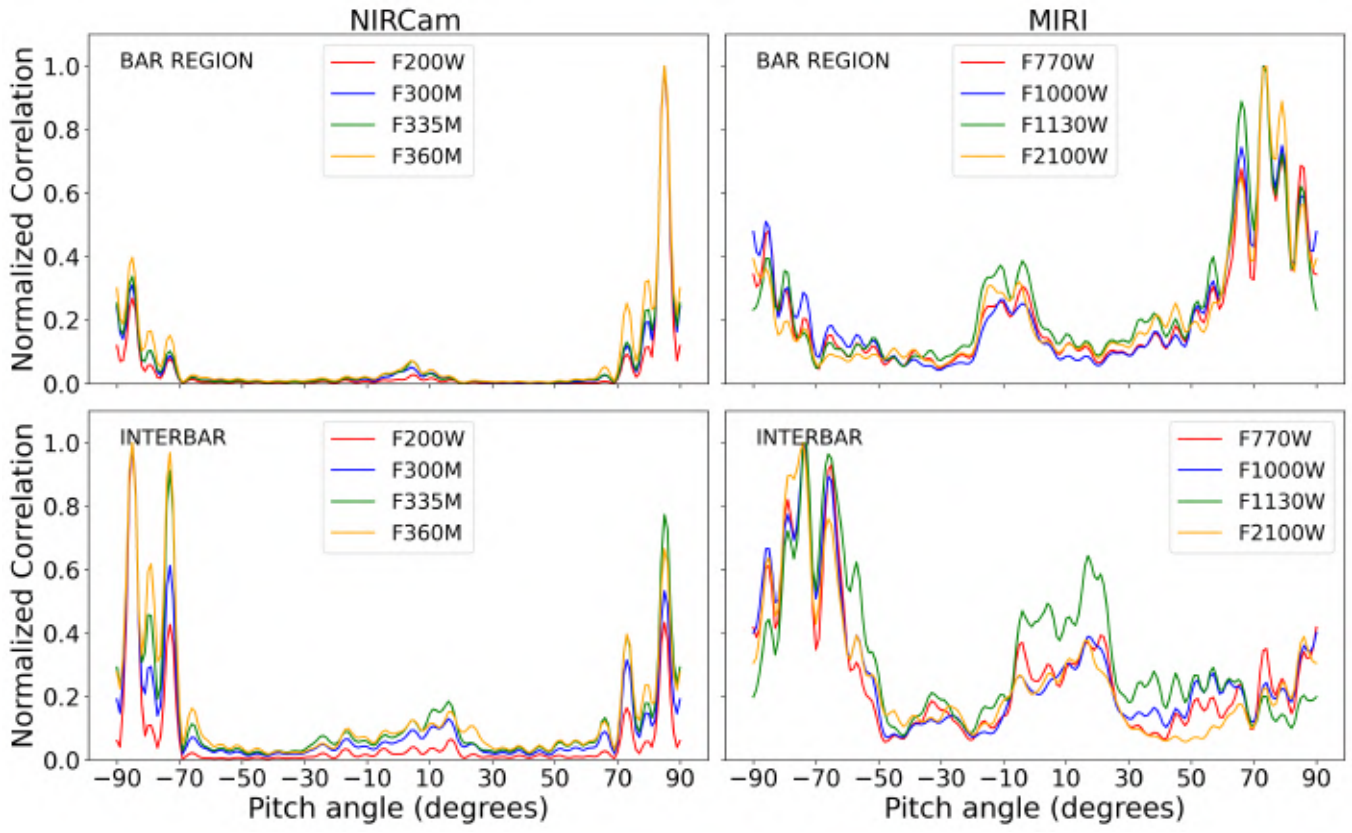

(c) Sum of the correlations calculated with $D_w$ = 0.75 for the bar and interbar regions of NGC 1512.

**Figure A7.** Results for the sum of the correlations in the analyzed regions of NGC 1512. On the left, plots from NIRCam images with filters $F200W$ (red), $F300M$ (blue), $F335M$ (green), and $F360M$ (yellow), where the upper panel corresponds to the bar region and the lower panel to the interbar region. On the right, plots from MIRI images with filters $F770W$ (red), $F1000W$ (blue), $F1130W$ (green), and $F2100W$ (yellow), following the same division between bar (upper) and interbar (lower) regions.

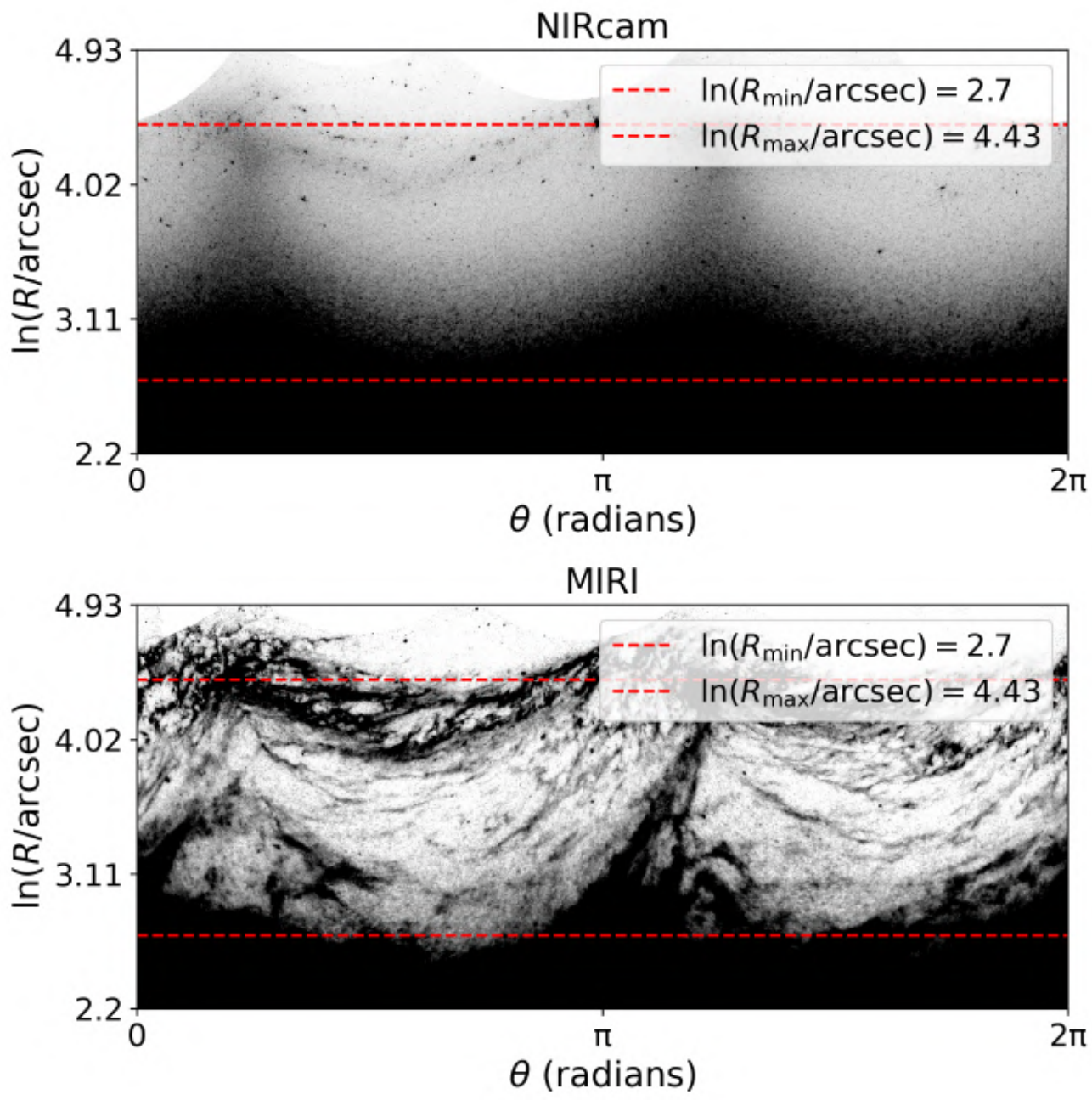

**Figure A6.** Diagrams $\ln(R/\mathrm{arcsec})$ vs. $\theta$ for the $F335M$ (NIRCam) and $F1130W$ (MIRI) images of NGC 1512. The dashed red lines indicate the values of $\ln(R_{min}/\mathrm{arcsec})$ and $\ln(R_{max}/\mathrm{arcsec})$ adopted in the analysis.

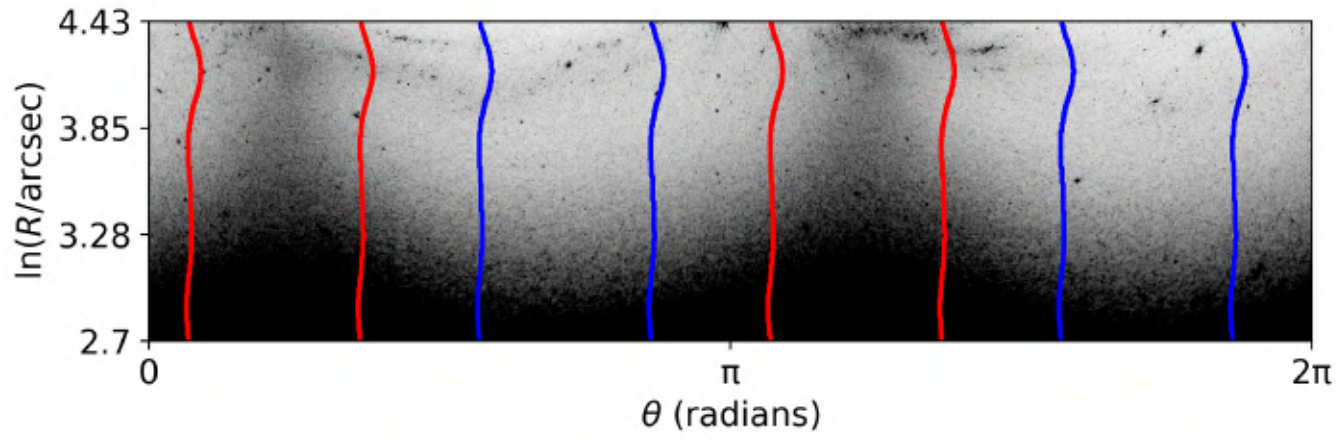

(a) Regions delimited in the $F335M$ image of the galaxy NGC 1512 from the 2D-FT/2D-IFT mask. The bar regions are located between the red lines, and the interbar regions are between the blue lines.

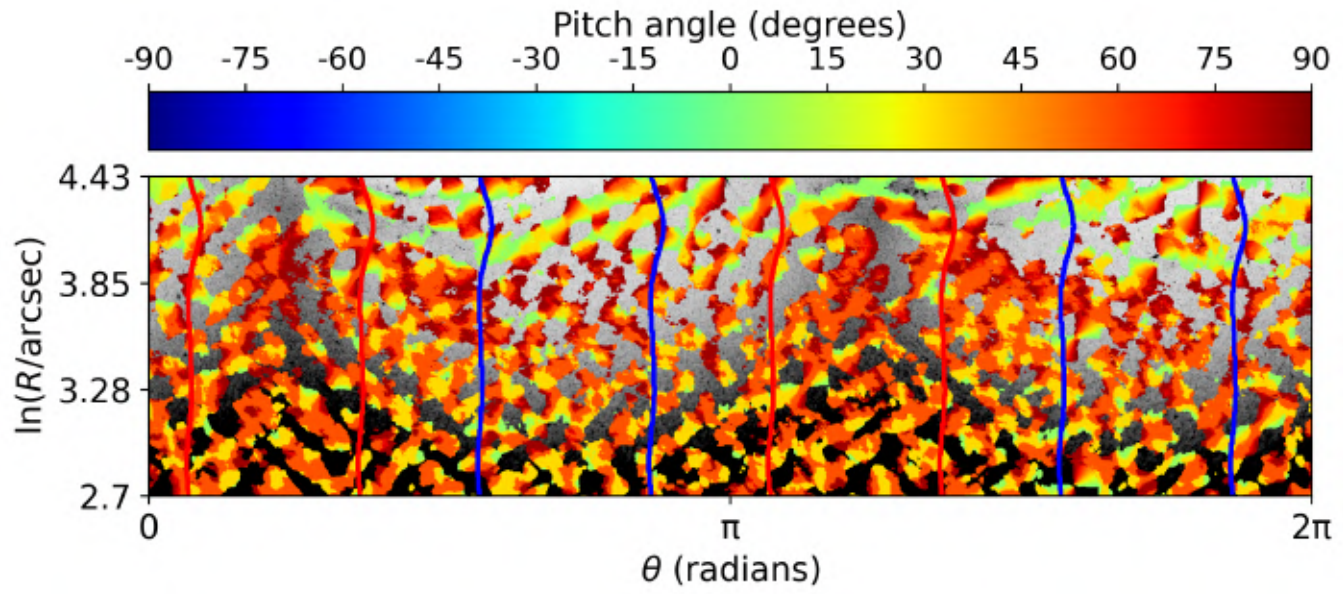

(b) Distribution of pitch angles obtained with the $D_w = 0.25$ filter.

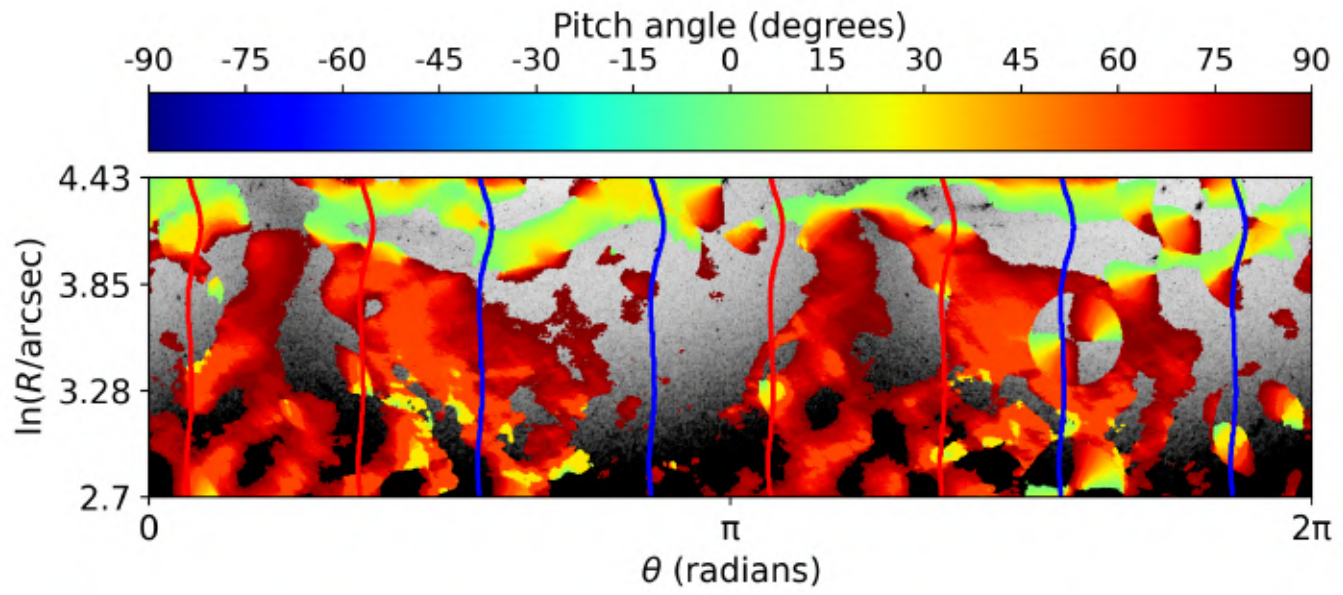

(c) Distribution of pitch angles obtained with the $D_w = 0.5$ filter.

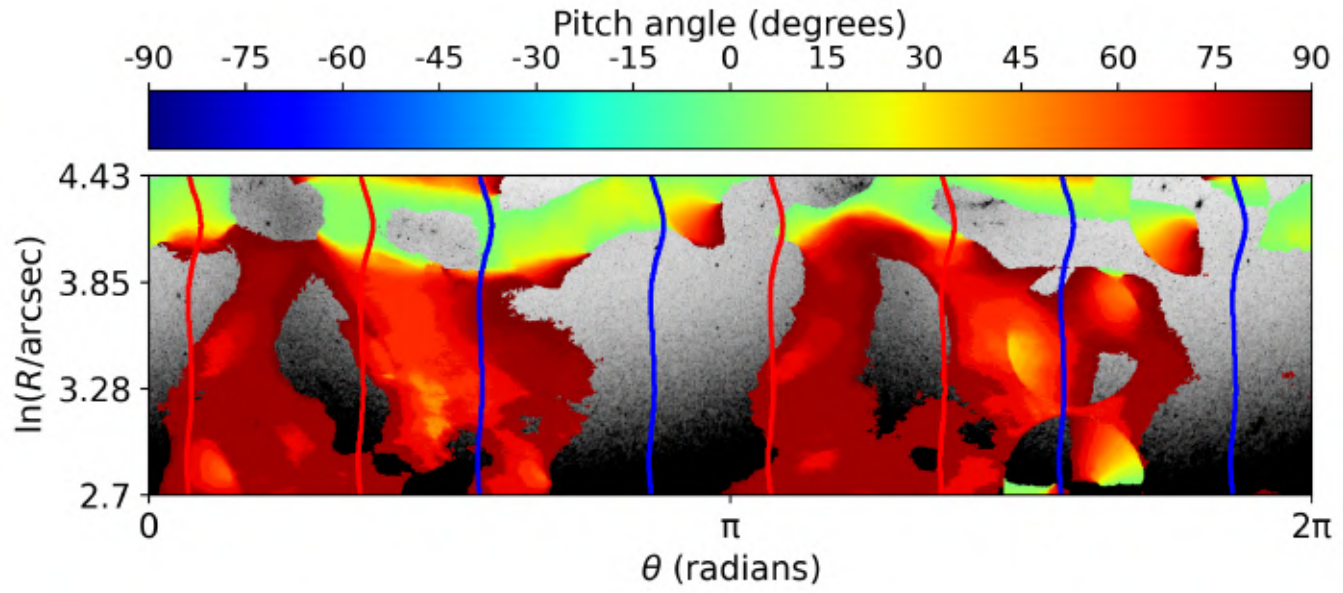

(d) Distribution of pitch angles obtained with the $D_w = 0.75$ filter.

**Figure A8.** Distribution of pitch angles obtained from the highest correlations at each position $(\ln(R/\text{arcsec}), \theta)$ in the $F335M$ image of NGC 1512. The colorbar represents pitch angle values, ranging from $-90°$ (dark blue) to $90°$ (dark red). The colors of the points indicate the pitch angle associated with the maximum correlation, based on the idea that the strongest correlations occur when the structure has the same pitch angle as the filter at the corresponding position.

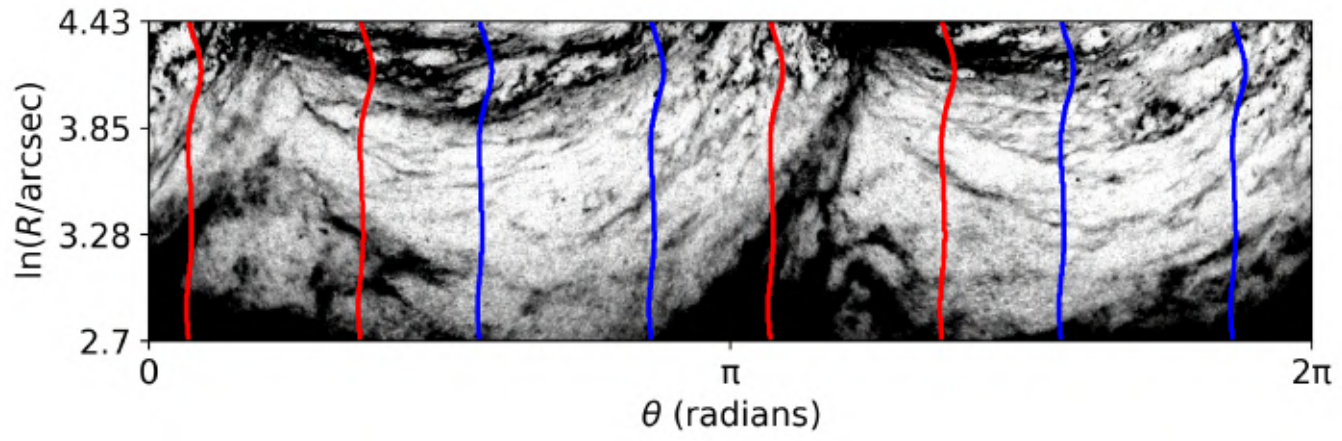

(a) Regions delimited in the $F1130W$ image of the galaxy NGC 1512 from the 2D-FT/2D-IFT mask. The bar regions are located between the red lines, and the interbar regions are between the blue lines.

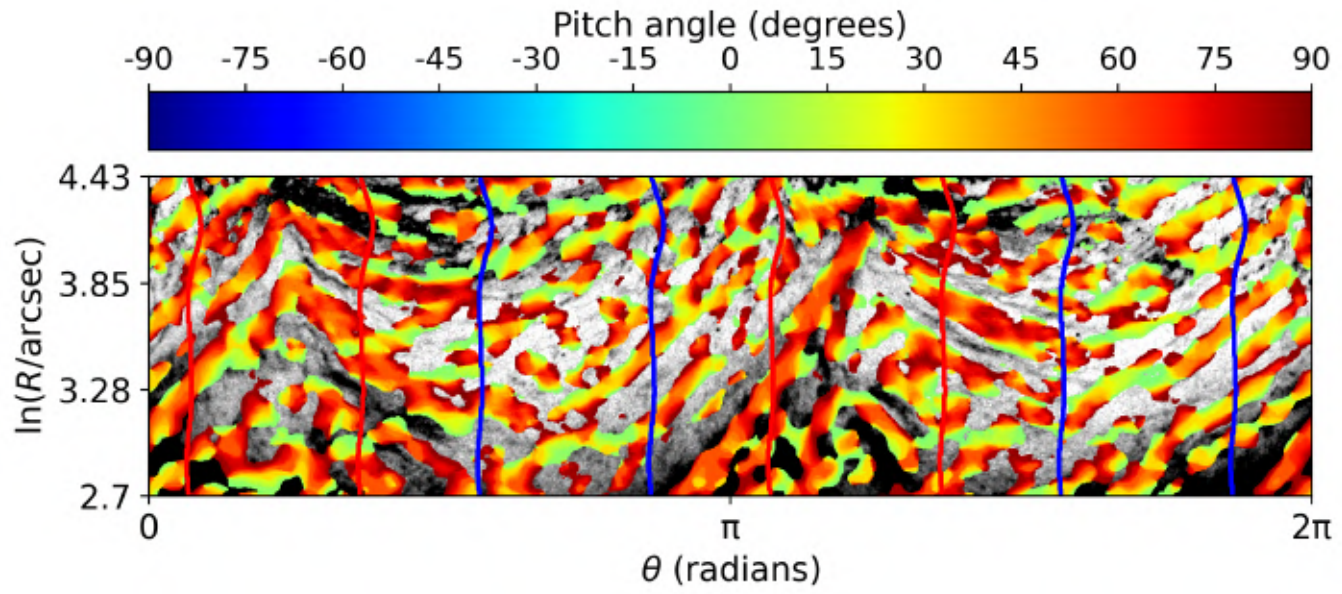

(b) Distribution of pitch angles obtained with the $D_w = 0.25$ filter.

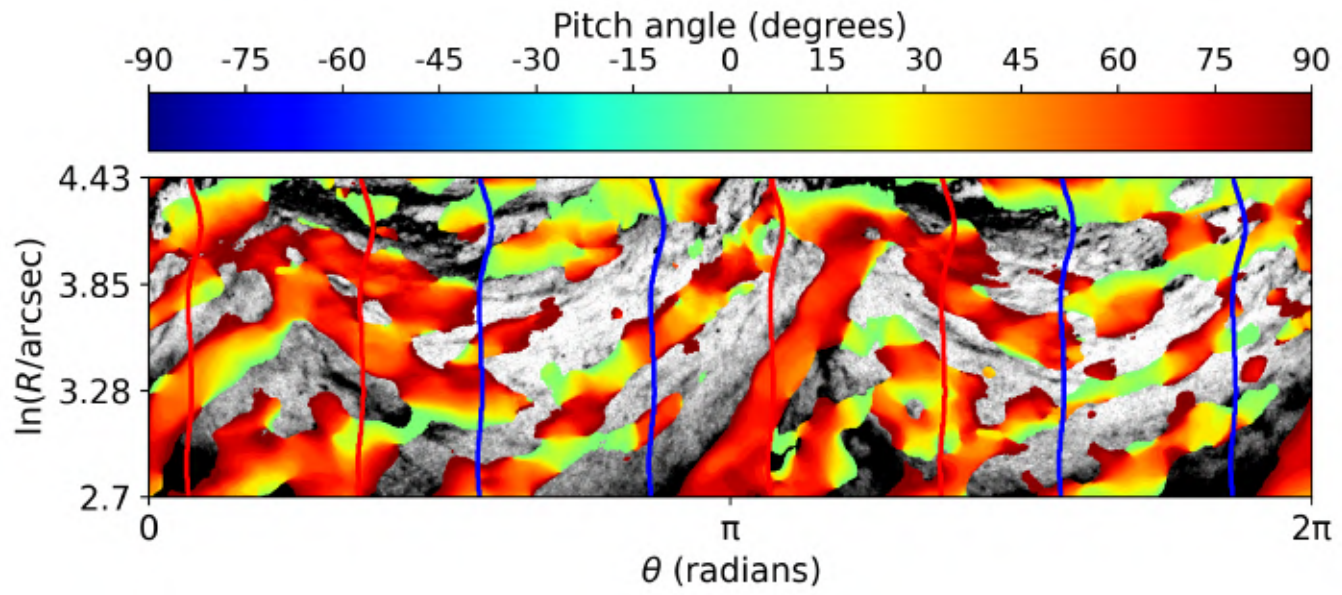

(c) Distribution of pitch angles obtained with the $D_w = 0.5$ filter.

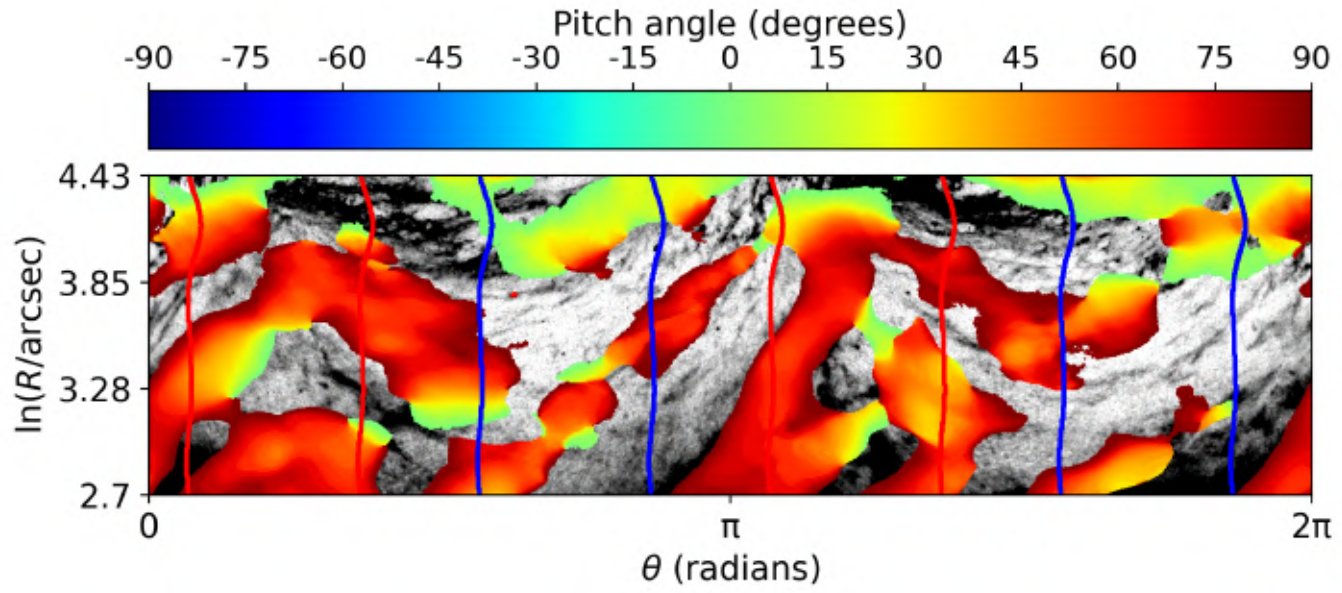

(d) Distribution of pitch angles obtained with the $D_w = 0.75$ filter.

**Figure A9.** Distribution of pitch angles obtained from the highest correlations at each position $(\ln(R/\text{arcsec}), \theta)$ in the $F1130W$ image of NGC 1512. The colorbar represents pitch angle values, ranging from $-90°$ (dark blue) to $90°$ (dark red). The colors of the points indicate the pitch angle associated with the maximum correlation, based on the idea that the strongest correlations occur when the structure has the same pitch angle as the filter at the corresponding position.

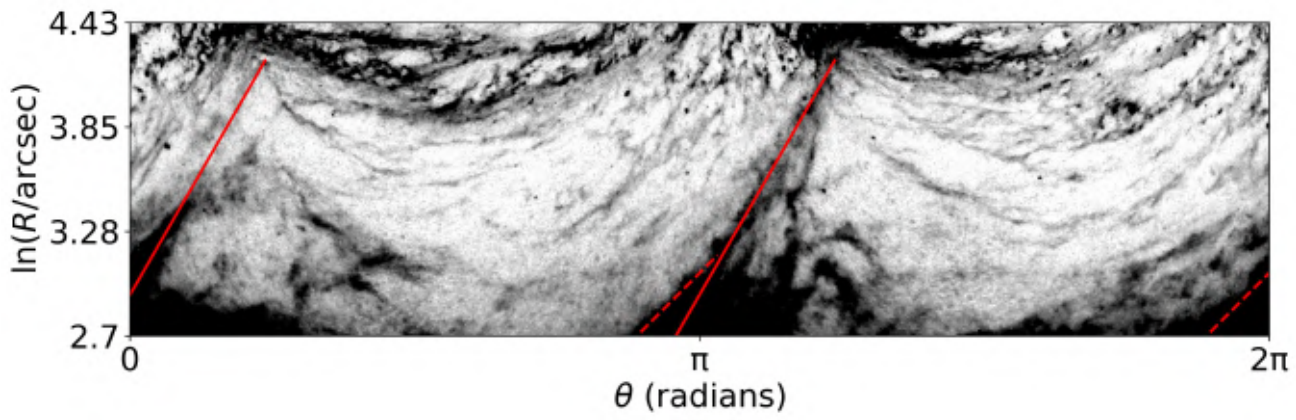

**Figure A10.** Lines overlaid on the $F1130W$ image of NGC 1512, indicating the curvature of the dust lanes. The solid red line corresponds to 60°, and the dashed red line to 45°.

### A3 NGC 2835

NGC 2835 differs from the other galaxies in the barred sample because the analysed region is dominated by spiral-like structures rather than by classical bar dust lanes. The representative pitch angles, approximately 15° and 27°, therefore fall in the range characteristic of spiral arms rather than the larger values measured for the dust lanes in the rest of the sample.

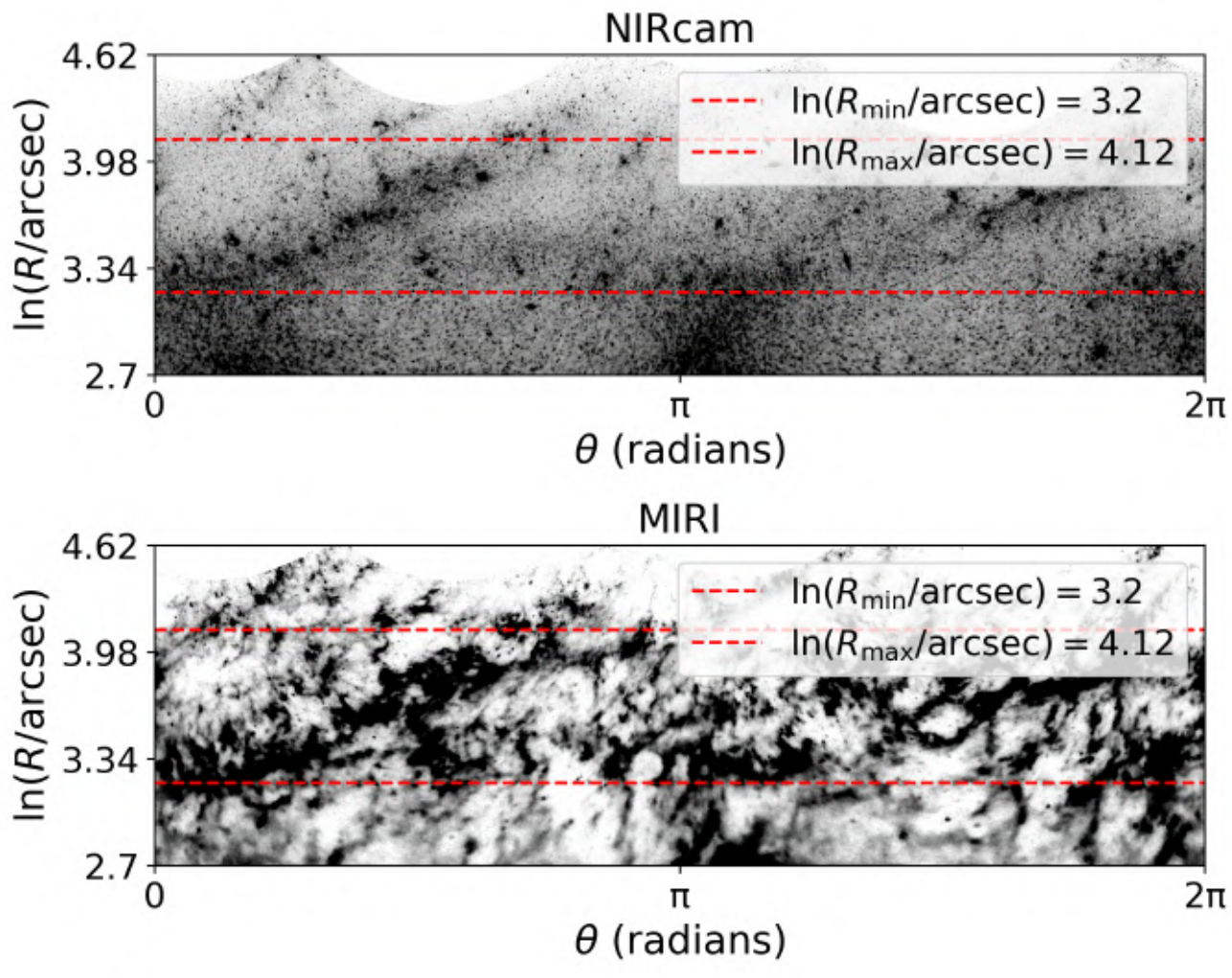

**Figure A11.** Diagrams ln($R$/arcsec) vs. $\theta$ for the $F335M$ (NIRCam) and $F1130W$ (MIRI) images of NGC 2835. The dashed red lines indicate the values of ln($R_{min}$/arcsec) and ln($R_{max}$/arcsec) adopted in the analysis.

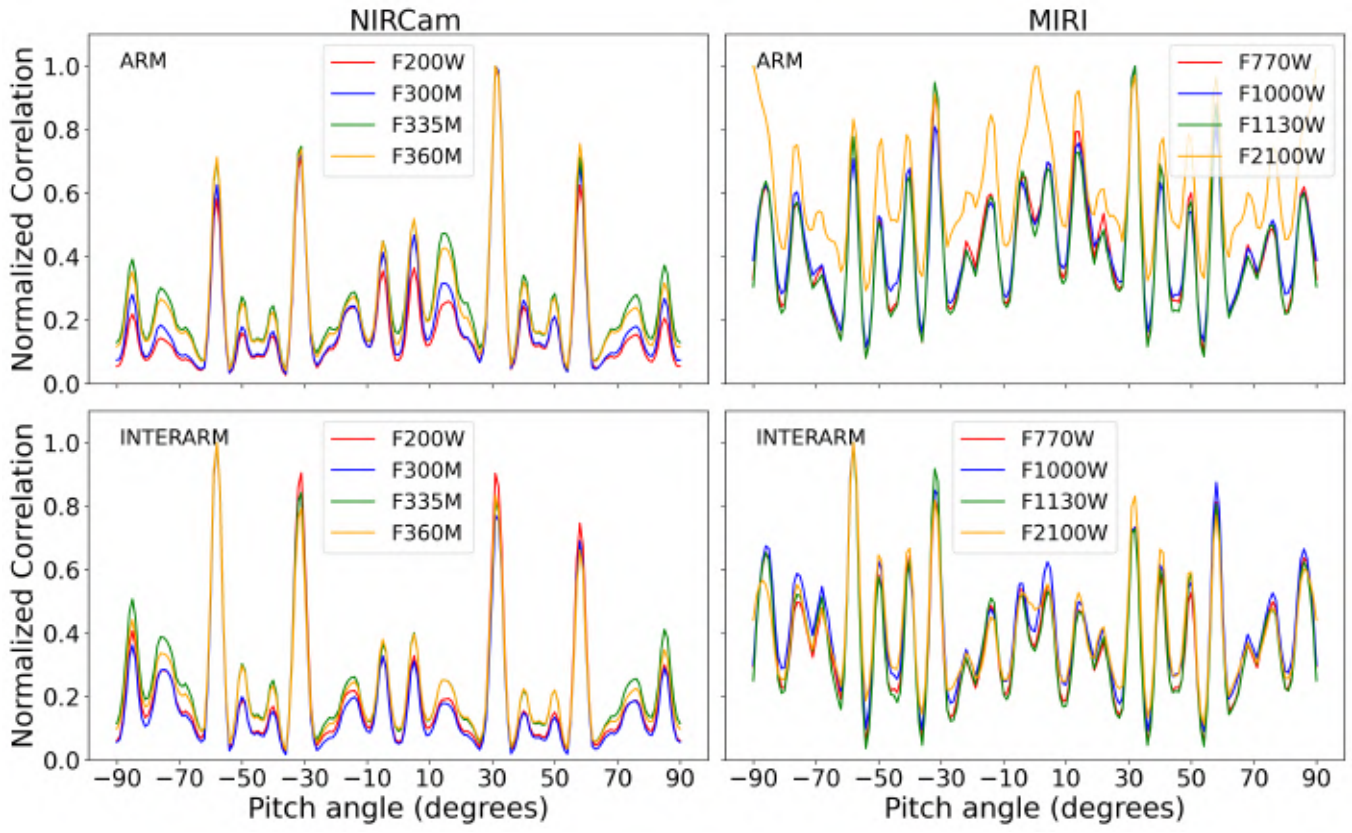

(a) Sum of the correlations calculated with $D_w$ = 0.25 for the arm and interarm regions of NGC 2835.

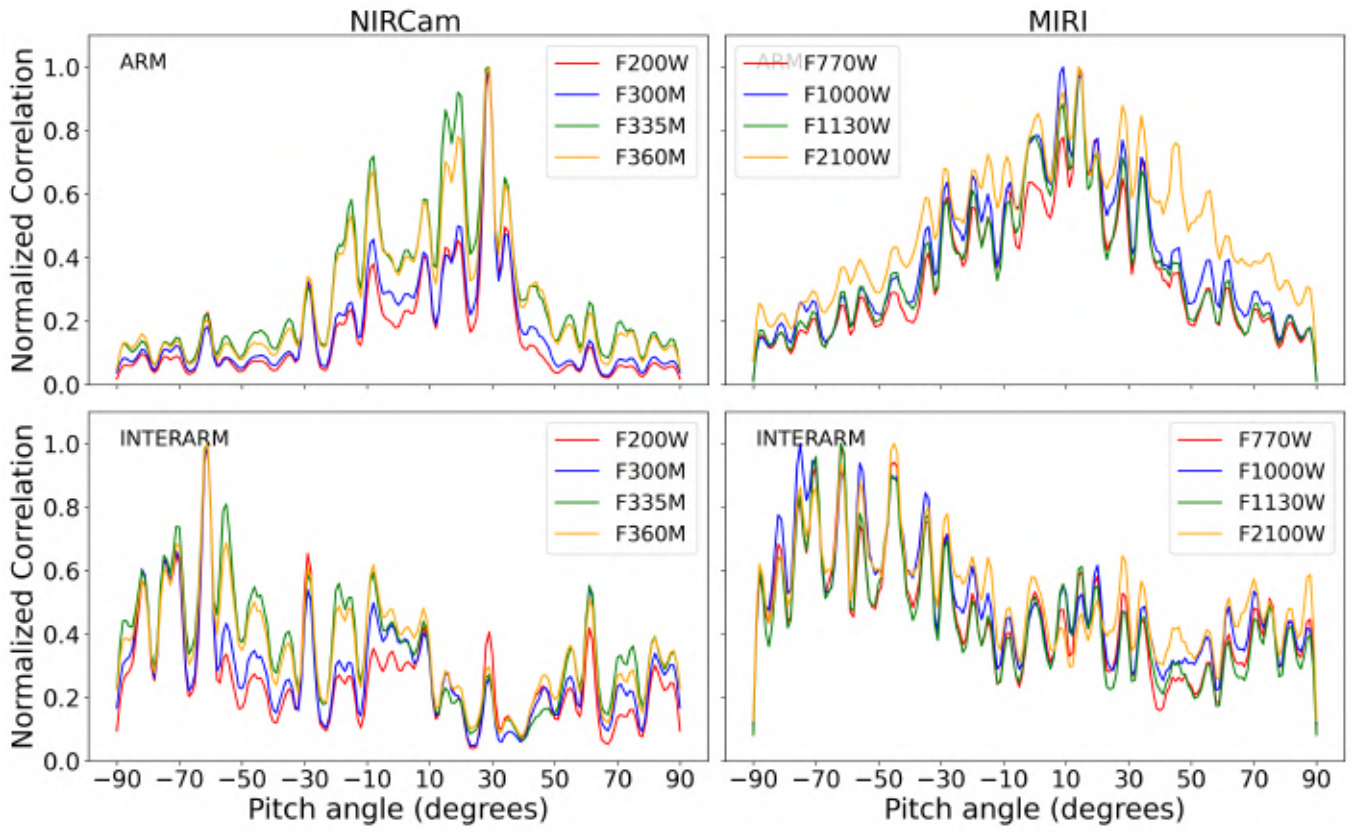

(b) Sum of the correlations calculated with $D_w$ = 0.5 for the arm and interarm regions of NGC 2835.

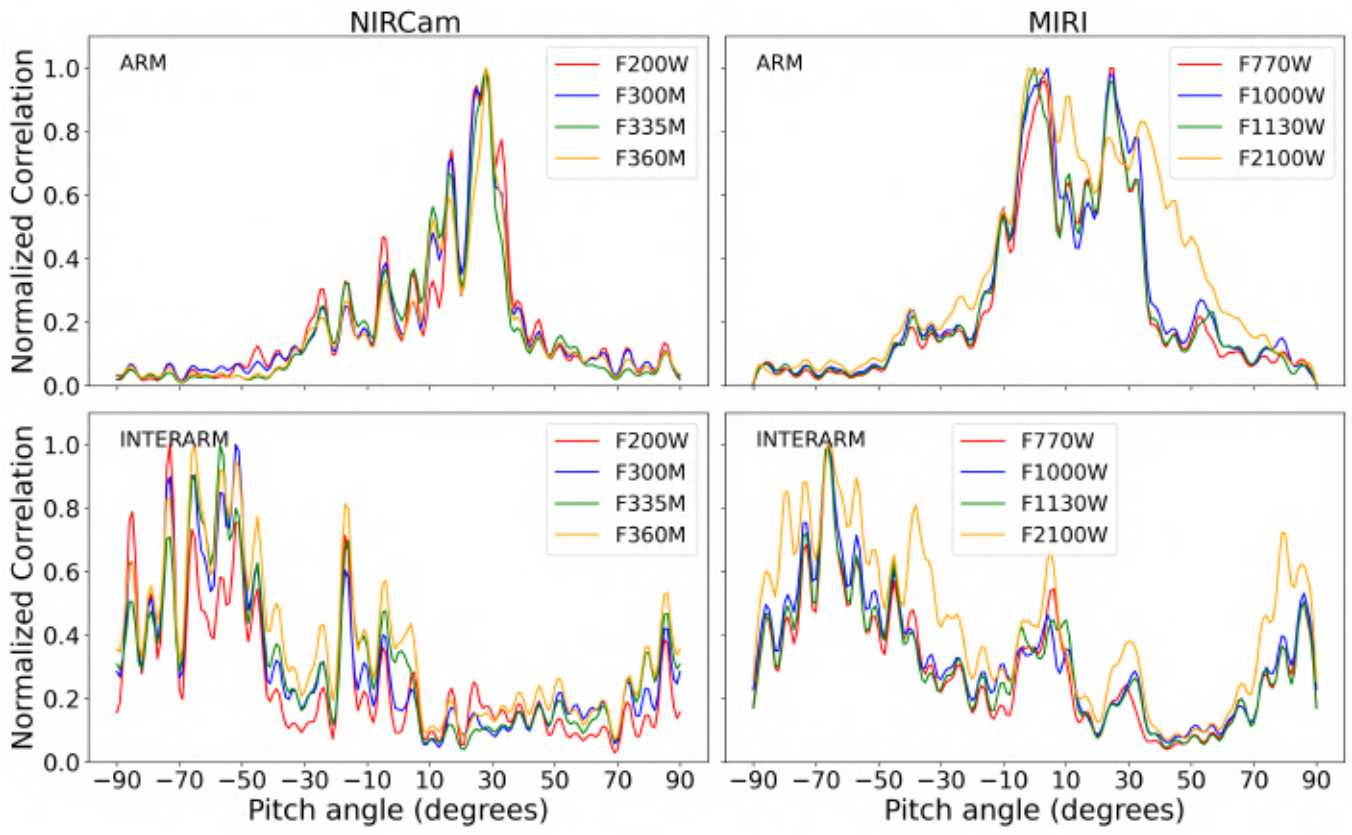

(c) Sum of the correlations calculated with $D_w$ = 0.75 for the arm and interarm regions of NGC 2835.

**Figure A12.** Results for the sum of the correlations in the analyzed regions of NGC 2835. On the left, plots from NIRCam images with filters $F200W$ (red), $F300M$ (blue), $F335M$ (green), and $F360M$ (yellow), where the upper panel corresponds to the arm region and the lower panel to the interarm region. On the right, plots from MIRI images with filters $F770W$ (red), $F1000W$ (blue), $F1130W$ (green), and $F2100W$ (yellow), following the same division between arm (upper) and interarm (lower) regions.

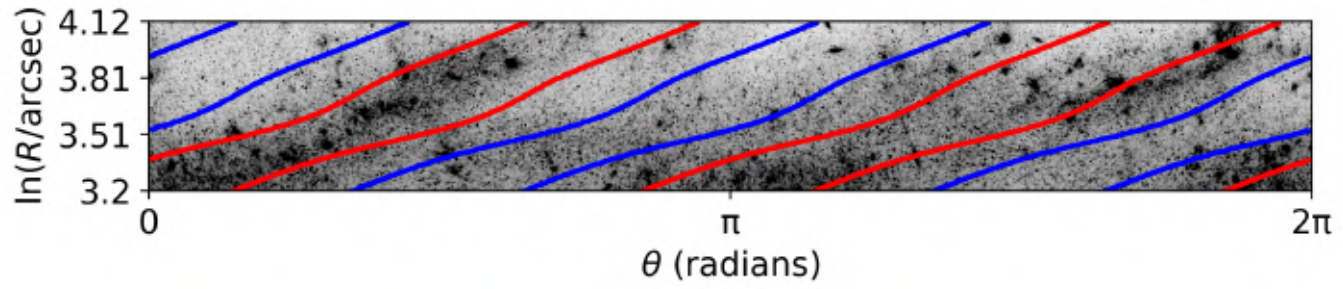

(a) Regions delimited in the $F335M$ image of the galaxy NGC 2835 from the 2D-FT/2D-IFT mask. The bar regions are located between the red lines, and the interbar regions are between the blue lines.

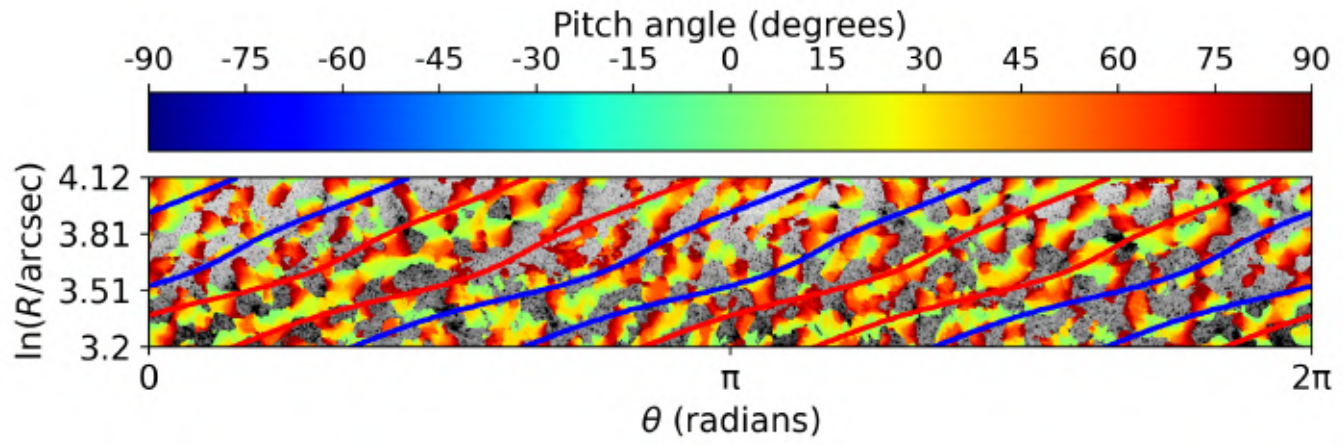

(b) Distribution of pitch angles obtained with the $D_w = 0.25$ filter.

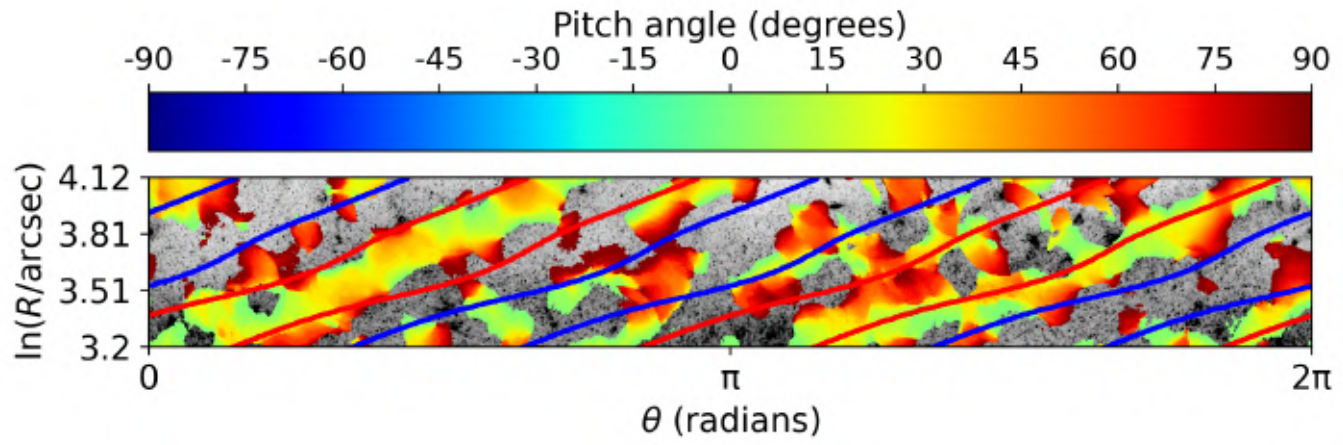

(c) Distribution of pitch angles obtained with the $D_w = 0.5$ filter.

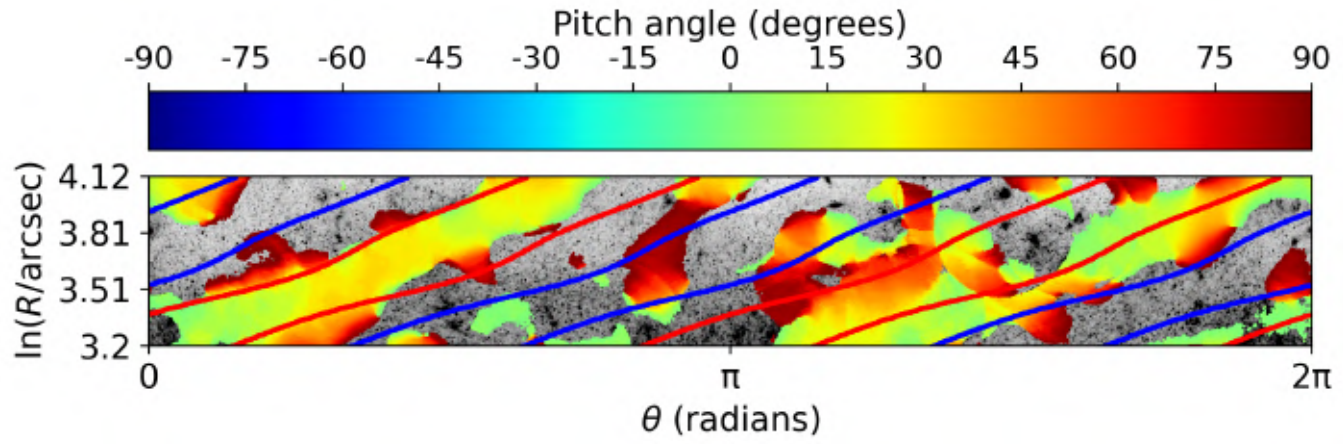

(d) Distribution of pitch angles obtained with the $D_w = 0.75$ filter.

**Figure A13.** Distribution of pitch angles obtained from the highest correlations at each position $(\ln(R/\mathrm{arcsec}), \theta)$ in the $F335M$ image of NGC 2835. The colorbar represents pitch angle values, ranging from $-90°$ (dark blue) to $90°$ (dark red). The colors of the points indicate the pitch angle associated with the maximum correlation, based on the idea that the strongest correlations occur when the structure has the same pitch angle as the filter at the corresponding position.

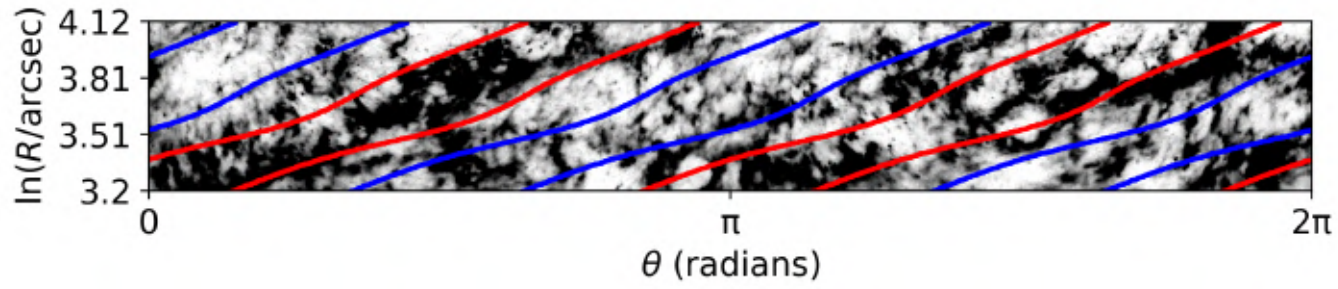

(a) Regions delimited in the $F1130W$ image of the galaxy NGC 2835 from the 2D-FT/2D-IFT mask. The bar regions are located between the red lines, and the interbar regions are between the blue lines.

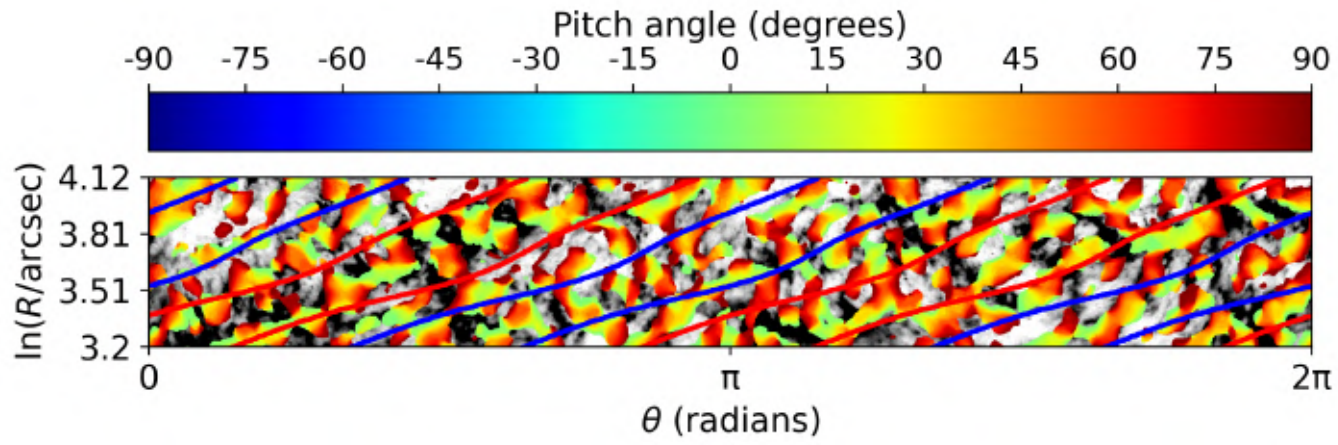

(b) Distribution of pitch angles obtained with the $D_w = 0.25$ filter.

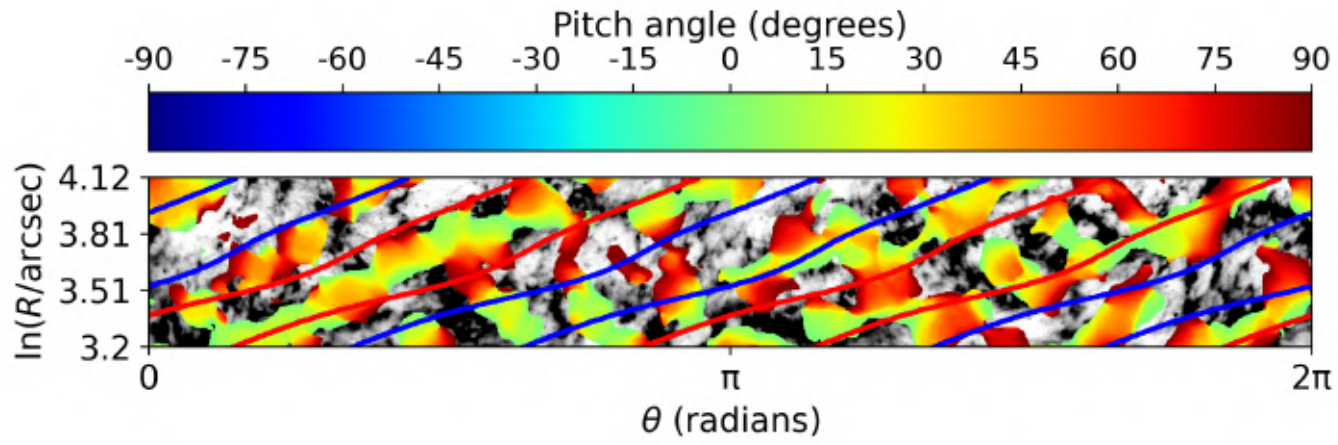

(c) Distribution of pitch angles obtained with the $D_w = 0.5$ filter.

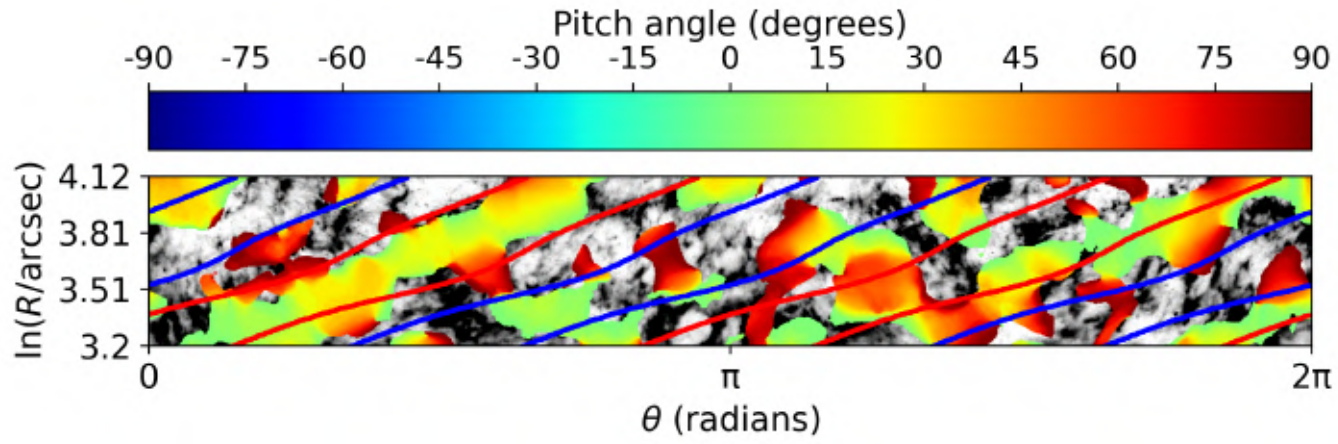

(d) Distribution of pitch angles obtained with the $D_w = 0.75$ filter.

**Figure A14.** Distribution of pitch angles obtained from the highest correlations at each position $(\ln(R/\mathrm{arcsec}), \theta)$ in the $F1130W$ image of NGC 2835. The colorbar represents pitch angle values, ranging from $-90°$ (dark blue) to $90°$ (dark red). The colors of the points indicate the pitch angle associated with the maximum correlation, based on the idea that the strongest correlations occur when the structure has the same pitch angle as the filter at the corresponding position.

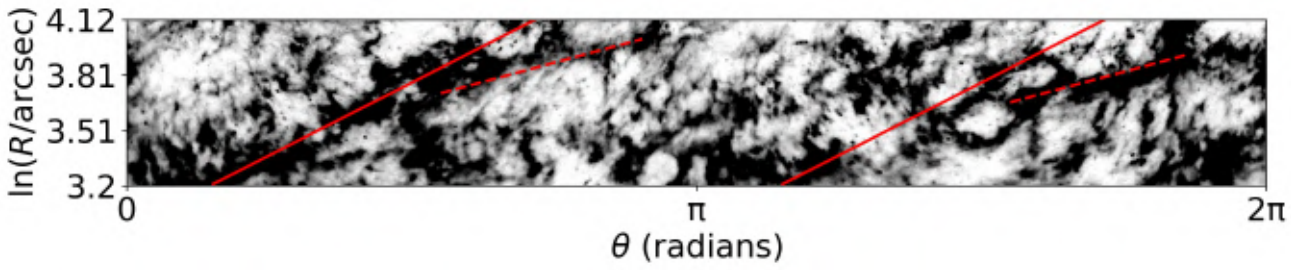

**Figure A15.** Lines overlaid on the $F1130W$ image of NGC 2835, indicating the pitch angles of the spiral structures. The solid red line corresponds to $27°$, and the dashed red line to $15°$.

### A4 NGC 3351

NGC 3351 shows a gradual change in the orientation of its curved dust lanes, with representative values near $\alpha \sim -46^{\circ}$ and $\alpha \sim -62^{\circ}$. The corresponding spatial maps show that these values trace different portions of the same extended structure.

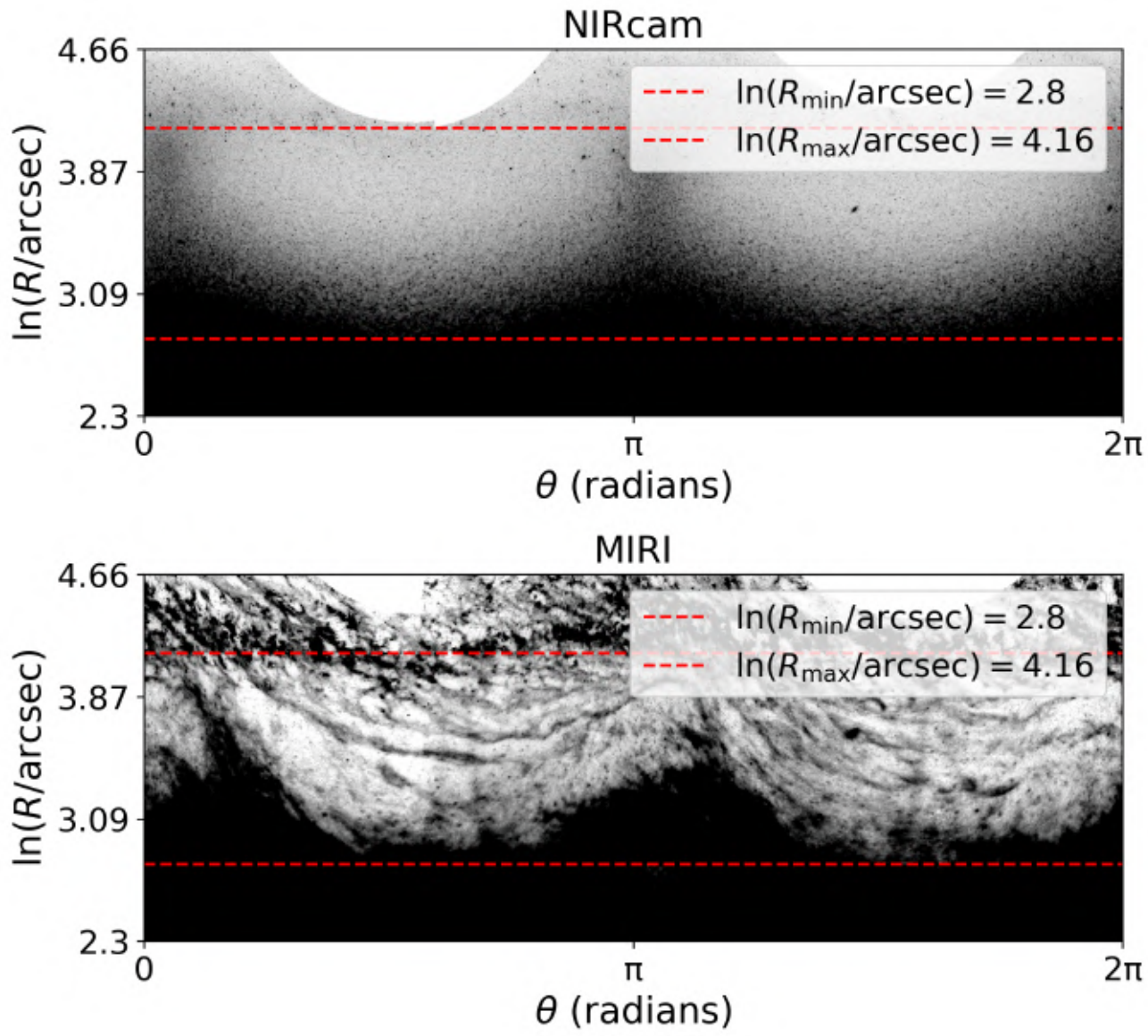

**Figure A16.** Diagrams $\ln(R/\text{arcsec})$ vs. $\theta$ for the $F335M$ (NIRCam) and $F1130W$ (MIRI) images of NGC 3351. The dashed red lines indicate the values of $\ln(R_{min}/\text{arcsec})$ and $\ln(R_{max}/\text{arcsec})$ adopted in the analysis.

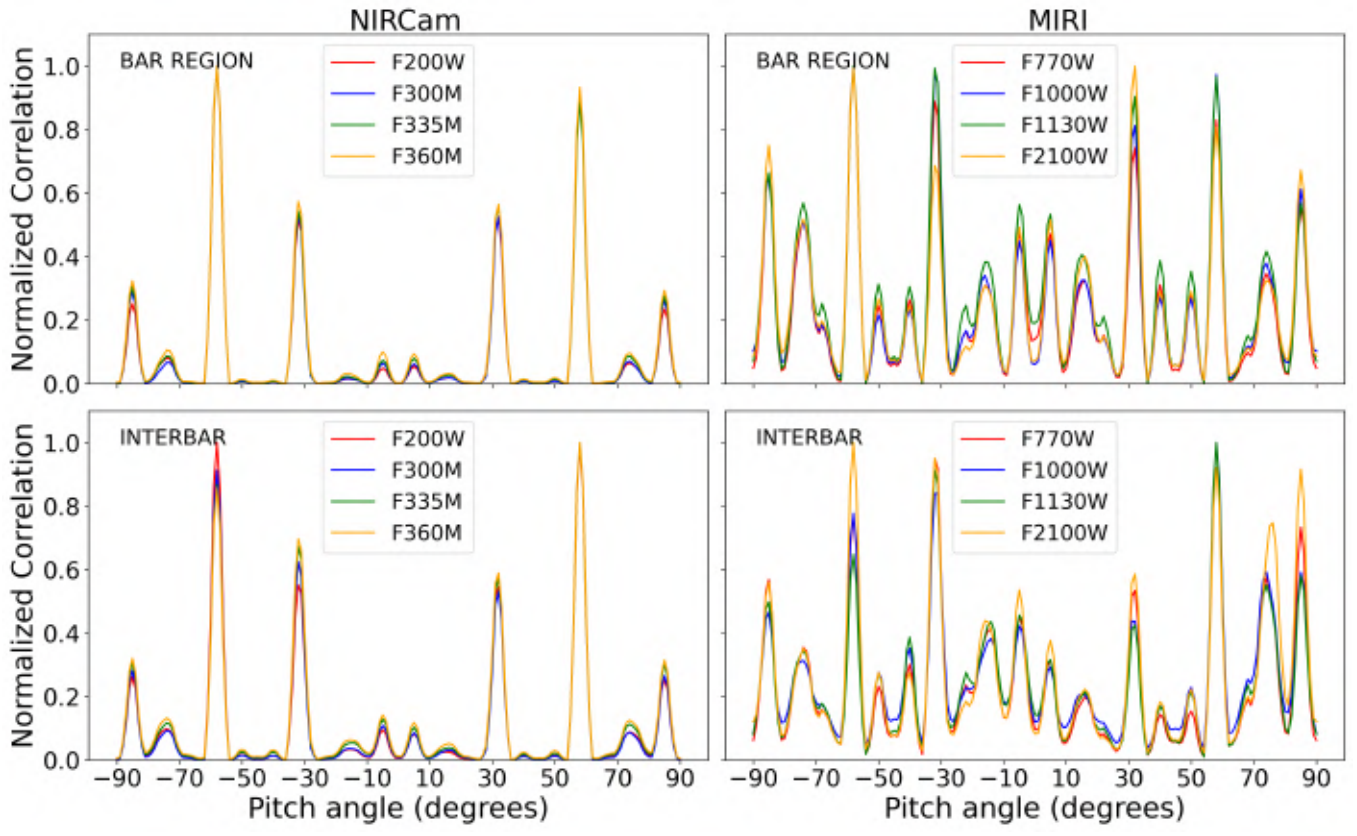

(a) Sum of the correlations calculated with $D_w = 0.25$ for the bar and interbar regions of NGC 3351.

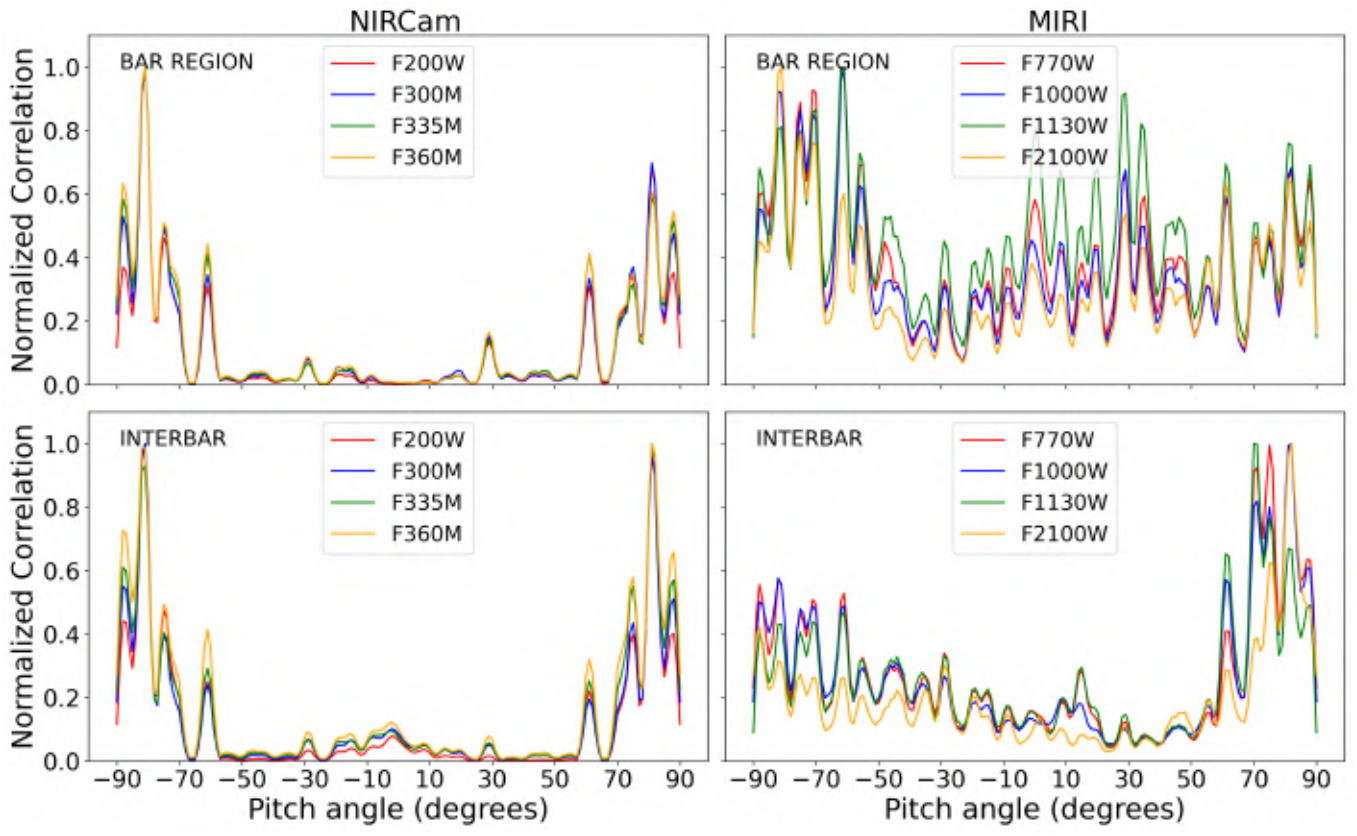

(b) Sum of the correlations calculated with $D_w = 0.5$ for the bar and interbar regions of NGC 3351.

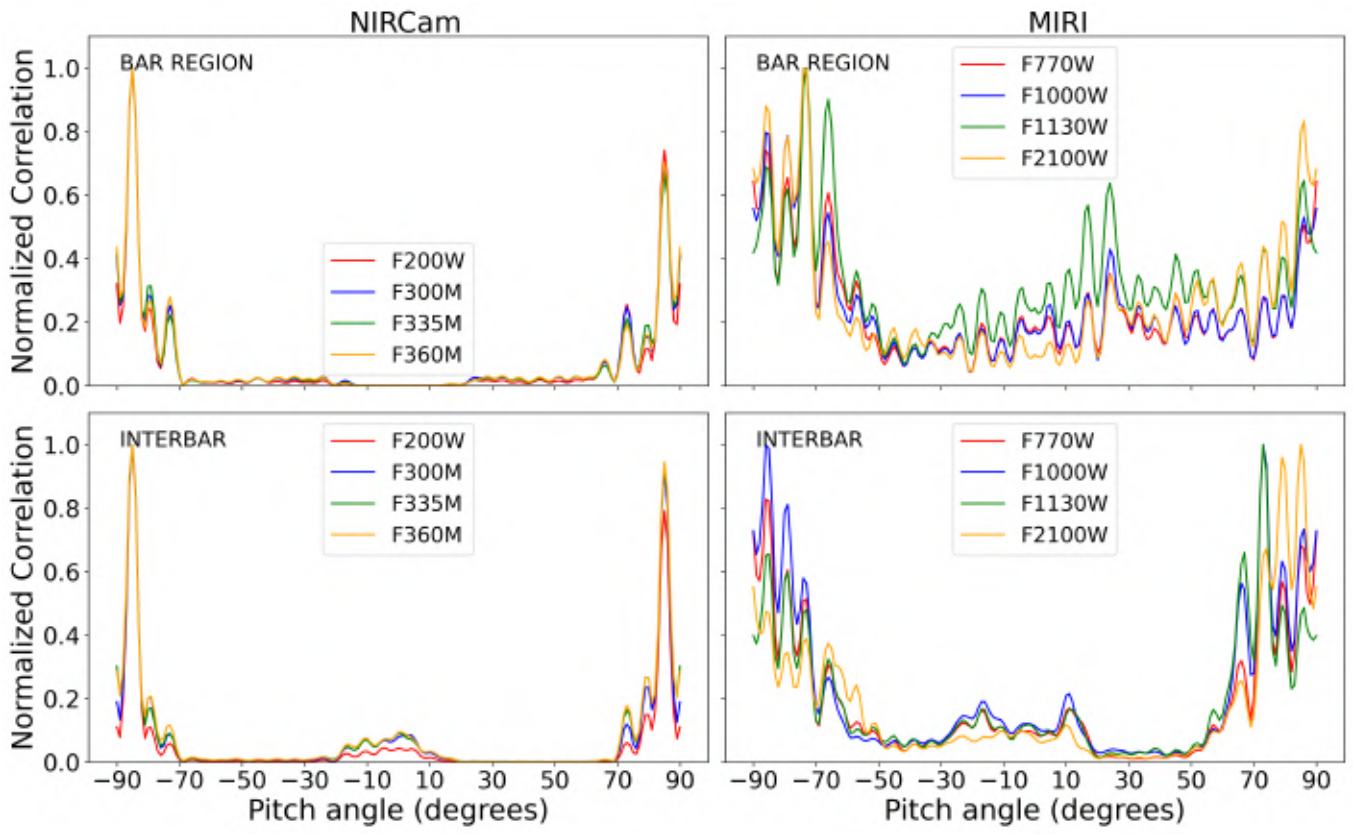

(c) Sum of the correlations calculated with $D_w = 0.75$ for the bar and interbar regions of NGC 3351.

**Figure A17.** Results for the sum of the correlations in the analyzed regions of NGC 3351. On the left, plots from NIRCam images with filters $F200W$ (red), $F300M$ (blue), $F335M$ (green), and $F360M$ (yellow), where the upper panel corresponds to the bar region and the lower panel to the interbar region. On the right, plots from MIRI images with filters $F770W$ (red), $F1000W$ (blue), $F1130W$ (green), and $F2100W$ (yellow), following the same division between bar (upper) and interbar (lower) regions.

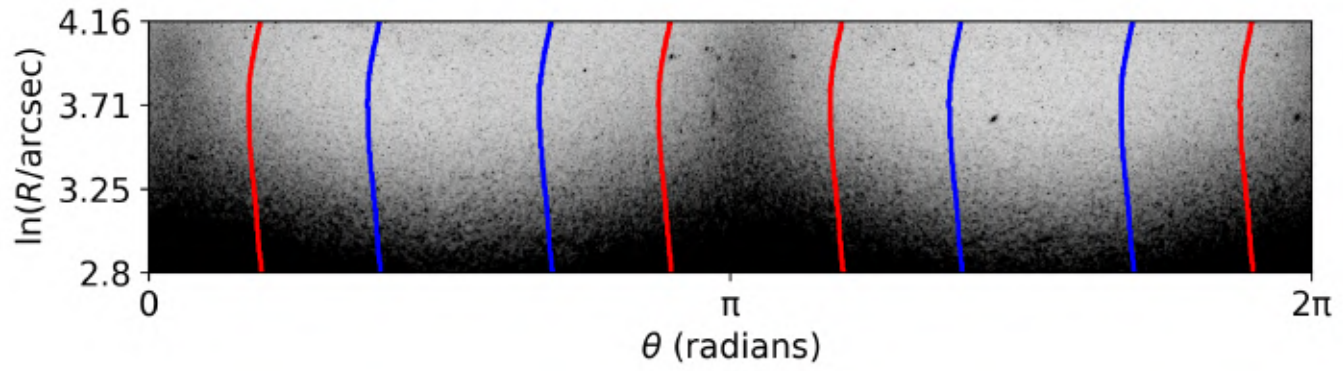

(a) Regions delimited in the $F335M$ image of the galaxy NGC 3351 from the 2D-FT/2D-IFT mask. The bar regions are located between the red lines, and the interbar regions are between the blue lines.

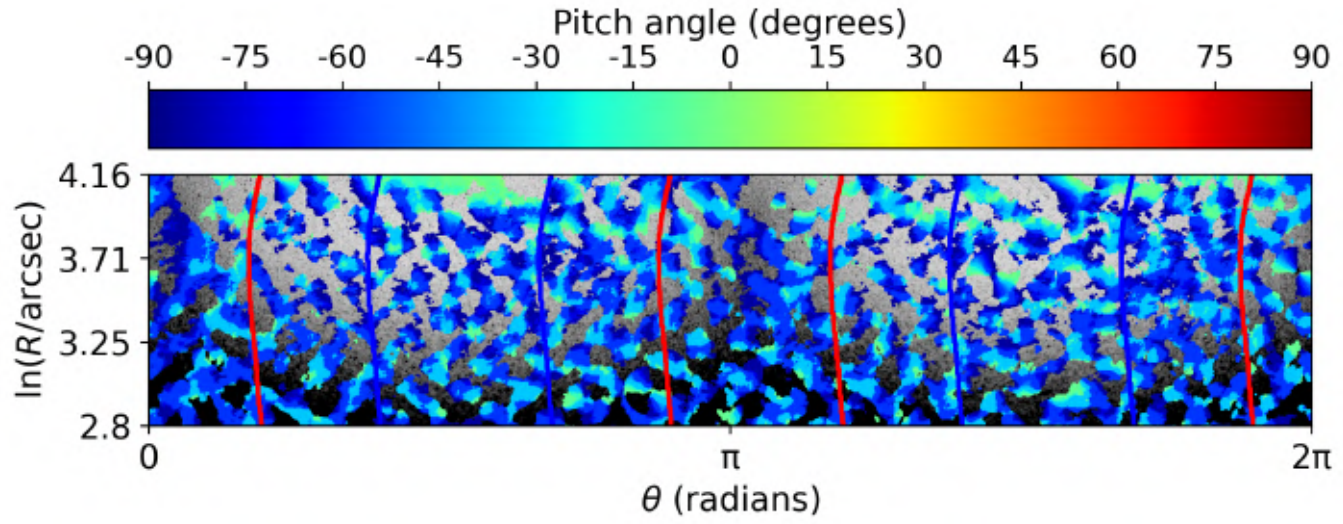

(b) Distribution of pitch angles obtained with the $D_w = 0.25$ filter.

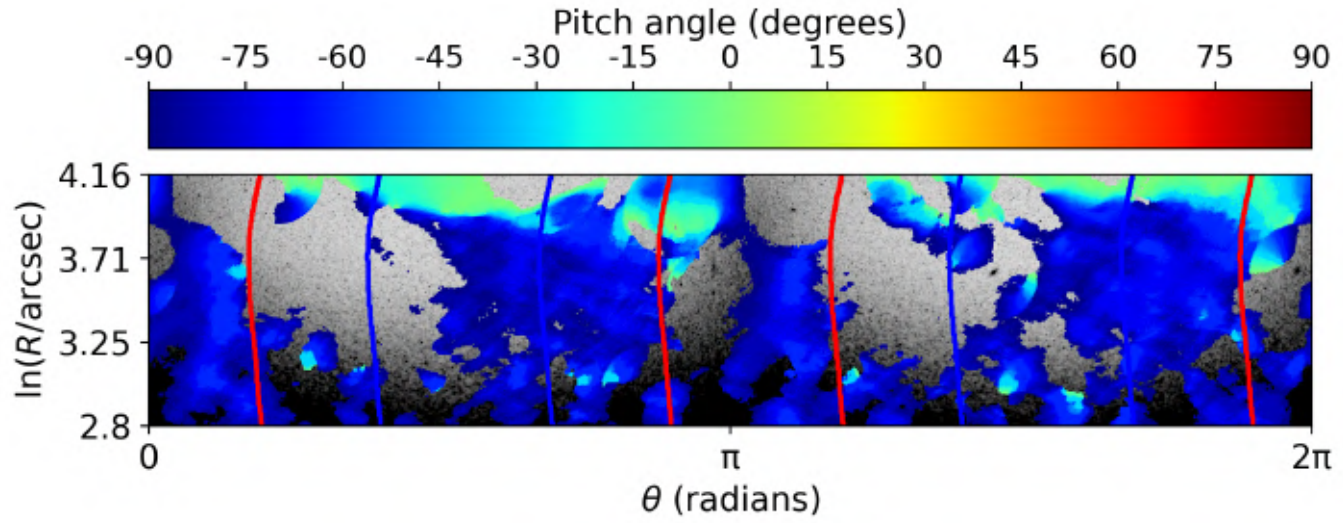

(c) Distribution of pitch angles obtained with the $D_w = 0.5$ filter.

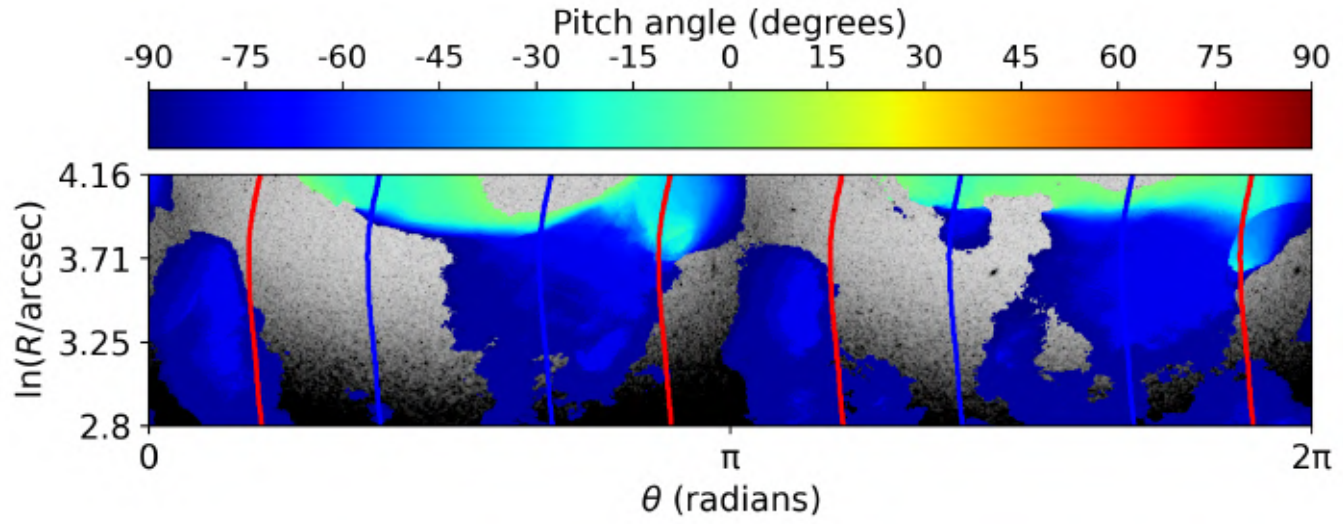

(d) Distribution of pitch angles obtained with the $D_w = 0.75$ filter.

**Figure A18.** Distribution of pitch angles obtained from the highest correlations at each position $(\ln(R/\mathrm{arcsec}), \theta)$ in the $F335M$ image of NGC 3351. The colorbar represents pitch angle values, ranging from $-90°$ (dark blue) to $90°$ (dark red). The colors of the points indicate the pitch angle associated with the maximum correlation, based on the idea that the strongest correlations occur when the structure has the same pitch angle as the filter at the corresponding position.

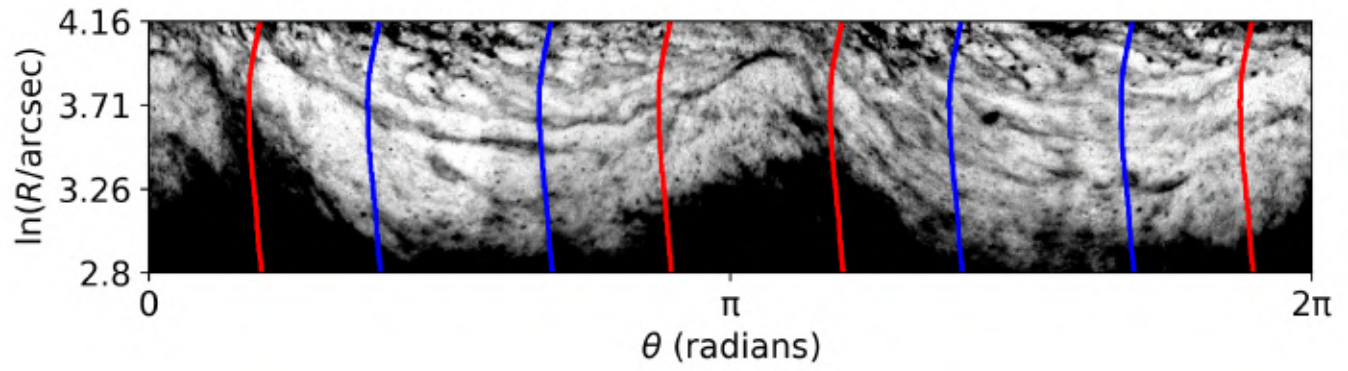

(a) Regions delimited in the $F1130W$ image of the galaxy NGC 3351 from the 2D-FT/2D-IFT mask. The bar regions are located between the red lines, and the interbar regions are between the blue lines.

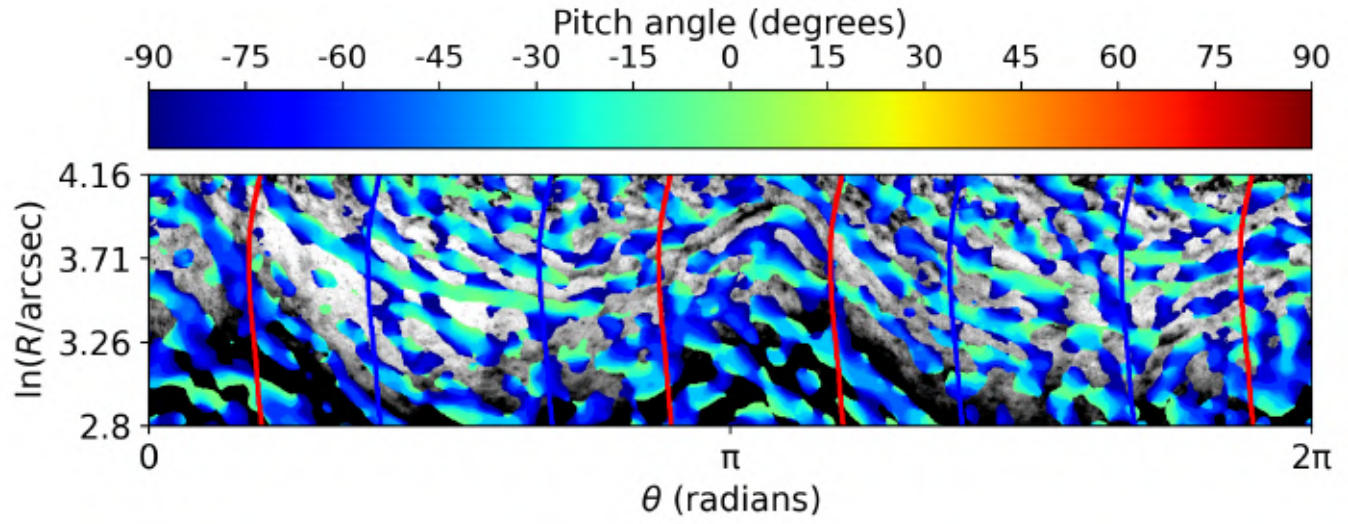

(b) Distribution of pitch angles obtained with the $D_w = 0.25$ filter.

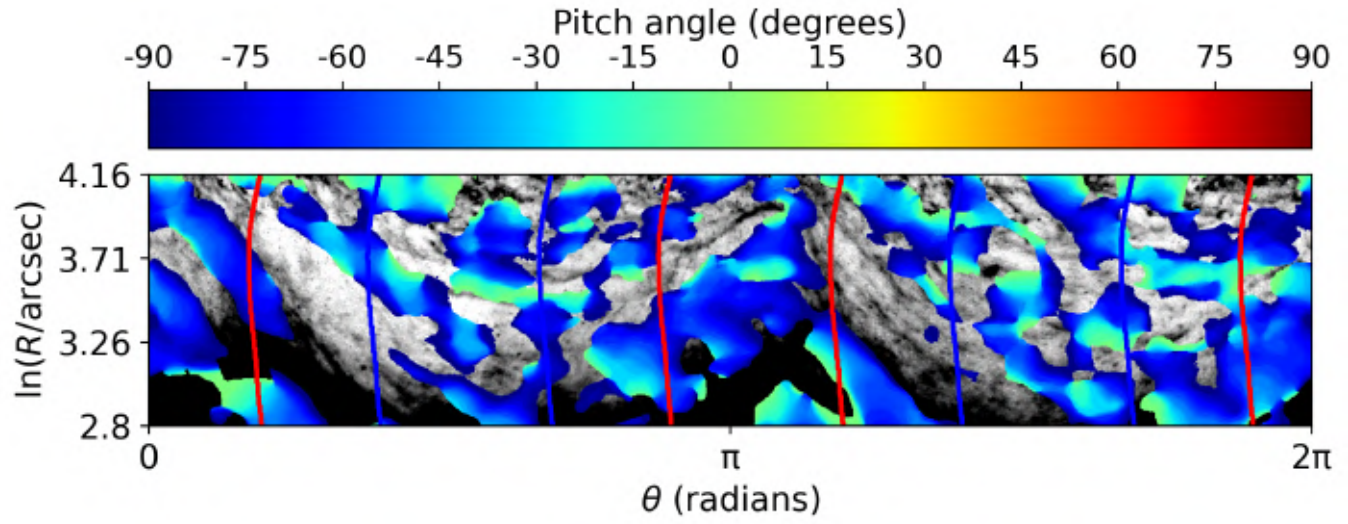

(c) Distribution of pitch angles obtained with the $D_w = 0.5$ filter.

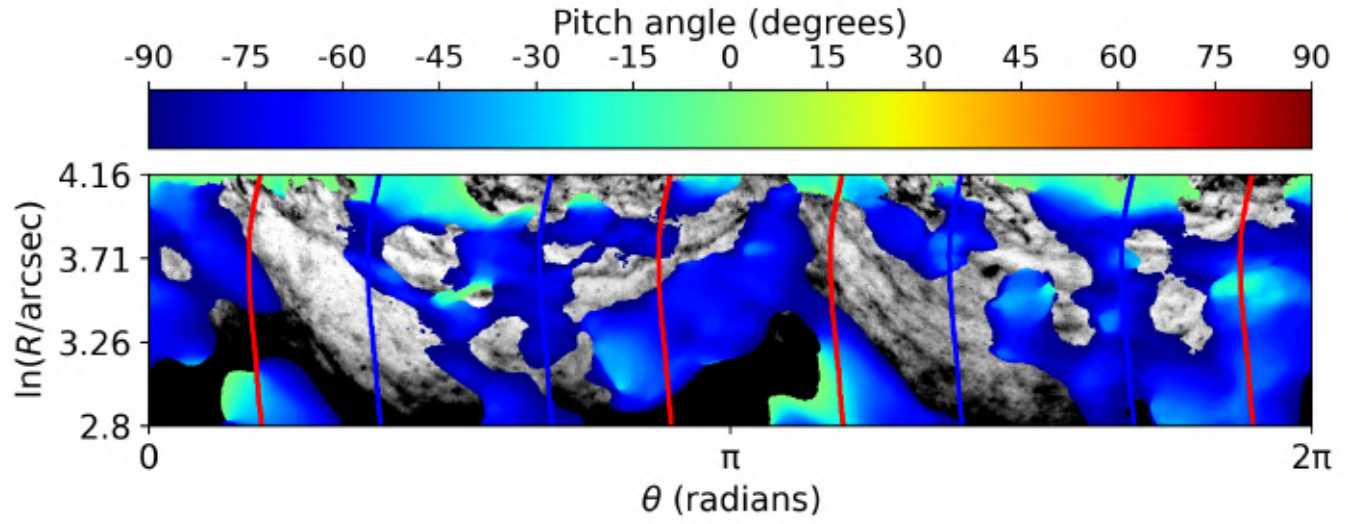

(d) Distribution of pitch angles obtained with the $D_w = 0.75$ filter.

**Figure A19.** Distribution of pitch angles obtained from the highest correlations at each position $(\ln(R/\mathrm{arcsec}), \theta)$ in the $F1130W$ image of NGC 3351. The colorbar represents pitch angle values, ranging from $-90°$ (dark blue) to $90°$ (dark red). The colors of the points indicate the pitch angle associated with the maximum correlation, based on the idea that the strongest correlations occur when the structure has the same pitch angle as the filter at the corresponding position.

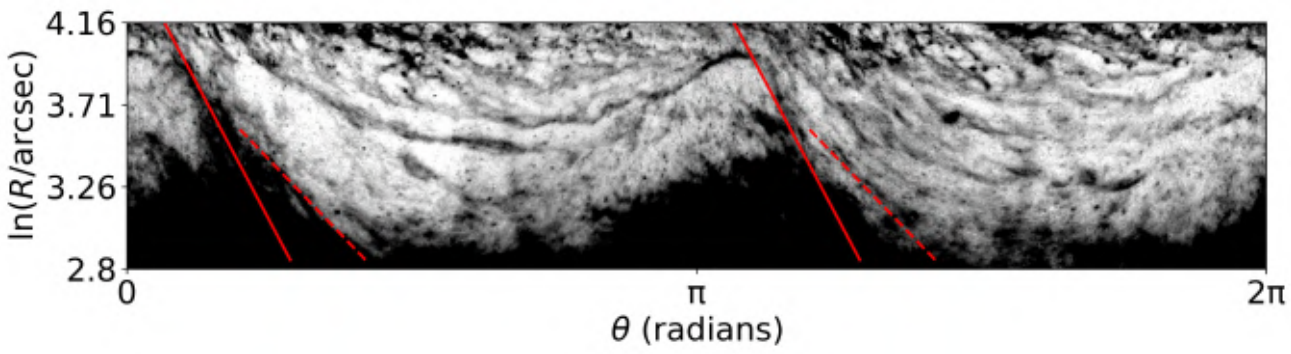

**Figure A20.** Lines overlaid on the $F1130W$ image of NGC 3351, indicating the curvature of the dust lanes. The solid red line corresponds to $-62°$, and the dashed red line to $-46°$.

### A5 NGC 7496

NGC 7496 exhibits a comparatively complex bar–spiral transition. Representative pitch-angle components near $\alpha \sim 72^{\circ}$ and $\alpha \sim 48^{\circ}$ describe, respectively, the straighter portion of the dust lane and the more curved region approaching the onset of the spiral structure. The multi-scale maps below illustrate this variation.

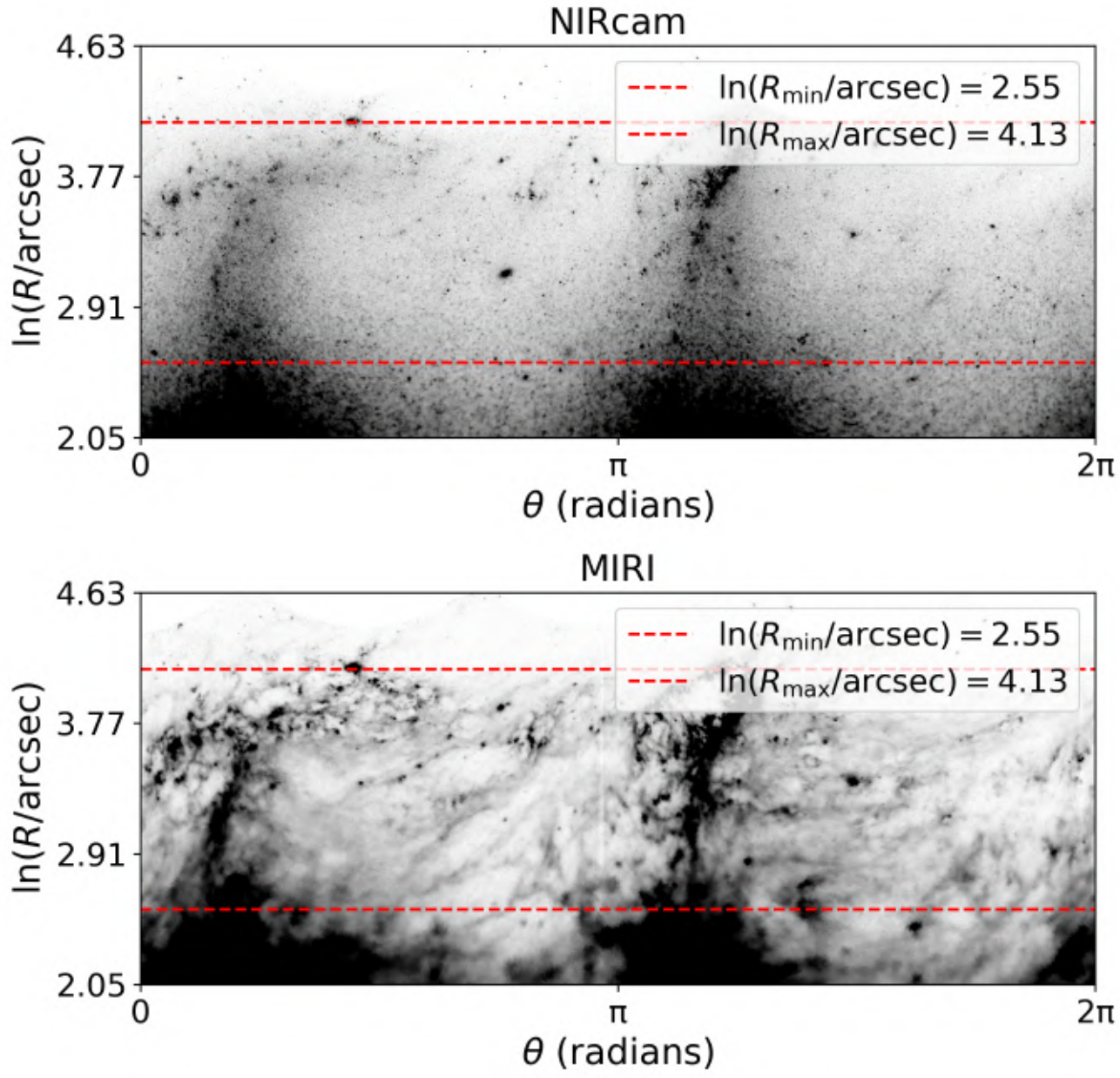

**Figure A21.** Diagrams $\ln(R/\mathrm{arcsec})$ vs. $\theta$ for the $F335M$ (NIRCam) and $F1130W$ (MIRI) images of NGC 7496. The dashed red lines indicate the values of $\ln(R_{min}/\mathrm{arcsec})$ and $\ln(R_{max}/\mathrm{arcsec})$ adopted in the analysis.

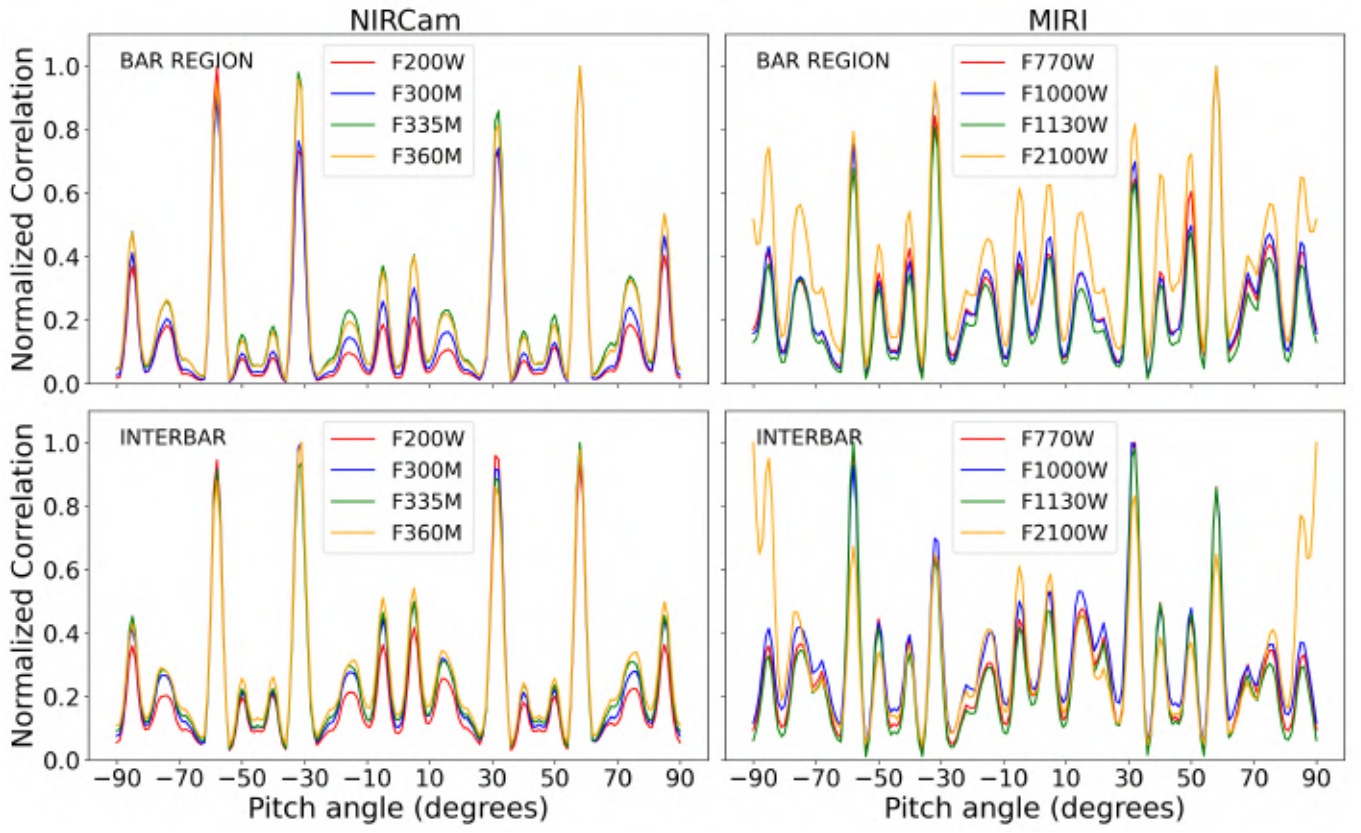

(a) Sum of the correlations calculated with $D_w = 0.25$ for the bar and interbar regions of NGC 7496.

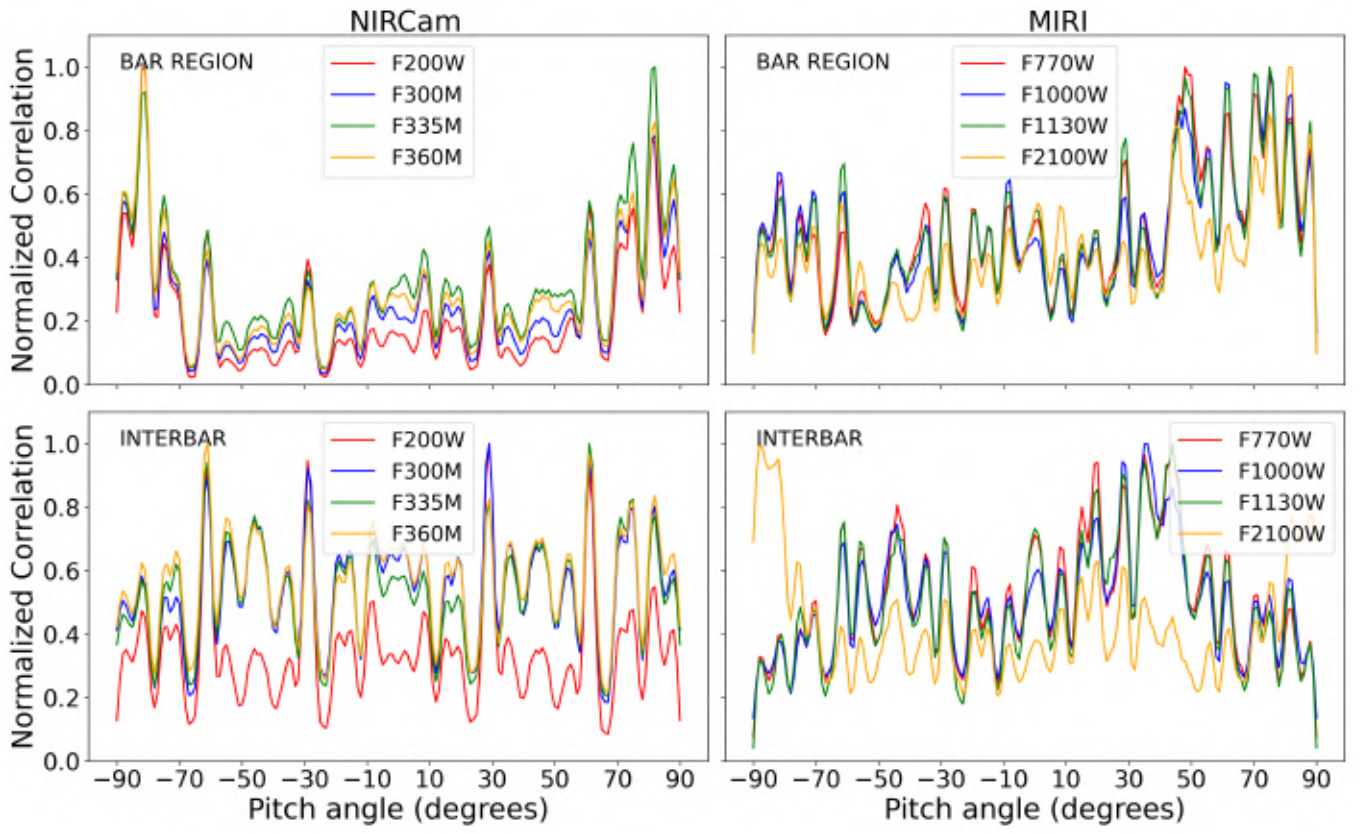

(b) Sum of the correlations calculated with $D_w = 0.5$ for the bar and interbar regions of NGC 7496.

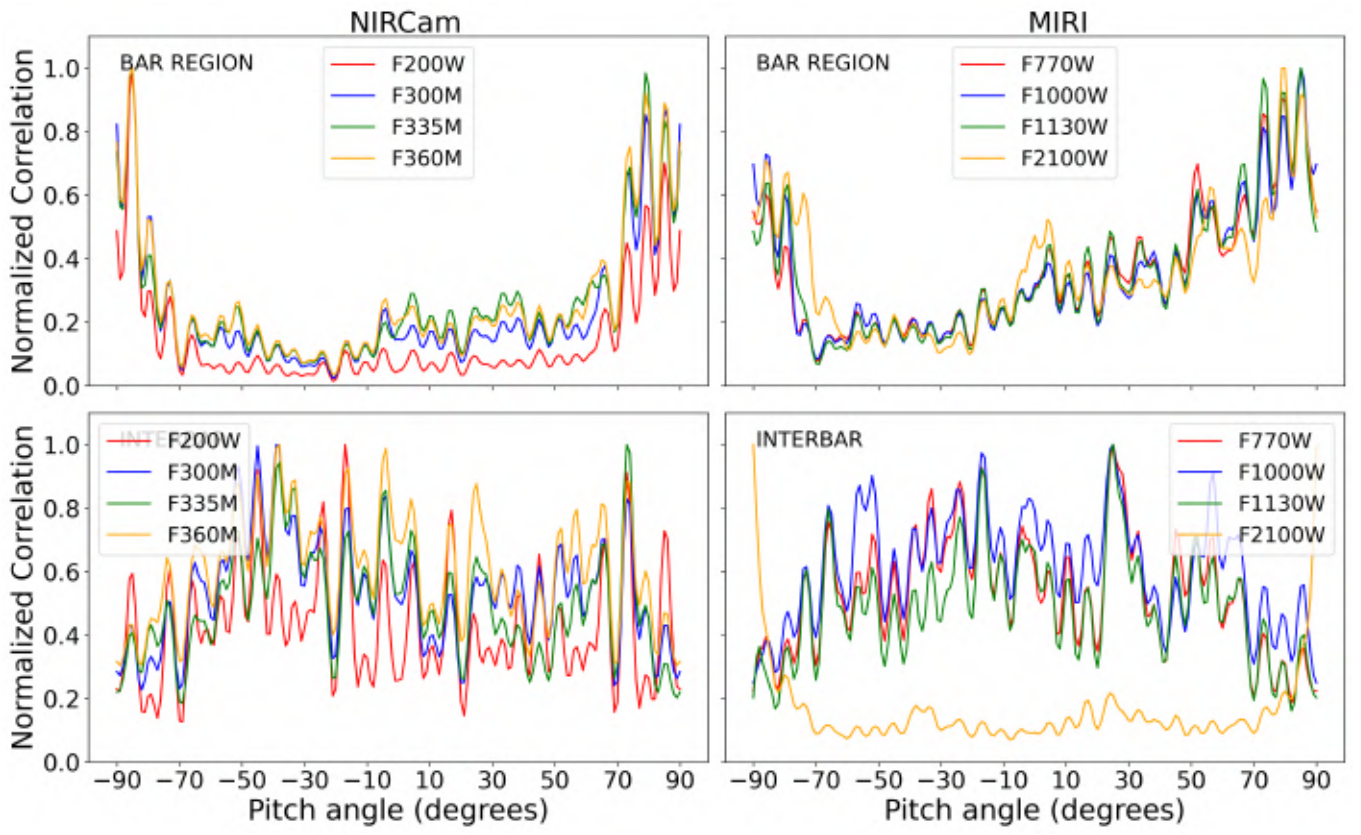

(c) Sum of the correlations calculated with $D_w = 0.75$ for the bar and interbar regions of NGC 7496.

**Figure A22.** Results for the sum of the correlations in the analyzed regions of NGC 7496. On the left, plots from NIRCam images with filters $F200W$ (red), $F300M$ (blue), $F335M$ (green), and $F360M$ (yellow), where the upper panel corresponds to the bar region and the lower panel to the interbar region. On the right, plots from MIRI images with filters $F770W$ (red), $F1000W$ (blue), $F1130W$ (green), and $F2100W$ (yellow), following the same division between bar (upper) and interbar (lower) regions.

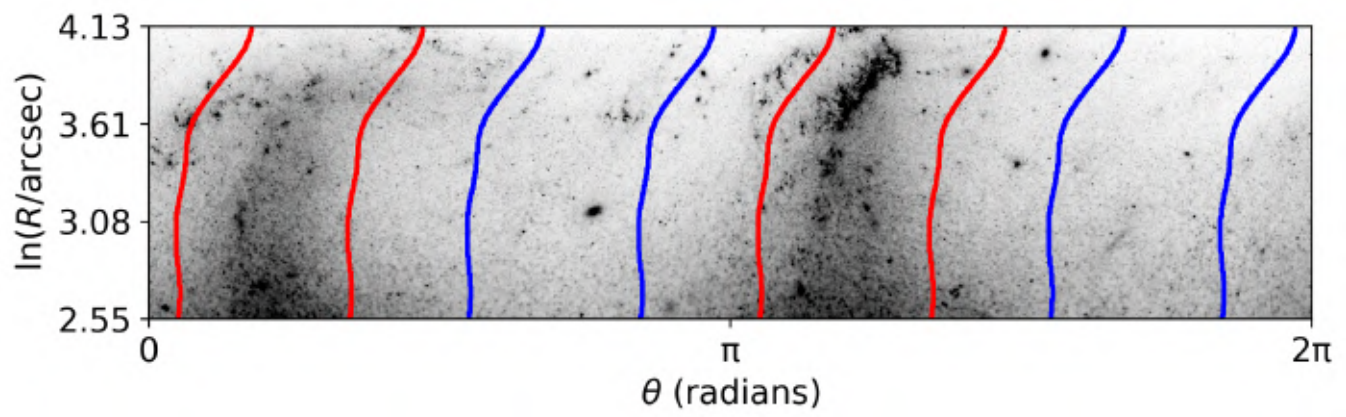

(a) Regions delimited in the $F335M$ image of the galaxy NGC 7496 from the 2D-FT/2D-IFT mask. The bar regions are located between the red lines, and the interbar regions are between the blue lines.

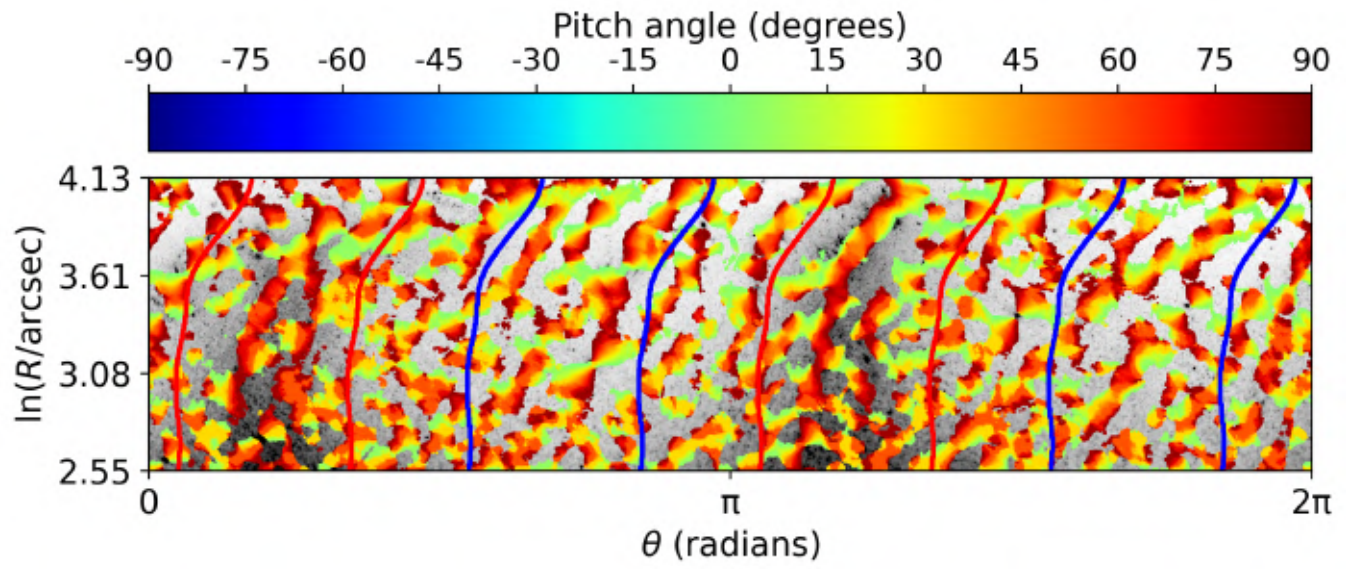

(b) Distribution of pitch angles obtained with the $D_w = 0.25$ filter.

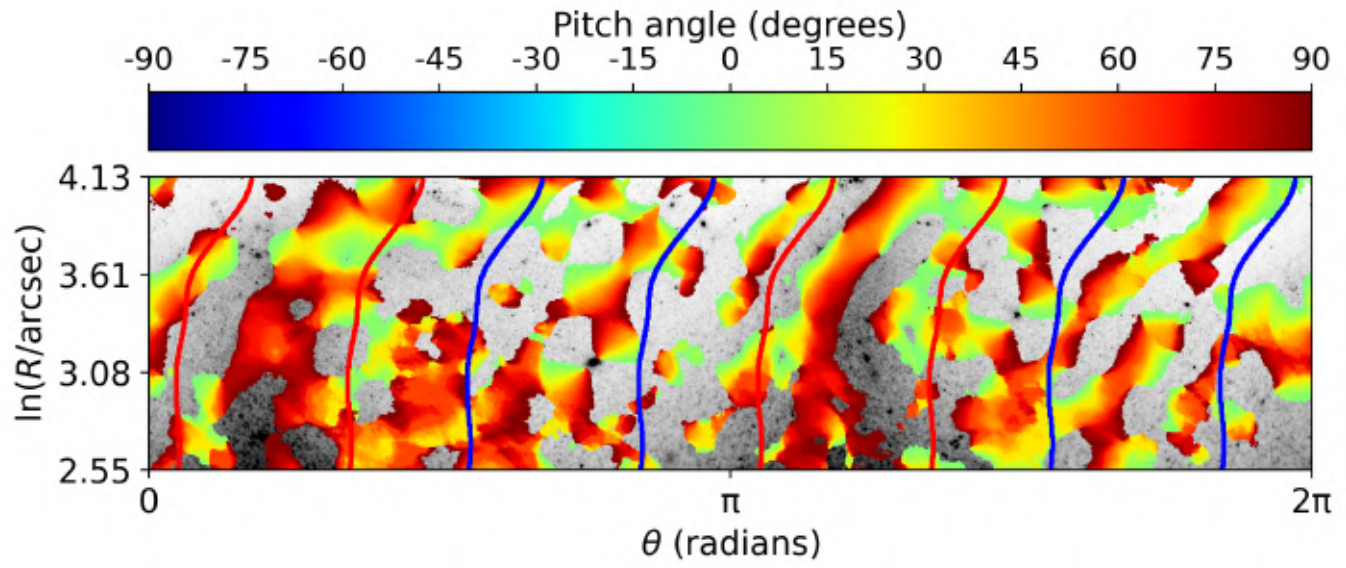

(c) Distribution of pitch angles obtained with the $D_w = 0.5$ filter.

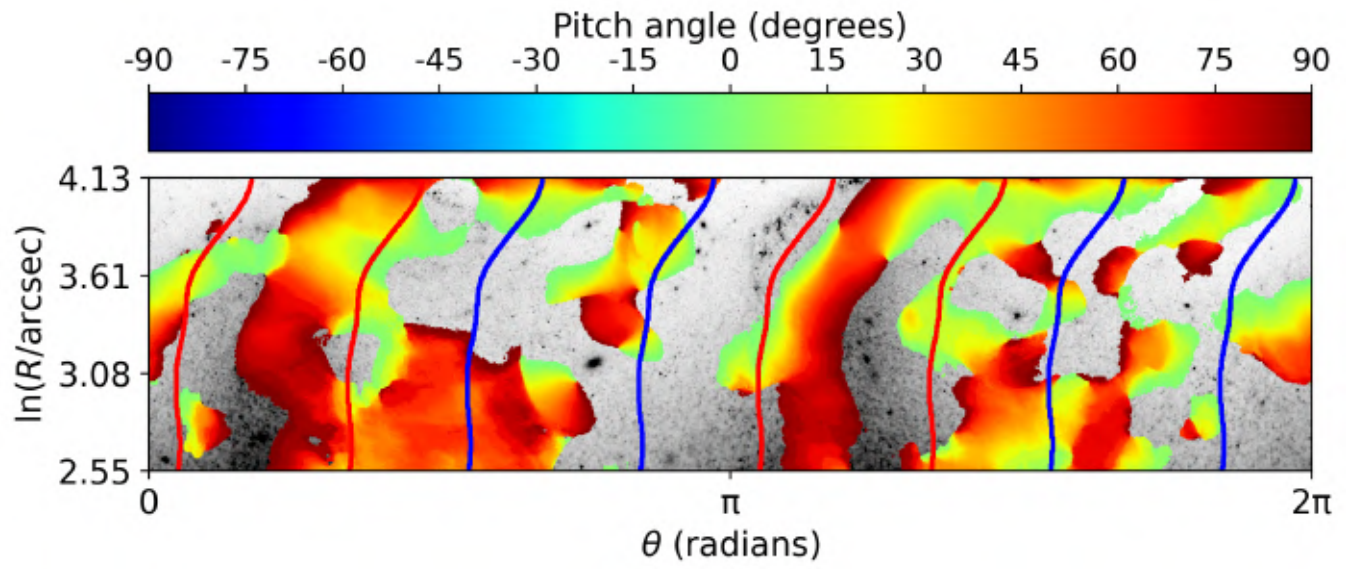

(d) Distribution of pitch angles obtained with the $D_w = 0.75$ filter.

**Figure A23.** Distribution of pitch angles obtained from the highest correlations at each position $(\ln(R/\text{arcsec}), \theta)$ in the $F335M$ image of NGC 7496. The colorbar represents pitch angle values, ranging from $-90°$ (dark blue) to $90°$ (dark red). The colors of the points indicate the pitch angle associated with the maximum correlation, based on the idea that the strongest correlations occur when the structure has the same pitch angle as the filter at the corresponding position.

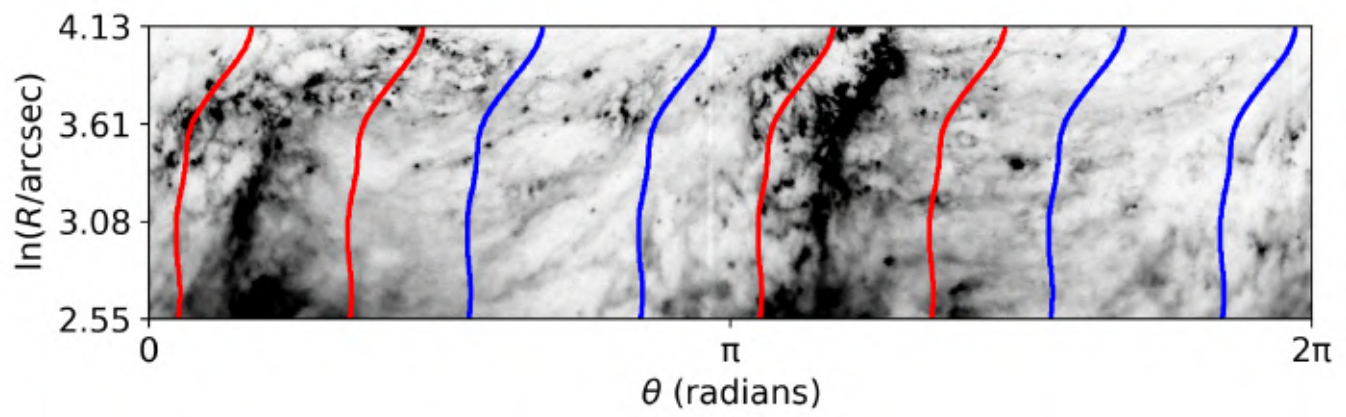

(a) Regions delimited in the $F1130W$ image of the galaxy NGC 7496 from the 2D-FT/2D-IFT mask. The bar regions are located between the red lines, and the interbar regions are between the blue lines.

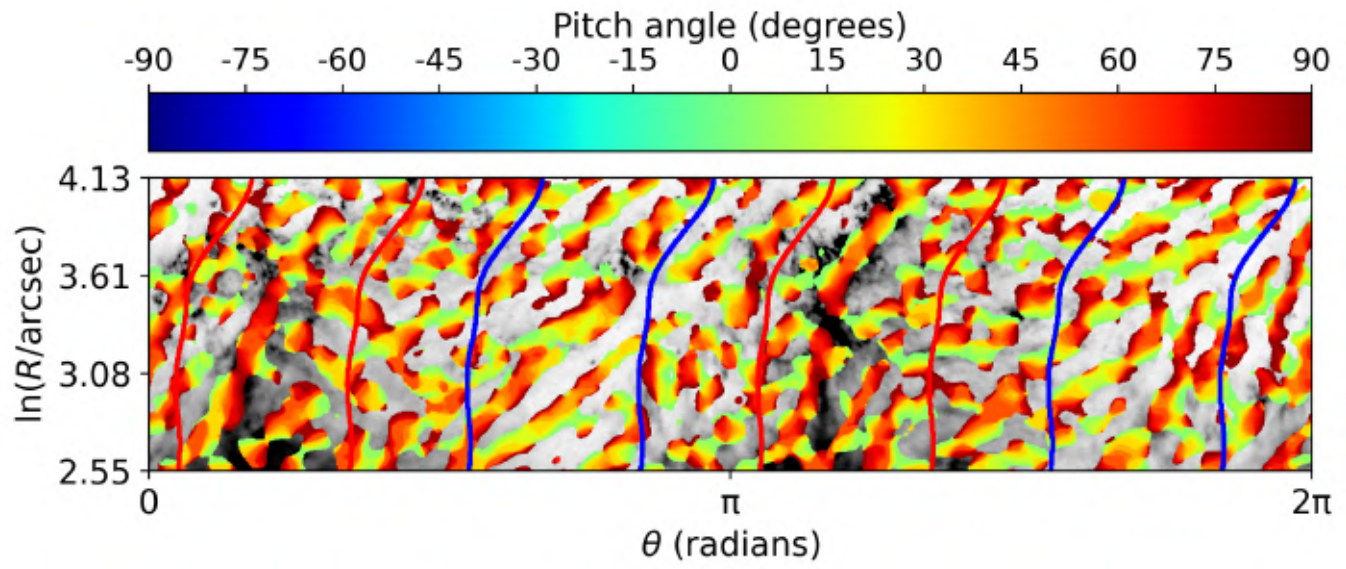

(b) Distribution of pitch angles obtained with the $D_w = 0.25$ filter.

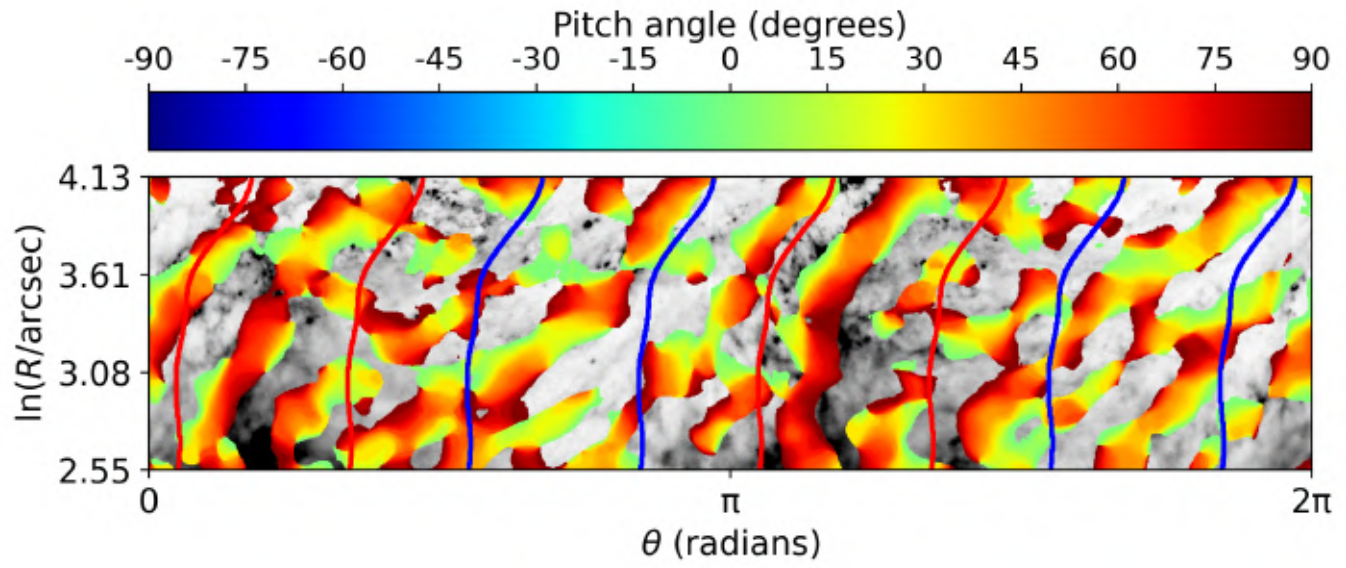

(c) Distribution of pitch angles obtained with the $D_w = 0.5$ filter.

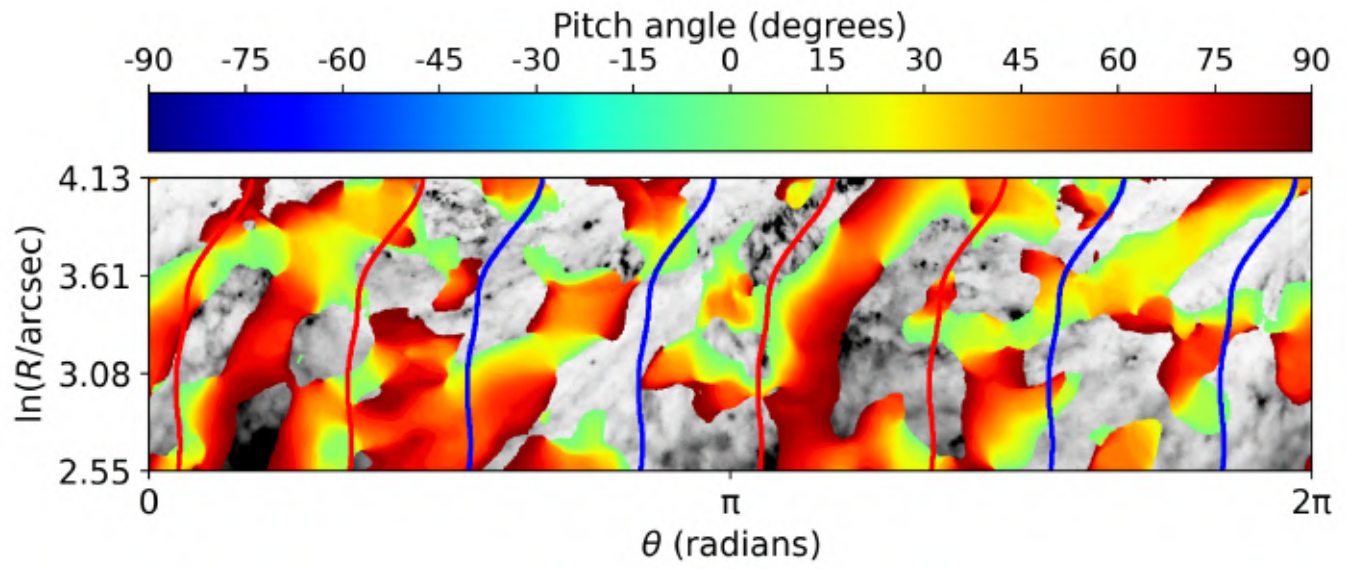

(d) Distribution of pitch angles obtained with the $D_w = 0.75$ filter.

**Figure A24.** Distribution of pitch angles obtained from the highest correlations at each position $(\ln(R/\text{arcsec}), \theta)$ in the $F1130W$ image of NGC 7496. The colorbar represents pitch angle values, ranging from $-90°$ (dark blue) to $90°$ (dark red). The colors of the points indicate the pitch angle associated with the maximum correlation, based on the idea that the strongest correlations occur when the structure has the same pitch angle as the filter at the corresponding position.

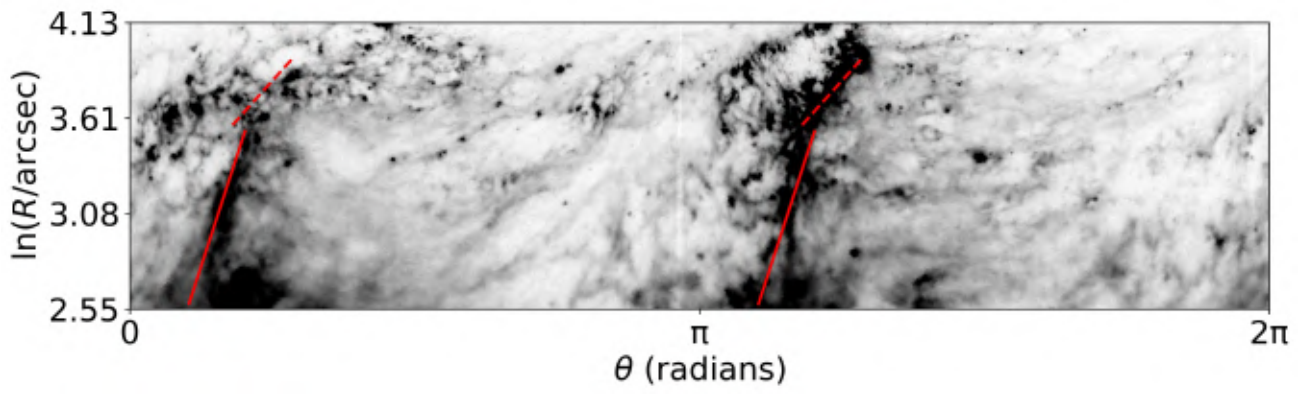

**Figure A25.** Lines overlaid on the $F1130W$ image of NGC 7496, indicating the curvature of the dust lanes. The solid red line corresponds to 72°, and the dashed red line to 48°.